\documentclass[11pt]{article}
\usepackage{amssymb,amsthm,amsmath,float}

\usepackage[total={6.5in,8.75in}, top=1.2in, left=0.9in, includefoot]{geometry}
\usepackage{graphicx}
\usepackage[pdftex,bookmarks, colorlinks, citecolor =blue, breaklinks]{hyperref}
\usepackage{url}

\newcommand{\Eq}[1]{(\ref{eq:#1})}

\newcommand{\Sec}[1]{\S \ref{sec:#1}}
\newcommand{\Fig}[1]{Fig.~\ref{fig:#1}}
\newcommand{\Tbl}[1]{Table~\ref{tbl:#1}}

\newcommand{\InsertFig}[4]
{\begin{figure}[h!t]
       \centerline{
         \includegraphics[width=#4]{./figures/#1}
       }
       \caption{{\footnotesize  #2}
       \label{fig:#3}}
\end{figure}}

\newcommand{\InsertFigTwo}[5] {
\begin{figure}[h!t]
       \centerline{
         \includegraphics[width=#5]{./figures/#1}
         \hskip 0.5in
         \includegraphics[width=#5]{./figures/#2}
       }
       \caption{{\footnotesize  #3}
       \label{fig:#4}}
\end{figure}}

\newcommand{\InsertFigThree}[6] {
\begin{figure}[h!t]
       \centerline{
         \includegraphics[width=#6]{./figures/#1}
         \hskip 0.3in
         \includegraphics[width=#6]{./figures/#2}
         \hskip 0.3in
         \includegraphics[width=#6]{./figures/#3}
       }
       \caption{{\footnotesize  #4}
       \label{fig:#5}}
\end{figure}}

\newcommand{\InsertFigFourB}[7] {
\begin{figure}[h!t]
       \centerline{
\renewcommand{\arraystretch}{0.01}
         \begin{tabular}{cccc}
         \includegraphics[width=#7]{./figures/#1}&  \includegraphics[width=#7]{./figures/#2}&
        \includegraphics[width=#7]{./figures/#3}  &  \includegraphics[width=#7]{./figures/#4}
        \end{tabular}
       }
       \caption{{\footnotesize  #5}
       \label{fig:#6}}
\end{figure}}

\newcommand{\InsertFigSix}[9] {
\begin{figure}[h!t]
       \centerline{
\renewcommand{\arraystretch}{0.01}
         \begin{tabular}{ccc}
         \includegraphics[width=#9]{./figures/#1}&  \includegraphics[width=#9]{./figures/#2}&  \includegraphics[width=#9]{./figures/#3}\\
         \includegraphics[width=#9]{./figures/#4}&  \includegraphics[width=#9]{./figures/#5}&  \includegraphics[width=#9]{./figures/#6}
        \end{tabular}
       }
       \caption{{\footnotesize  #7}
       \label{fig:#8}}
\end{figure}}

\newcommand{\bC}{{\mathbb{ C}}}

\newcommand{\bR}{{\mathbb{ R}}}
\newcommand{\bS}{{\mathbb{ S}}}
\newcommand{\bT}{{\mathbb{ T}}}
\newcommand{\bZ}{{\mathbb{ Z}}}

\newcommand{\cF}{{\cal F}}

\newcommand{\cO}{{\cal O}}
\newcommand{\cP}{{\cal P}}

\newcommand{\eps}{\varepsilon}

\newcommand{\va}{\textbf{a}}
\newcommand{\vb}{\textbf{b}}

\newcommand{\mean}[1]{\left< #1 \right>}
\newcommand{\Var}[1] {\mbox{Var}({#1})}
\newcommand{\Sob}[1]{\|{#1}\|_{-1}}		

\newcommand{\beq}[1]{\begin{equation}\label{eq:#1}}
\newcommand{\eeq}{\end{equation}}

\newenvironment{se}[1]{\equation\label{eq:#1}\aligned}{\endaligned\endequation}
\newcommand{\bsplit}[1]{\begin{se}{#1}}
\newcommand{\esplit}{\end{se}}

\newenvironment{example}[1][]
  {
	\setlength \leftmargini {1.0em}		
	\setlength \topsep {0.5em}			
	\begin{quote}
	{\it Example#1} }
	{\end{quote}
  }
\newcommand{\bexam}[1][:]{\begin{example}[#1]}
\newcommand{\eexam}{\end{example}}

\title{Designing a Finite-Time Mixer: Optimizing Stirring for Two-Dimensional Maps}
\author{R.A.~Mitchell and J.D.~Meiss \thanks
      {
       The authors acknowledge support from NSF grant DMS-1211350.
        Useful conversations with James Curry, Robert Easton and Kathryn Padberg-Gehle are gratefully acknowledged. 
      }
    \\
 \begin{tabular}{c}
Department of Applied Mathematics\\
University of Colorado \\
Boulder, CO 80309-0526 \\
Rebecca.A.Mitchell@colorado.edu\\
James.Meiss@colorado.edu\\
\end{tabular}
}
\date{\today}
\begin{document}
\maketitle

\begin{abstract}
\vspace*{1ex}
\noindent

Mixing of a passive scalar in a fluid flow results from a two part process in which large gradients are first created by advection and then smoothed by diffusion. We investigate methods of designing efficient stirrers to optimize mixing of a passive scalar in a two-dimensional nonautonomous, incompressible flow over a finite time interval. The flow is modeled by a sequence of area-preserving maps whose parameters change in time, defining a mixing protocol. Stirring efficiency is measured by a negative Sobolev seminorm; its decrease implies creation of fine scale structure. A Perron-Frobenius operator is used to numerically advect the scalar for two examples: compositions of Chirikov standard maps and of Harper maps. In the former case, we find that a protocol corresponding to a single vertical shear composed with horizontal shearing at all other steps is nearly optimal. For the Harper maps, we devise a predictive, one-step scheme to choose appropriate fixed point stabilities and to control the Fourier spectrum evolution to obtain a near optimal protocol.
\end{abstract}

\section{Introduction}\label{sec:Intro}

The design of an efficient mixer for a passive scalar in an incompressible flow requires effective mechanical ``stirring" of the fluid that eventually will lead to homogenization through diffusion. The point is that when stirring gives rise to fine scale filaments, even a small diffusivity can be highly effective in homogenization. 
As was noted by Spencer and Wiley \cite{Spencer51}:
\begin{quote}
When the materials to be mixed are very viscous liquids\ldots neither diffusion nor turbulence can assist very much in mixing. Rather, mixing must be effected by some complex, continuous deformation which serves to disperse the components to the desired degree.
\end{quote}
Thus, as a first step in designing a mixer one can ignore diffusion and study only advection by an incompressible flow. 

In this paper we are interested in the finite-time case: how can one design the best mixer for a given initial state and a given finite time? For example, the partitioned-pipe mixer \cite{Galaktionov03} and the rotated-arc mixer (RAM) \cite{Speetjens14} are finite-time mixers in which fluid is injected at one end of a cylindrical pipe, flows past a finite number of mixing elements, and is emitted at the other end. Other examples of this type of mixing include ``making up rubber formulations on compounding rolls, using an extruder as a mixer, kneading bread dough, and pulling taffy" \cite{Spencer51}. Our goal is to develop methods for optimizing the geometry of the elements to maximize mixing. 

Most previous studies of mixing by stirring consider autonomous dynamics and concepts appropriate for the infinite-time limit, i.e., positive Lyapunov exponents or measure-theoretic mixing. Indeed, it was noted by Aref that chaotic advection can lead to effective stirring \cite{Aref84}. Examples of the optimal design of a periodic sequence of obstacles in a channel or pipe were studied in \cite{Galaktionov03, ChunPing03, Gibout06, Hsiao14}. However, it is not clear what role chaos plays in the finite-time, nonautonomous case. Froyland \textit{et al} do study the finite-time case in a flow with diffusion, but assume the advective part is fixed, only modifying diffusion via local perturbations to enhance mixing \cite{Froyland16}. 

We view the stirring process as a finite sequence of steps; each corresponding to one element of the mixer. 
The design corresponds mathematically to the selection of a finite sequence $f_t: D \to D$, $t = 1,2,\ldots, T$, of maps on a domain $D$. Each map is a fixed-time map of an incompressible fluid flow so that each $f_t$ is a volume-preserving diffeomorphism. We will study several families of maps, each a composition of shears, see \Sec{Shears}. The mixer acts on a passive scalar, represented by a density $\rho: D \to \bR$. Our goal is to find a sequence $\{f_t\}$ that takes a fixed ``unmixed" initial state, $\rho^0$, to one that is ``optimally mixed" after flowing through the full system, i.e., under the composition
\beq{MapSeq}
	F_T = f_T \circ f_{T-1}\circ \ldots \circ f_2 \circ f_1.
\eeq
Ignoring diffusion, sources, and reactions, the evolution of the density is determined by the Perron-Frobenius operator,
\beq{Perron}
	\rho^{t}(z) = \cP_{f_t}[\rho^{t-1}](z) = \rho^{t-1}(f_{t}^{-1}(z)) , \quad t = 1, 2, \ldots, T.
\eeq
In order to select an ``optimal" mixing protocol, we must use a ``mix-norm" to define the stirring effectiveness; our choice, following \cite{Thiffeault12}, is a negative Sobolev seminorm; see \Sec{MixingMeasures}.

We are primarily interested in the design of efficient, finite-time mixers in a three-dimensional domain. However, in this paper we treat a simpler problem: we assume that the axial velocity is constant and that $f$ corresponds to the map from a two-dimensional cross-section, through a mixing element, to another such cross-section. Since the geometry of the mixing elements will vary, subsequent maps will differ. Each of the $f_t$ will be chosen from a given family that satisfies some geometric constraints. For example the RAM consists of an inner pipe surrounded by a spinning outer pipe, and the mixing elements correspond to windows cut into the inner pipe. Each map represents the effect of the angularly sheared flow induced by the boundary drag. The maps, $f_t$, will depend on parameters that represent the window width, length and rotation rate. 

In this paper we will study two families, Chirikov and Harper maps, see \Sec{Shears}. Each map depends on one or more parameters, say $a_t$. Optimization then corresponds to selecting the best sequence from a given family, that is, to vector $\va = (a_1,a_2,\ldots a_T)$ of parameters for maps in the family. As a constraint we assume that the total energy expended in the mixing process is bounded. The simple version of this is to assume that some norm, $\|\va\|$, is fixed. When the norm is small, any fixed map in one of our families is only weakly chaotic on a small subset of phase space. However, varying the parameters from step-to-step can nevertheless result in effective stirring.

The simplest method to find an optimal protocol is to exhaustively search all possible cases; this corresponds to selecting an optimal parameter vector $\va$ that satisfies an energy constraint. Thus we can simulate the optimization process by a random optimization method: simply select the best result from a large number of trials that are equi-distributed on the sphere $\|\va\|^2 = E$, corresponding to a fixed energy.

Our investigation seeks to understand mixing by focusing only on advection created by a sequence of stirring events, accounting for diffusion in the mixing measure itself. Through numerical simulations using simple area-preserving maps and through analytical considerations based on Fourier analysis and key characteristics of the maps, we gain some understanding of how to design efficient stirrers.

In \Sec{Shears} we introduce families of maps, consisting of a composition of shears, to describe a stirring mechanism. In \Sec{MixingMeasures} we discuss the use of negative Sobolev seminorms as a measure of mixing. In \Sec{Numerics} we explain how the Perron-Frobenius method is used to apply the map to a passive scalar. We present results from numerical experiments with Chirikov's standard map in \Sec{StdMap}, and the Harper map in \Sec{HarperMap}.

\section{Compositions of Shears}\label{sec:Shears}

As a simple model, we study a family of two-dimensional maps given by the composition of a pair of shears on a periodic domain. Taking the domain to be the torus with unit period, $D = \bT^2 = (\bR/\bZ)^2$, and $z = (x,y) \in \bT^2$, the vertical and horizontal shears have the form
\bsplit{Shears}
	s_y(x,y)&=(x\,, y+X(x))  , \\
	s_x(x,y)&=(x+Y(y)\,, y) ,
\esplit
where $X, Y: \bS \to \bS$ are circle maps. Each such shear is the time-one map of a simple incompressible velocity field; for example,
$s_x(x,y)$ is the time-one flow map for the \textsc{ode}
\beq{fluidVelocity}
	(\dot{x},\dot{y}) = v_x(x,y) = (Y(y),0).
\eeq
Composing the shears \Eq{Shears} gives the one-element map
\beq{GenStdMap}
	f(x,y) = s_x \circ s_y(x,y) = (x+Y(y+X(x)) \,,\, y + X(x)), 
\eeq
that could be realized by imposing a vector field $v_y = (0,X(x))$  for $0 < t < 1$ followed by the vector field \Eq{fluidVelocity} for $1 < t < 2$, i.e., by two successive shears.
When the shears are smooth, \Eq{GenStdMap} is an area-preserving diffeomorphism with the inverse
\[
	f^{-1}(x,y) = (x-Y(y)\,, y-X(x-Y(y))) .
\]
This family of maps includes many well-studied cases, including Arnold's cat map \cite{Arnold68}
with $X(x) = x$ and $Y(y) = y$, Chirikov's standard map \cite{Chirikov79a, Meiss92} for which
\bsplit{StdMap}
	X(x) &= -a \sin(2\pi x) ,\\
	Y(y) &= y ,
\esplit
and the Harper map \cite{Artuso94, Shinohara02} for which
\bsplit{HarperMap}
	X(x) &= -a \sin(2\pi x) ,\\
	Y(y) &= b \sin(2\pi y) .
\esplit

The design of a mixer consists of the choice of a mixing protocol, that is, to a sequence of $T$  maps \Eq{MapSeq}. Formally, $T$ may be infinite, but we will typically take $T = \cO(10)$. The goal is to optimize the mixing for some maximal total ``energy" of the flow. As a measure of this energy we take a squared norm of the shear
\beq{EnergyStd}
	E_s = \sup_{D} \| s(x,y) -(x,y) \|^2 .
\eeq
For example for the Harper maps, $E_{s_x} = b^2$, $E_{s_y} = a^2$.
Since $s_x$ corresponds to the velocity \Eq{fluidVelocity}, the energy is proportional to the kinetic energy of the fluid. For the sequence \Eq{MapSeq}, the parameters $(a_t,b_t)$ of the Harper maps are constrained by
\beq{EnergyHarper}
	\sum_{t=1}^T (a_t^2 + b_t^2)  = E_H ,
\eeq
where $E_H$ is some given maximum energy. Note that the energy for a sequence of Chirikov maps includes that required to create the horizontal shear $s_x$, thus for \Eq{StdMap}
\beq{EnergyChirikov}
	\sum_{t=1}^T a_t^2 +T  = E_C .
\eeq

\section{Mixing Measures}\label{sec:MixingMeasures}

Mixing, in a measure theoretic sense, is defined through an infinite time limit  \cite{Lasota94}. A transformation $f: D \to D$ that preserves a measure $\mu$ is \textit{mixing} if for any two measurable sets $A$ and $B$, the measure of the overlap $\mu(A \cap f^{-t}(B)) \to \mu(A) \mu(B)/\mu(D)$ as $t \to \infty$. Equivalently, $f$ is mixing if for any initial density $\rho^0 \in L^1(D)$, $\rho^t(x)$, determined by \Eq{Perron}, converges \textit{weakly} as $t \to \infty$ to its mean
\beq{RhoMean}
	\mean{\rho} = \frac{1}{|D|} \int_D \rho \,d\mu ,
\eeq
where $|D| = \mu(D)$ is the volume of $D$.
Mixing is also implied by a positive Kolmogorov-Sinai entropy, computed---according to Pesin---as the mean of the sum of positive Lyapunov exponents of $f$. 

For our purposes, measure-theoretic mixing is not an appropriate yardstick for two reasons. Firstly, mixing implies that every $L^2$ initial state, not just the those of physical interest, approaches a uniform state. Moreover, even if the entropy is zero, a physical state may still develop fine-scale structure \cite{Mathew05}. Secondly, we are interested in a given state becoming ``mixed-up" after a \textit{finite} number of \emph{different} transformations, not in the limit $t \to \infty$ for a single map.

As was noted by Danckwerts, in the presence of diffusion, the decrease of variance
\beq{Variance}
	\Var{\rho} = \mean{\rho^2} - \mean{\rho}^2, 
\eeq
is one measure of mixing \cite{Danckwerts52}. Indeed, diffusion will cause $\Var{\rho}$ to decrease at a rate proportional to $\|\nabla \rho\|^2$ \cite{Thiffeault12}. Thus to be effective, a stirrer should increase gradients so that diffusion is activated. This ``intensity of segregation" has been used as to optimize mixing in diffusive flows \cite{Kruijt01, Galaktionov03, Gibout06}. However, in the absence of diffusivity and sources, the variance (and indeed each $L^p$-norm) of the density is constant. 


One class of mixing measures include the mix-norm of Mathew \cite{Mathew05} and equivalent Sobolev seminorms \cite{Thiffeault12}. For functions that have zero mean, the squared $H^q$-seminorm is defined by
\beq{Sobolev}
	\| \rho \|^2_{H^q} = \frac{1}{|D|} \int_D |(-L^2\Delta)^{q/2} \rho|^2 \,d\mu ,
\eeq
where $L$ is a normalizing length scale. When $q/2$ is not a positive integer, the Laplacian $\Delta$ is to be interpreted by its Fourier symbol. For the case $D = \bT^2$,  we set $L = 1/(2\pi)$ and \Eq{Sobolev} reduces to
\beq{FourierSobolev}
	\|\rho\|^2_{H^q} =  \sum_{m \in \bZ^2\setminus \{0\}}  |m|^{2q} |\hat{\rho}_m|^2 ,
\eeq
where the Fourier coefficient of $\rho$ for mode $m = (m_1,m_2) \in \bZ^2$ is
\beq{Fourier}
   \hat{\rho}_m = \int_{\bT^2} \rho(z) e^{-2\pi i m \cdot z} \, d^2z.
\eeq
An effective stirring mechanism will increase $\|\rho\|^2_{H^1} = \|\nabla \rho\|^2$; however, this norm is difficult to compute effectively since it relies on accurate evaluation of high Fourier modes. However, following \cite{Thiffeault12}, note that the Cauchy-Schwarz inequality implies that
\[
	\| \rho \|^2  = - \int_D \nabla \rho \cdot \nabla^{-1} \rho 
	              \le \|\rho\|_{H^1} \|\rho\|_{H^{-1}},
\]
assuming $\mean{\rho} = 0$.
Since the $L^2$ norm is invariant under the Perron-Frobenius dynamics \Eq{Perron}, if $\|\rho \|_{H^{-1}} \to 0$, then $\|\rho\|_{H^1} \to \infty$. Thus a measure of the effectiveness of the stirring is the degree to which the $H^{-1}$ seminorm decreases.
Since high Fourier modes are scaled by $|m|^{-1}$, the $H^{-1}$ norm will not be as sensitive to fine-scale structure, and thus is more amenable to computation. 
It is possible to show that any negative Sobolev seminorm has similar properties \cite{Lin11},
and that  if $\Sob{\rho^t} \to 0$ as $t \to \infty$ for all $ \rho^0 \in L^2$, then $f$ is mixing \cite{Mathew05}.

Below, we use the $H^{-1}$ seminorm as our primary measure of mixing.

\section{Computing the Perron-Frobenius Operator}\label{sec:Numerics}

To compute the image $\rho^t(x,y)$ and the seminorm \Eq{Sobolev} numerically, one must represent the density in some finite basis set. One possible technique is the ``mapping"  \cite{Spencer51} or ``Ulam" method \cite{Ulam60, Froyland05, Froyland09, Frahm10}: choose a grid of $N\times N$ points $(x_i,y_j) = (\tfrac{i}{N},\tfrac{j}{N})$ on the torus, with $\rho_{ij} = \rho(x_i,y_j)$ and approximate the operator $\cP$ \Eq{Perron} by an $N^2 \times N^2$ stochastic matrix $P$,
\[
	\rho^t_{i,j} = P_{ij,i'j'}\rho^{t-1}_{i'j'} .
\]
Here $P_{ij,i'j'}$ is the fraction of the $i'j'$ cell that is mapped onto the $ij$ cell by $f$. This can be estimated---for a fixed map $f$---by a one step iteration of a large number of trial points in each cell. Such a discretization introduces an effective diffusion that will decrease the variance \Eq{Variance} \cite{Kruijt01}, and explicit diffusion can also be included in the process \cite{Gorodetskyi12}. However, this method is not practical for our purposes, since the stochastic matrix would need to be recomputed for each new map in the sequence \Eq{MapSeq}.

A second discrete method is to use a Fourier basis. This seems natural since the Sobolev norm that we seek to minimize is most easily evaluated in Fourier space \Eq{FourierSobolev}. Each shear \Eq{Shears} has a relatively simple representation in Fourier space. Using \Eq{Fourier} with
$m = (m_1,m_2)$, we obtain
\bsplit{PerronFourier}
	\hat{\cP}_{s_y}[\rho])_m &= \int_{\bT^2} \rho(x,y-X(x)) e^{-2\pi i m \cdot z} d^2z
	         = \sum_{n\in \bZ} \hat{\rho}_{n,m_2} \cF_{n-m_1,m_2}[X] ,\\
	(\hat{\cP}_{s_x}[\rho])_m &= \int_{\bT^2} \rho(x-Y(y),y) e^{-2\pi i m \cdot z} d^2 z
	         = \sum_{n\in \bZ} \hat{\rho}_{m_1,n} \cF_{n-m_2,m_1}[Y] ,        
\esplit
where for each $(j,k) \in \bZ^2$, we define a functional $\cF_{j,k}: C^0(\bS) \to\bC$ that gives a Fourier coefficient of the exponential of a function:
\[
	\cF_{j,k}[g] \equiv \int_0^1e^{2\pi i(jx-kg(x))}dx .
\]
For a linear shear $id(x) = x$ this gives
\beq{FourierShear}
	\cF_{j,k}[id] = \delta_{j,k},
\eeq
the Kronecker-delta, while for a sinusoidal shear such as $X(x)$ in \Eq{StdMap} and \Eq{HarperMap} we have
\beq{FourierSine}
	\cF_{j,k}[-a \sin(2\pi x)] = J_j(-2\pi k a) = J_{-j}(2\pi ka),
\eeq
using the standard identity
\[
	e^{iz \sin \theta} = \sum_{j \in \bZ} J_j(z) e^{i j \theta} 
\]	
for the Bessel function. Note that in general $\cF_{j,0}[g] = \delta_{j,0}$, which implies, in particular, that the $(0,0)$-Fourier mode is preserved---as expected for any volume-preserving map.

Composing the two results \Eq{PerronFourier} gives the general formula for updating Fourier coefficients for the map \Eq{GenStdMap}:
\beq{FourierIteration}
	\hat{\rho}_m^t= \hat{\cP}_{s_x} \circ \hat{\cP}_{s_y}[\rho^{t-1}] 
	       = \sum_{n\in\bZ^2}\hat{\rho}_{n}^{t-1}\cF_{n_1-m_1,n_2}[X]\cF_{n_2-m_2,m_1}[Y]
\eeq
since $f^{-1} = s_y^{-1} \circ s_x^{-1}$. For example for the Chirikov map, \Eq{FourierIteration} becomes, using \Eq{FourierSine},
\beq{StdFourierUpdate}
	\hat{\rho}_m^t =  \sum_{n\in\bZ^2}\hat{\rho}_n^{t-1}J_{m_1-n_1}(2\pi n_2a)\delta_{n_2-m_2,m_1}
	=\sum_{k \in \bZ} \hat{\rho}_{k,m_1+m_2}^{t-1} J_{m_1-k}\left(2\pi(m_1+m_2)a\right) .\\
\eeq
For the Harper map, \Eq{FourierIteration} becomes
\beq{HarperFourierUpdate}
\hat{\rho}_m^T=\sum_{n\in\bZ^2}\hat{\rho}_n^{t-1}J_{m_1-n_1}(2\pi n_2 a)J_{n_2-m_2}(2\pi m_1 b).
\eeq

To implement this method numerically, one must truncate the Fourier basis, keeping say, the 
$N^2 = (2M+1)^2$ modes with $m_i \in [-M,M]$. Since the sums in \Eq{FourierIteration} naturally generate modes that are not in the basis set, these must be discarded. Conversely, modes that are not in the truncated basis also may contribute to those in the truncation, and these give errors in the numerical scheme.

A final technique for implementing \Eq{Perron} is to simply evaluate an analytically defined function $\rho^0$ at a point $f^{-1}(x,y)$. Given a grid of points $(x_i,y_j)$ we can define
\beq{GridIteration}
	\rho^t_{i,j} = \rho^t(x_i,y_j) = \rho^0(f_1^{-1}\circ f_2^{-1} \circ \ldots \circ f_t^{-1}(x_i,y_j))
\eeq
whenever the initial state $\rho^0$ is known everywhere. Numerically this is efficient if one is interested only in the final mixed state, $\rho^T$, but less so if one would also like intermediate states $t < T$ since $\rho^t_{ij}$ cannot be used to compute the next image. Nevertheless, to the extent that one can accurately compute the composition of the maps---subject to the growth of errors due to sensitive dependence on initial conditions---the resulting values of $\rho^t$ are accurate when $\rho^0$ is Lipchitz, though of course we only know the values on the selected grid.

We choose a regular grid of $N^2$ points, so that the Sobolev norm, \Eq{FourierSobolev} can be easily evaluated using a fast Fourier transform. Using discrete lattice with $N^2$ points, we have $z_k = \frac{k}{N}$ for
$k \in [0,N)^2$, and $\rho_k = \rho(z_k)$, with
\begin{align*}
	\hat{\rho}_m &= \sum_{k \in [0,N)^2}  \rho_k e^{-2\pi i \frac{m
	\cdot k}{N}} , \\
	\rho_k &= \frac{1}{N^2}\sum_{m\in [0,N)^2} \hat{\rho}_m e^{2\pi i \frac{m \cdot k}{N}} .
\end{align*}

\section{Chirikov Map}\label{sec:StdMap}

In this section we discuss various techniques to find optimal mixing protocols for a sequence of Chirikov maps \Eq{StdMap}. For most of the results in this section, we use an initial Gaussian profile,
\beq{InitialGaussian}
	\rho^0(z)=\exp\left(-\eps{\|z-z_o\|^2}\right), \quad
	z_o = (0.5,0.5), \quad \eps = 10.
\eeq
This distribution has standard deviation $\sigma=\sqrt{\frac{1}{2\eps}} \approx 0.2236$. We will also assume that 
\beq{StdChirikovEnergy}
	T = 12, \quad \mbox{and} \quad \sum_{t=1}^{12} a_t^2= 0.03, 
\eeq
so that the total energy \Eq{EnergyChirikov} is $E_C=12.03$. Unless otherwise specified, we use a $301\times301$ grid and choose the maximum Fourier mode $M = 150$ for the computations.

\subsection{Single Step Stirring}\label{sec:StdSingleStep}

If the initial state consists of a single, nonconstant Fourier mode, $\rho^0(x,y) = \hat{\rho}_{k,l}^0 e^{2\pi i (kx+ly)}$, then, using \Eq{StdFourierUpdate}, the squared seminorm \Eq{FourierSobolev} with $q=-1$   at $t=1$ is
\beq{StdOneMode}
	\|\rho^1\|_{-1}^2 =  |\hat{\rho}_{k,l}^0|^2 \sum_{m=-\infty}^\infty \frac{J^2_{m-k}(2\pi la_1)}{m^2+(m-l)^2}.
\eeq
This results in a function that asymptotically decreases as the shear amplitude, $a_1$, grows; however, it does not decrease monotonically because of the oscillations of the Bessel functions, see \Fig{StdMapNorms}(a). Indeed, if the total energy \Eq{EnergyStd} is small enough, then the optimally mixed protocol, after one step, may be that with $a_1 = 0$. 

When a full spectrum initial condition is used in \Eq{StdFourierUpdate}, the oscillations seen in \Fig{StdMapNorms}(a) are mostly averaged away. Figure \ref{fig:StdMapNorms}(b) shows the seminorm for Cauchy (black curve) and Gaussian (red curve) initial spectra. Note that the resulting curves are not symmetric about $a_1 = 0$; for these real initial spectra, a positive shear is more effective than a negative shear. Of course, even for $a_1=0$, the density is still sheared since $s_x$ is not the identity for \Eq{StdMap}. For large shear, the Sobolev norm decreases, but---though it is not seen in the figure---the decrease saturates. As $a_1 \to \infty$, $\Sob{\rho^1} \to 0.758$, for the Gaussian initial spectrum, and $\to 0.809$ for the Cauchy distribution. It is not possible to ``completely" mix after one step, even with arbitrarily large energy.

\InsertFigTwo{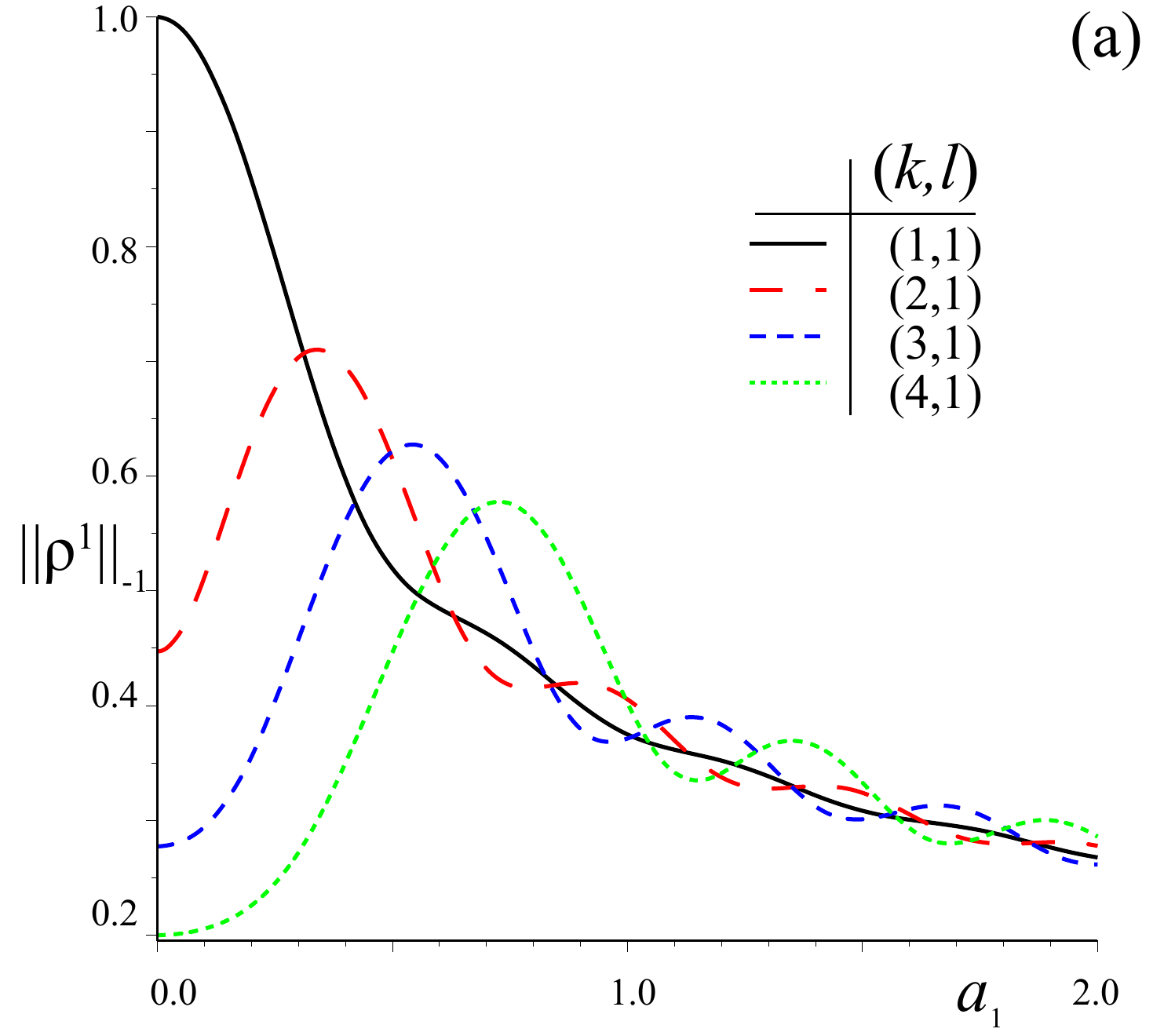}{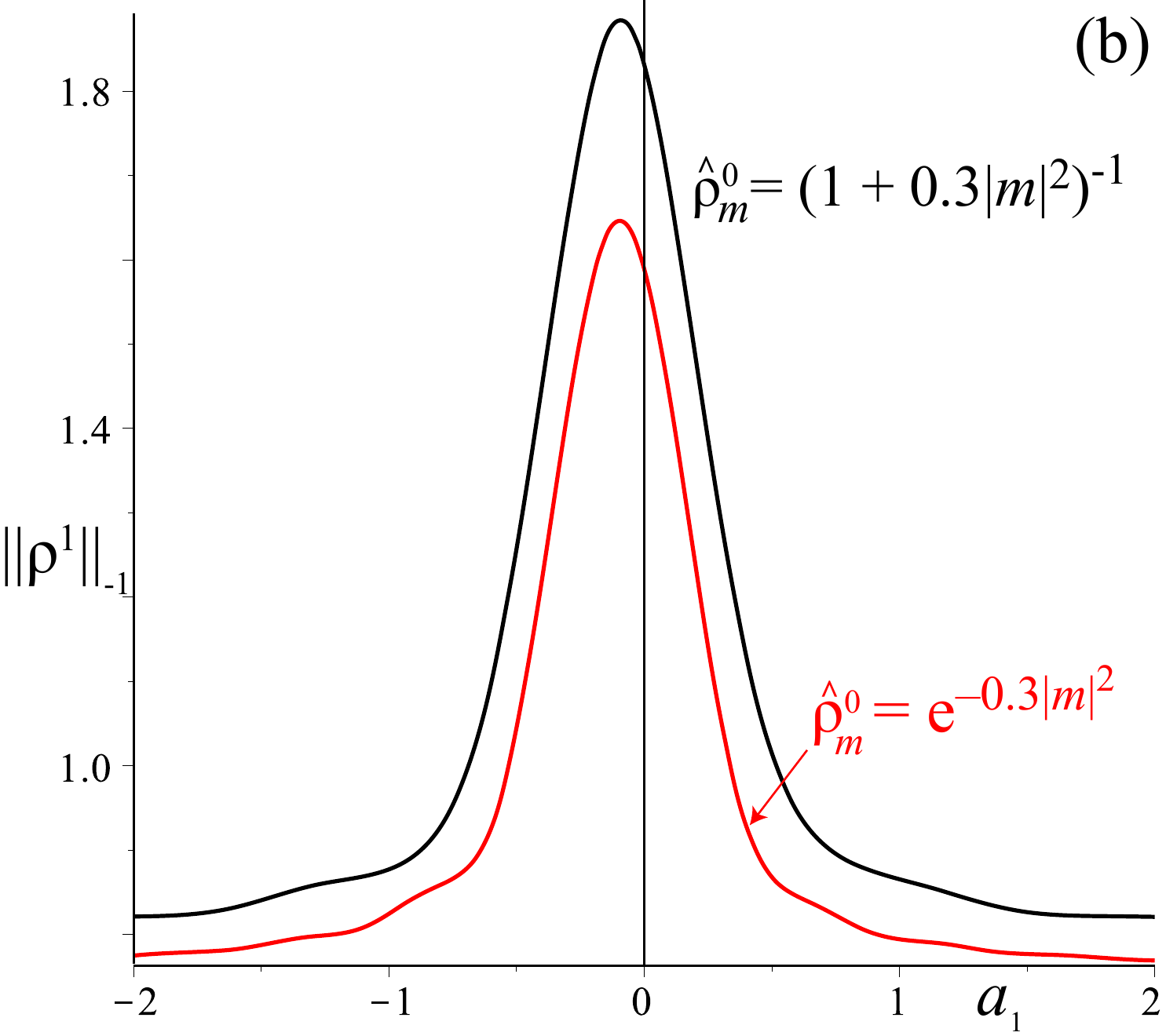}{The $H^{-1}$ seminorm \Eq{StdOneMode} after one step of the standard map for initial state $\rho^0(x,y) = \exp(2\pi i(kx+ly))$ (left panel) and one that has a Cauchy or Gaussian Fourier spectrum (right panel) as a function of the shear strength $a_1$. }{StdMapNorms}{3in}

\subsection{Constant Amplitude Stirring}\label{sec:StdConst}

When the amplitude is constant, $\va = (a_o,a_o,\ldots,a_o)$, and satisfies the energy constraint \Eq{StdChirikovEnergy}, then $a_0 = 0.05$. If we think of this as the autonomous standard map, then its phase portrait is almost completely regular; its only features are  elliptic islands around the fixed point at $(x,y) = (0,0)$ and the period-two orbit at $(\tfrac12, \tfrac12)$, see \Fig{StdMapConstPhase}. The initial density \Eq{InitialGaussian} is centered over this period-two orbit, and evolves as shown in \Fig{ChirikovConst}. The computations were done using the grid method \Eq{GridIteration} on an $N = 2M+1 = 301$ squared grid. It is clear from the figure that the stirring is dominated by the horizontal shear $s_x$ with $Y(y) = y$ from \Eq{StdMap}, though some stretching from the saddle fixed point at $(\tfrac12,0)$ can also be seen. 

\InsertFig{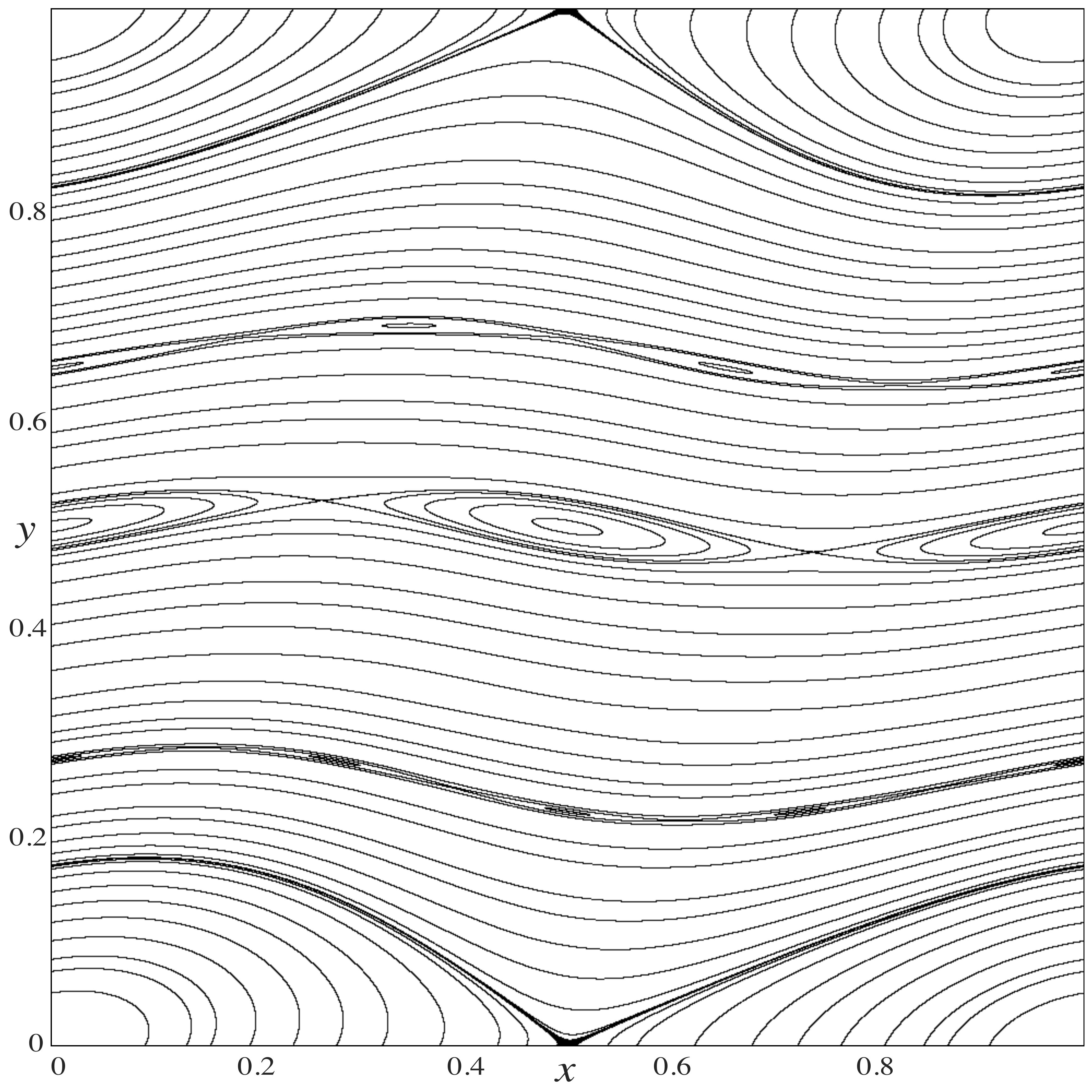}{Phase portrait of the autonomous Chirikov standard map with $a_t = a_0 = 0.05$.}{StdMapConstPhase}{2in}

\InsertFigFourB{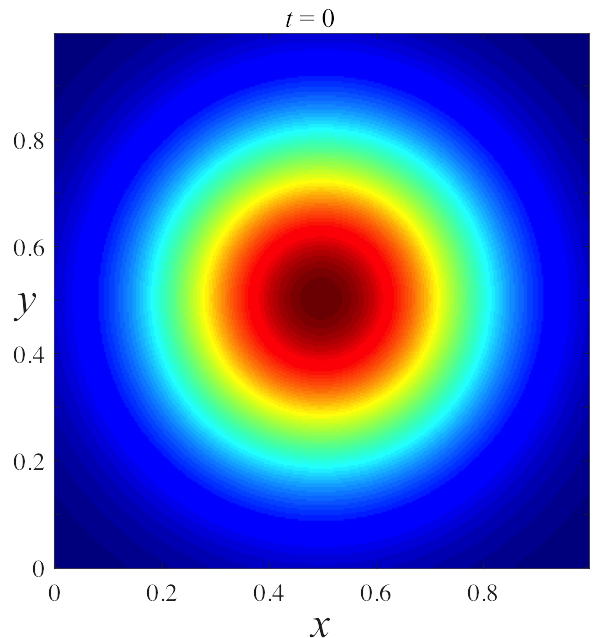}{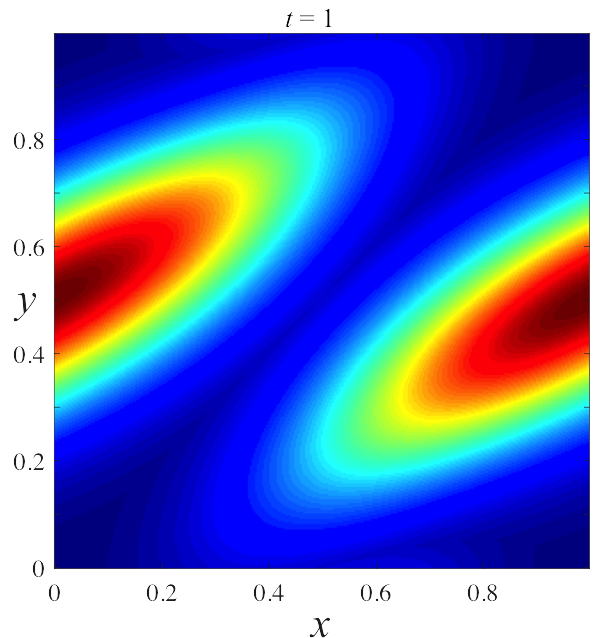}{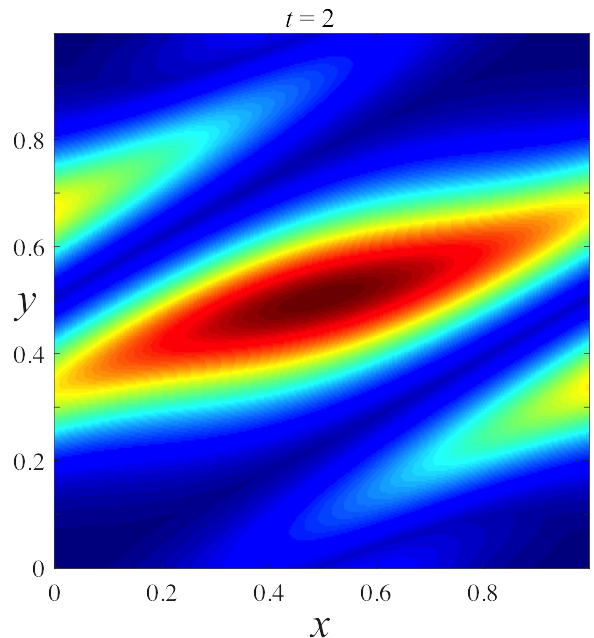}{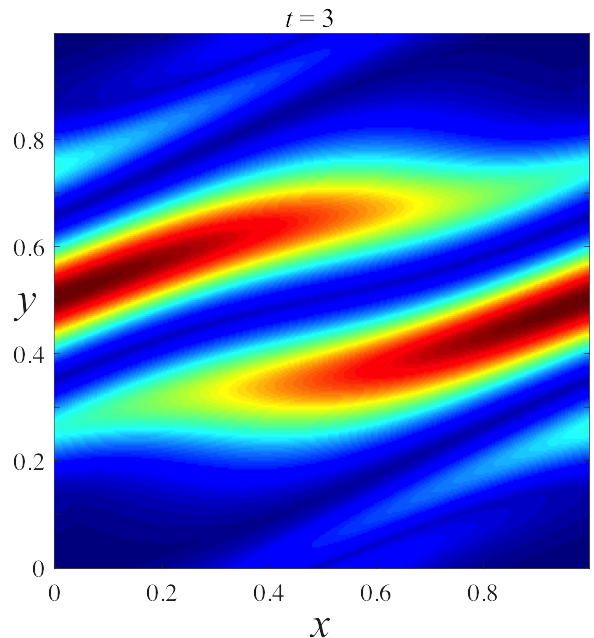}{Evolution of the density $\rho^t(x,y)$ for $t=0,1,2,3$ applying the standard map \Eq{StdMap} with constant amplitude $a_t = 0.05$ with initial condition \Eq{InitialGaussian} using $N = 301$ grid points. For a movie of all $12$ steps, see \url{http:/amath.colorado.edu/faculty/jdm/movies/ChirikovConst.mp4}.}{ChirikovConst}{1.5in}

Figure \ref{fig:ChirikovConstSob} shows the decay of the Sobolev norm \Eq{FourierSobolev} with time. Here the computations were done with a varying number of grid points $N$, as shown in the figure.  Note that as $N$ is increased, the computations quickly converge: the value of $\|\rho^{12}\|_{-1}$ varies by less than $10^{-3}$ for $N \ge 51$. This indicates that our standard choice $N=301$ gives sufficient accuracy for $T=12$.
The decrease in the mix-norm saturates with time, just as the single step case of \Sec{StdSingleStep} saturated with increasing amplitude. The point is that repeatedly reapplying the same shear amplitudes may not be the best mixing strategy. This constant case will serve as the benchmark of comparison for our optimization methods.

\InsertFig{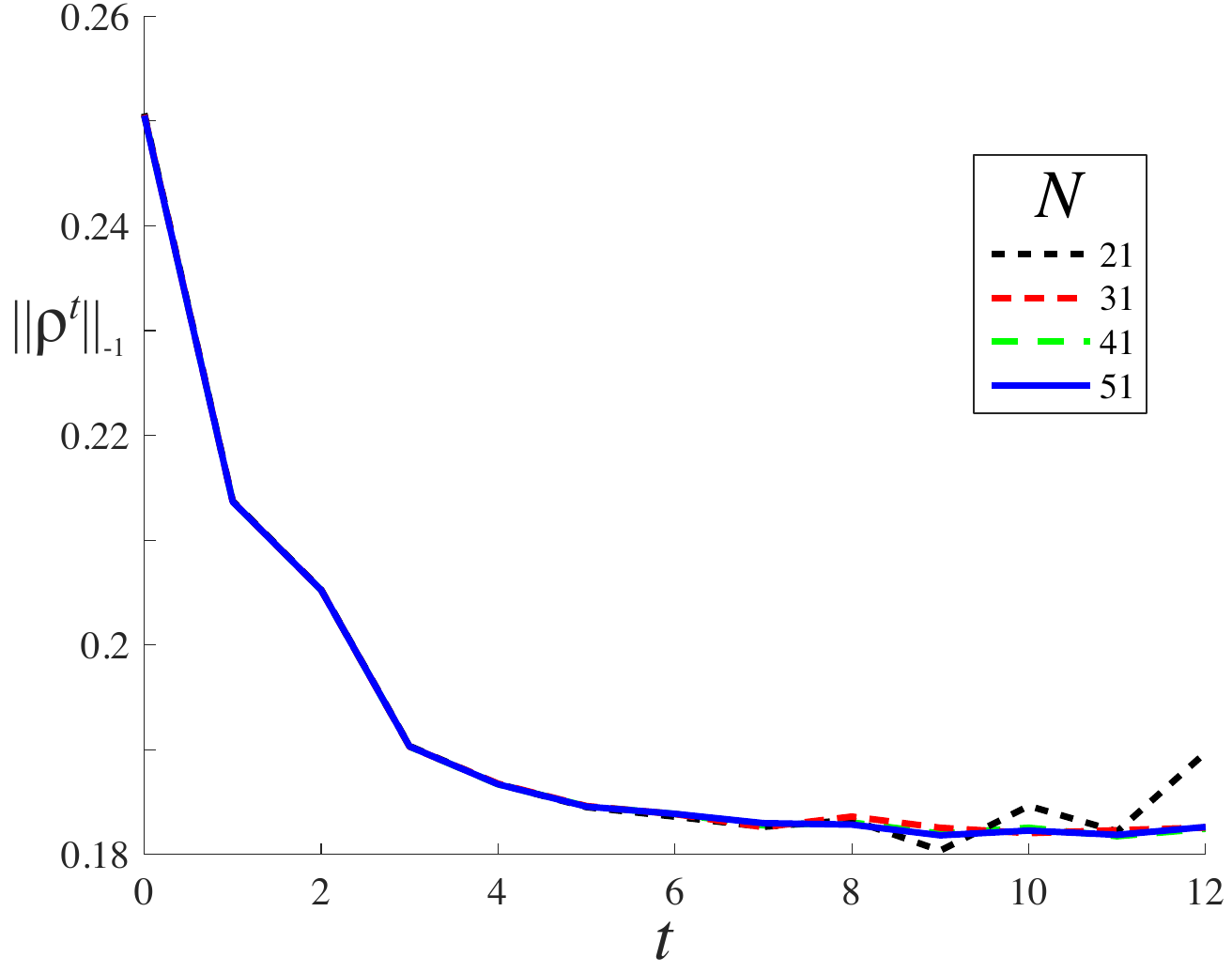}{Decay of $\Sob{\rho^t}$ for the standard map \Eq{StdMap} with constant amplitude $(a_0,a_0,\ldots,a_0)$, as in \Fig{ChirikovConst}. The different curves are computed with grid sizes $N$ as shown in the legend.}{ChirikovConstSob}{3in}

\subsection{Varying Amplitudes: Random Optimization Method}\label{sec:StdRO}

Choosing parameters at random from some distribution can serve as a crude optimization scheme: simply select the best protocol---the solution with smallest seminorm \Eq{FourierSobolev}---from the trials.
For the standard map \Eq{StdMap}, there is one parameter, $a_t$, for each step of the map, so the parameter space will have $n = T$ dimensions. For convenience we think of the stirring parameters as a vector, letting
\[
	\va=(a_1,a_2,\ldots,a_T)
\]
denote a given stirring protocol. The energy \Eq{EnergyChirikov} then becomes
\[
	T+\|\va\|^2=E_C.
\]

To satisfy the energy constraint, we must sample amplitude vectors from the surface of a ball in $\mathbb{R}^n$. To ensure uniform sampling on $\bS^{n-1}$, we randomly select $n$-dimensional vectors from a standard \textit{iid} normal distribution, scaling each realization to ensure it has the correct energy.

Figure \ref{fig:ChirikovHist}(a) shows the distribution of $\Sob{\rho^T}$ for some two million randomly chosen amplitude vectors with the fixed energy \Eq{StdChirikovEnergy} using the Gaussian initial profile \Eq{InitialGaussian}. Almost all of the trials, 94.8\%, perform better than the constant protocol, and the best case achieves the norm $\Sob{\rho^T} = 0.0545$, which is $3.35$ times smaller than the constant protocol. We will refer to this vector as the ``random optimal protocol" (RO) even though it is certainly not truly optimal. 

The evolution of $\Sob{\rho^t}$ is shown in  \Fig{ChirikovHist}(b) for the constant protocol, the RO protocol, and the worst performing solution of the random trials, which we call ``RW". Note that though the seminorm for the RW case decreases on the first step, it remains essentially constant after that. Aside from the first step of the map, the optimal trial significantly out-performs the constant case.

The twenty most successful amplitude vectors are shown in \Fig{ChirikovMC}. For these vectors, the largest magnitudes of $a_t$ occur primarily in steps $t=2$ or $3$; the first step and all following steps have smaller amplitudes.
This suggests a simple scheme: concentrate the energy in the vertical shear in a small number of key steps of the map. We investigate this next.

\InsertFigTwo{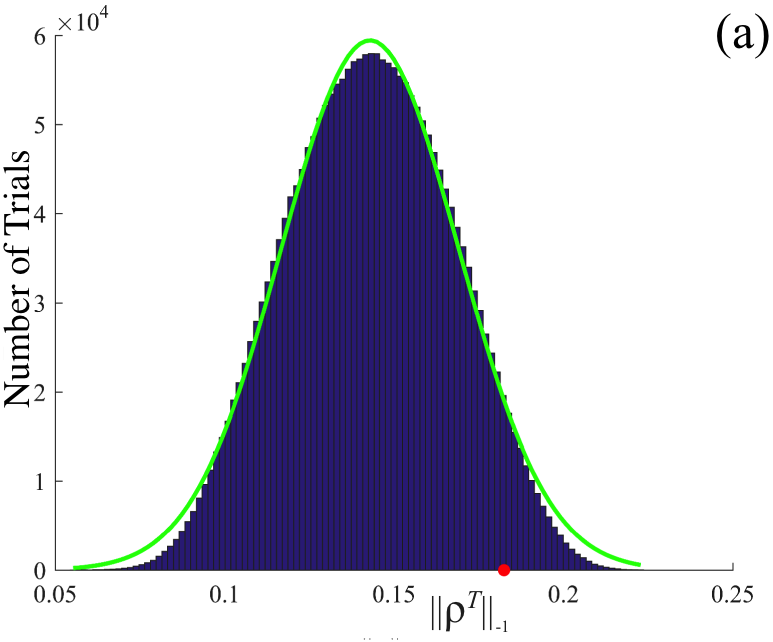}{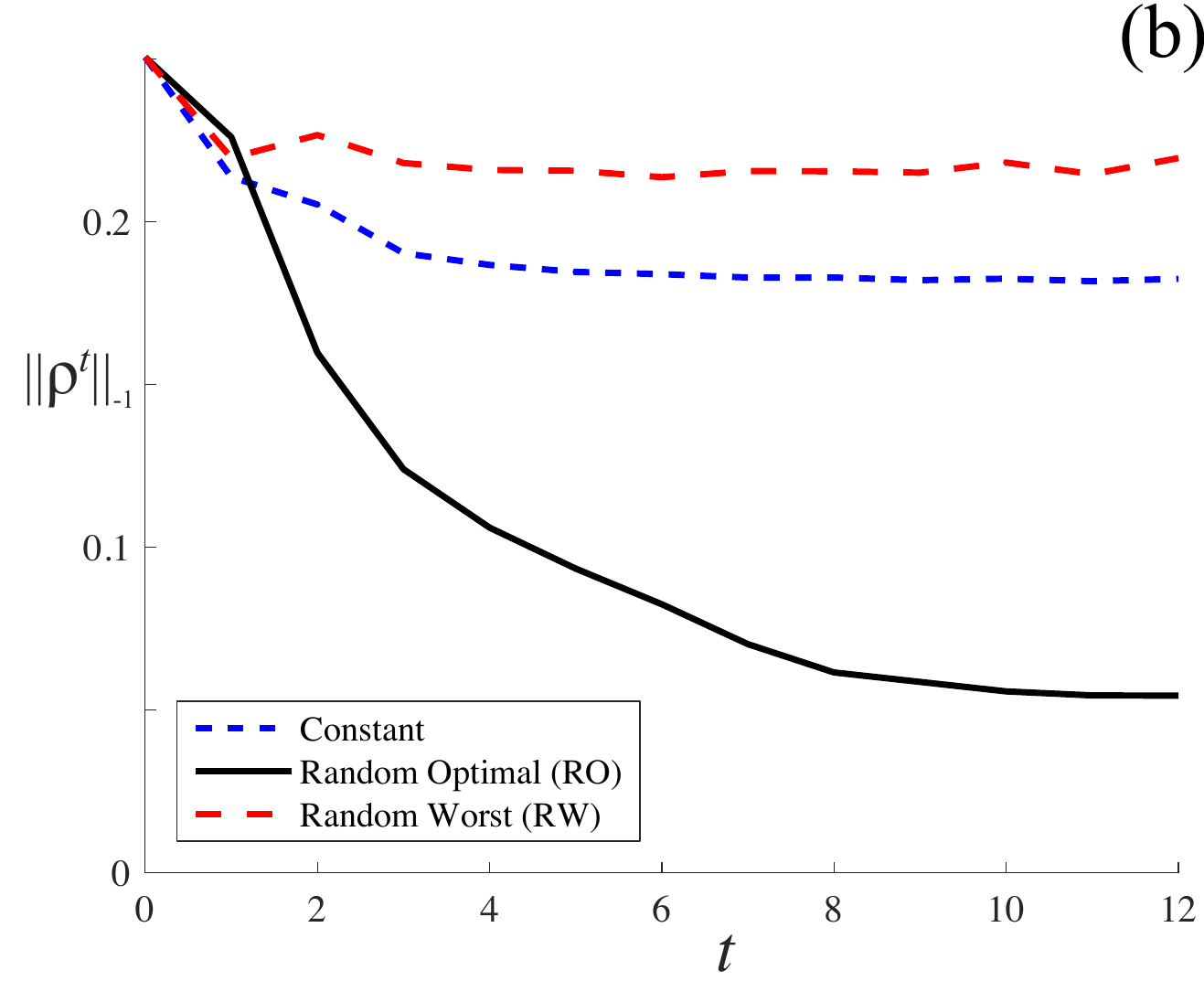}{(a) Histogram of $\Sob{\rho^T}$ for $\sim 2.2\times 10^6$ trials of the RO algorithm applied to the standard map \Eq{StdMap} with $E_C = T+0.03$, $T=12$, and initial condition \Eq{InitialGaussian}. The green curve indicates the best fit normal distribution with mean $0.1429$ and standard deviation $0.03704$.
The red point indicates the norm achieved for the constant case $(a_0,\ldots,a_0)$.
(b) Comparison between the mix-norms as a function of time for the constant case, the RO protocol and the RW protocol.}{ChirikovHist}{3in}

\InsertFig{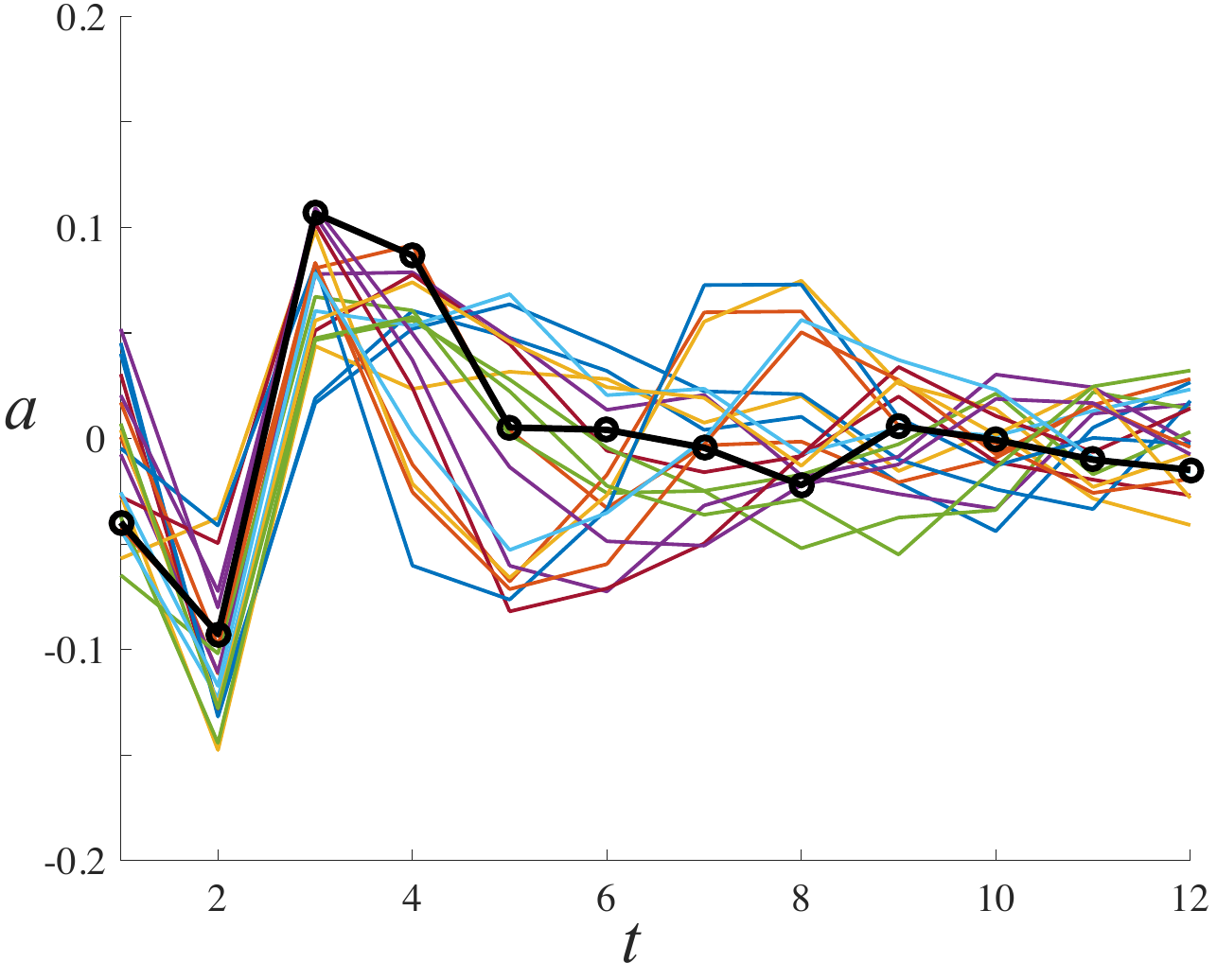}{Amplitude vectors for the best twenty protocols found from the RO algorithm described in \Fig{ChirikovHist}. 
The black curve indicates the RO protocol and a movie of the spatial evolution of this case is at \url{http:/amath.colorado.edu/faculty/jdm/movies/ChirikovMC.mp4}}{ChirikovMC}{3in}

\subsection{Sparse Stirring}\label{sec:StdSparse}

Since the most successful protocols found at random have energy concentrated in a small number of map steps, we can modify the optimization scheme to assume sparsity in $\va$. Fixing $d \le T$ to be the number of nonzero $a_t$, we first randomly select $d$ values from the set $\{1,2,\ldots,T\}$ to correspond to the nonzero $a_t$, and then perform the random optimization as described above, but with $n=d$. 
For $d=1$, there are simply $2\times{12} = 24$ discrete choices since exactly one of the $a_t = \pm 0.03$. For $d>1$, in addition to the discrete selection of $\binom{12}{d}$ shearing times, we must choose sparse vectors on $\bS^{d-1}$ to fix the energy. Consequently the number of trials should increase with $d$. The best amplitude vectors that we found for each case are shown in \Fig{ChirikovSparse}; the optimal protocol for $d=1$ corresponds to shearing on the second step with $a_2 = -\sqrt{0.03} \approx -0.173$.
The performance of the optimal protocols as $d$ varies is shown in \Fig{ChirikovSparseSob}; note that $\Sob{\rho^t}$ does not not depend significantly on $d$. Moreover, the optimal sparse protocols achieve even a slightly smaller mix-norm than the full set of random trials. This shows that the latter technique did not find the true optimum, even with more than two million trials.

\InsertFigFourB{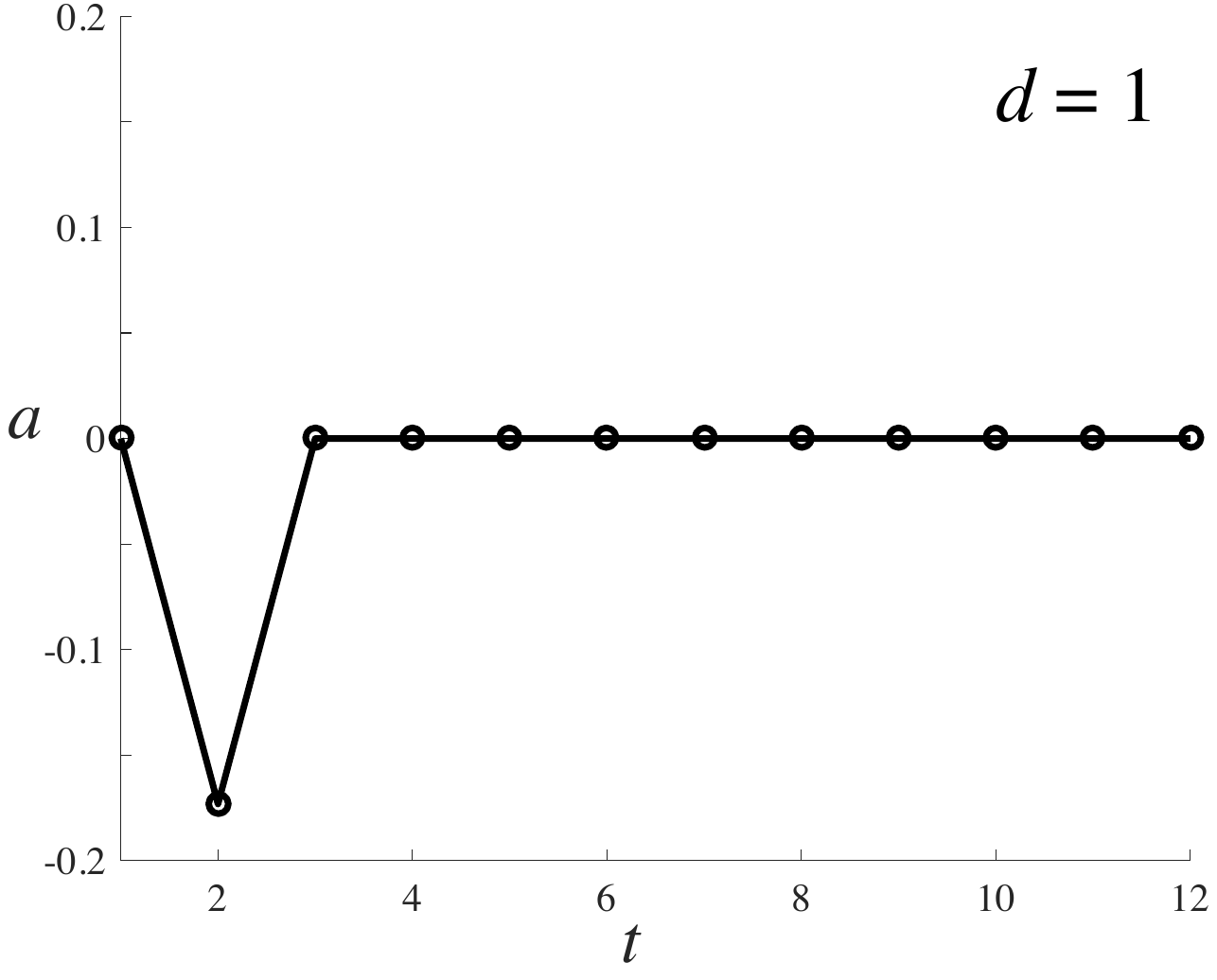}{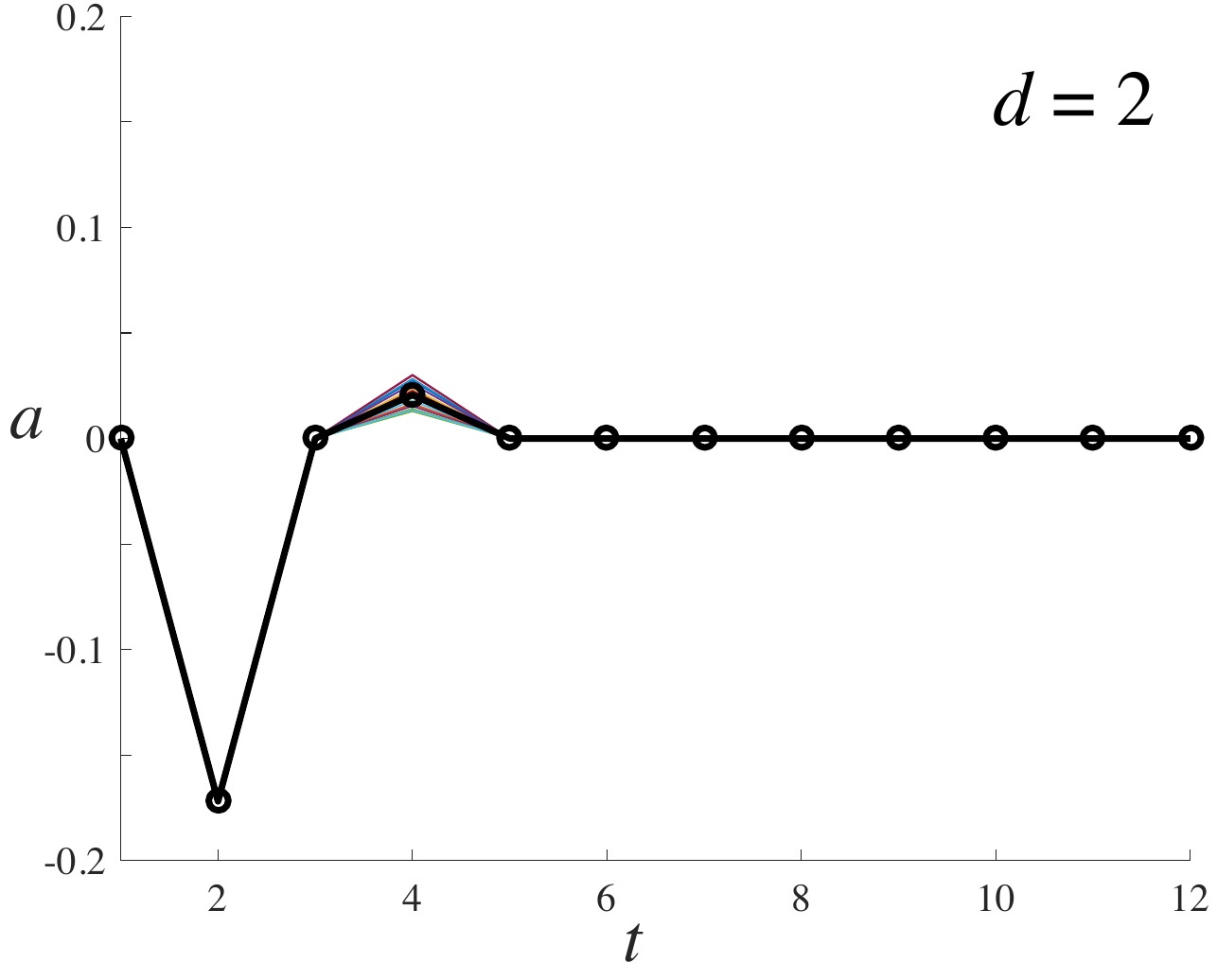}{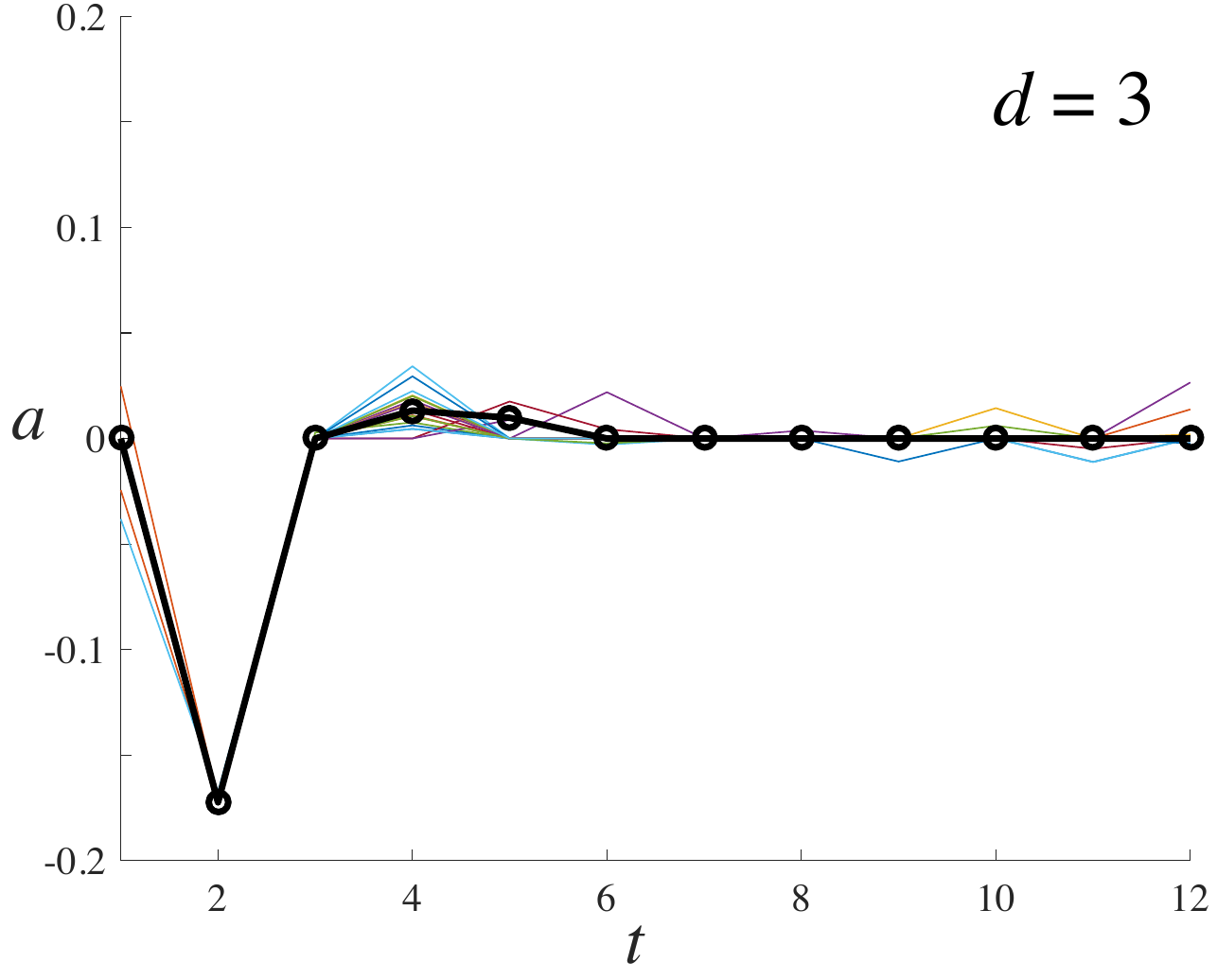}{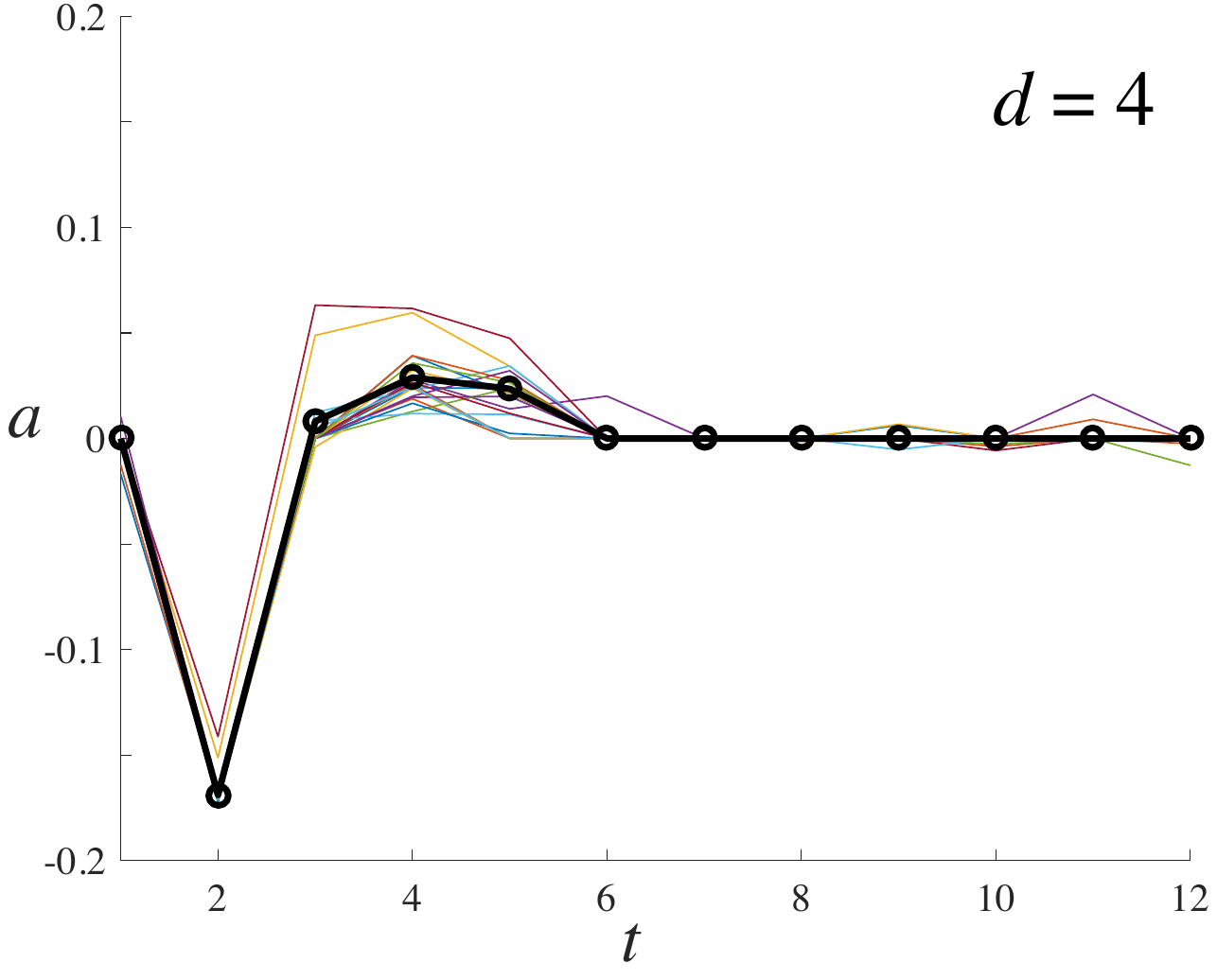}{Amplitude vectors of the best ($d=1$) and the best twenty ($d>˙1)$ protocols found from the sparse RO method. The open circles/black line indicates the optimal protocol for each case.}{ChirikovSparse}{1.5in}
\InsertFig{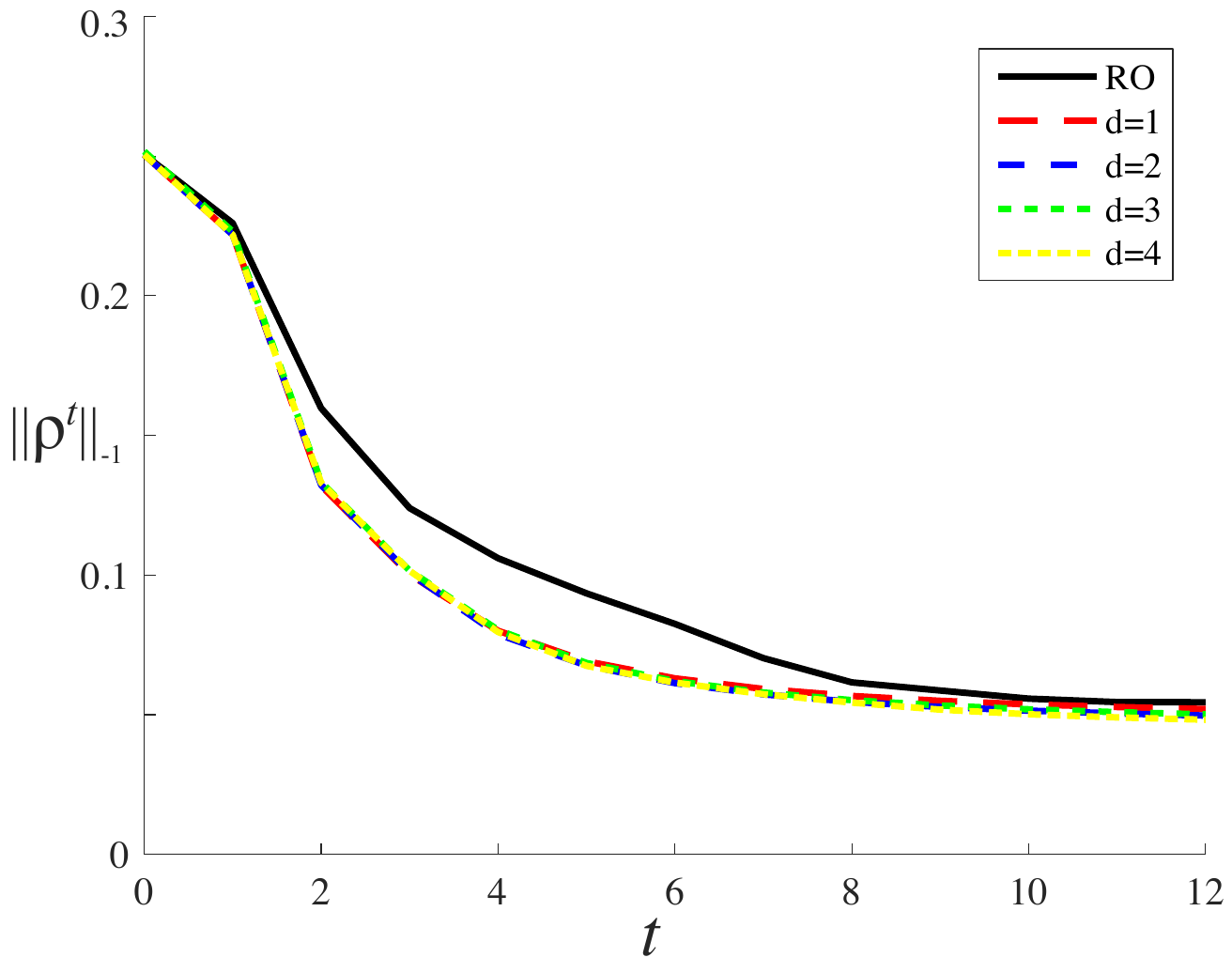}{Comparison of $\Sob{\rho^t}$ for the random optimal protocol with the sparse optimal protocols for $d=1,2,3,4$.}{ChirikovSparseSob}{3in}

If we think of the composite map \Eq{MapSeq} as a single, area-preserving map, for $d=1$ and $a_2 \neq 0$ it has the form
\beq{TStepSparse}
	 F_T(x,y) = \left\{\begin{array}{l}
	 			x + y+ (T-1) y' \\ 
				y - a_2 \sin(2\pi(x+y))
				\end{array} \right.
\eeq
when $T \ge 2$. Defining new variables $(\xi,\eta) = (x+y,Ty)$, this reduces to a single standard map in $(\xi,\eta)$ with effective amplitude $T a_2$. Several phase portraits of the map \Eq{TStepSparse} are shown in \Fig{TStepSparsePhase}. As $T$ increases the elliptic regions decrease in size, becoming  invisible for $T \ge 8$. Contrast this with the regularity of the constant amplitude case shown in \Fig{StdMapConstPhase}, which has the same total energy. Indeed when $T=12$ the dynamics are nearly uniformly chaotic since, $12 |a_2| \approx 2.08 \gg 0.972/(2\pi) \approx 0.155$, which is the threshold for global chaos for the standard map \cite{Meiss92}. This is one reason why the sparse case is such an effective stirrer.

\InsertFigThree{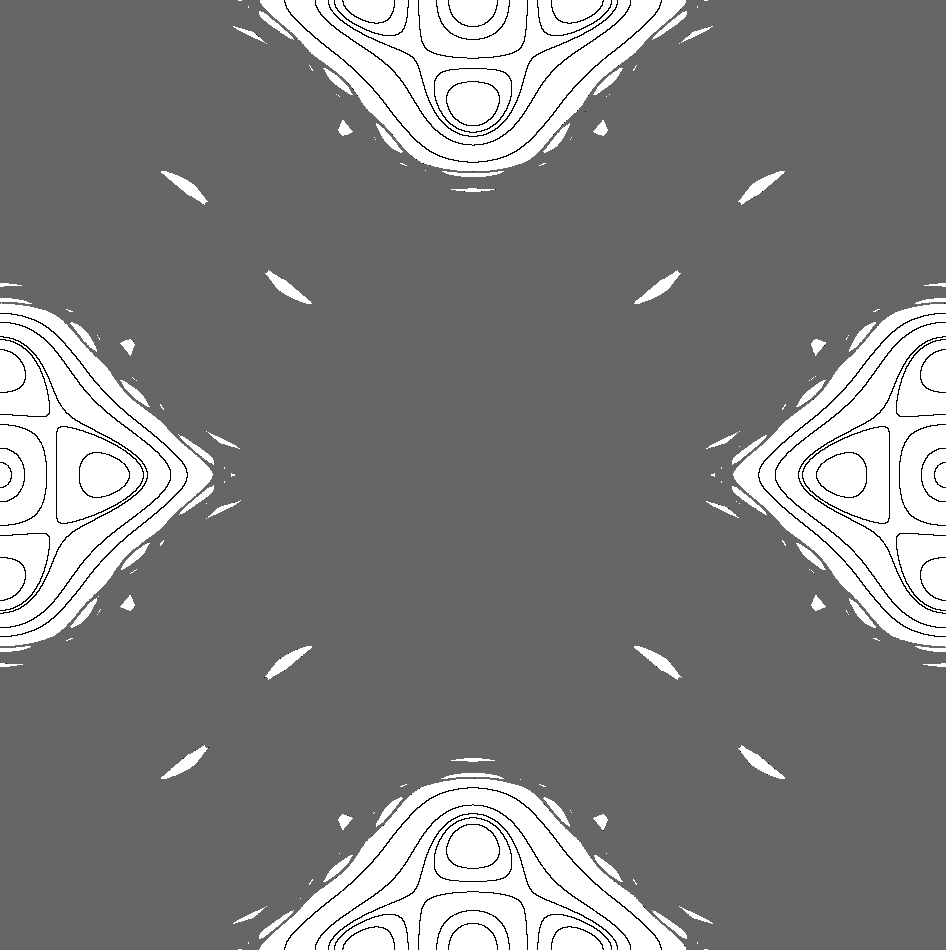}{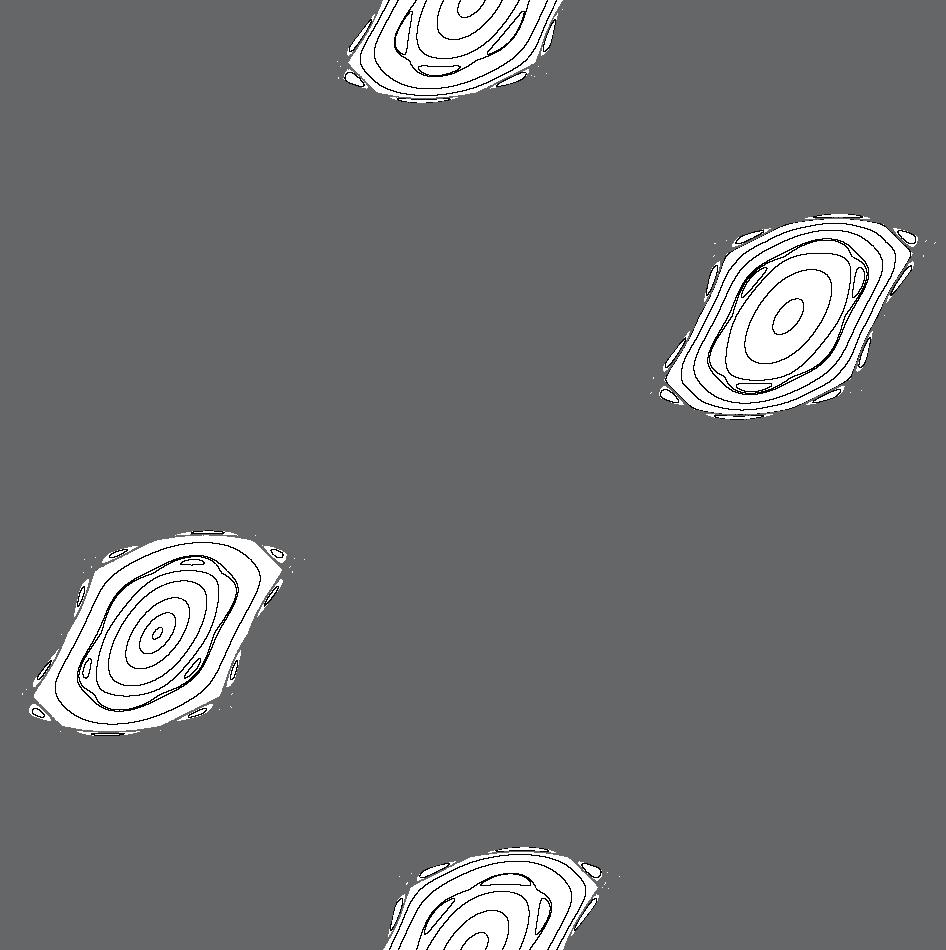}{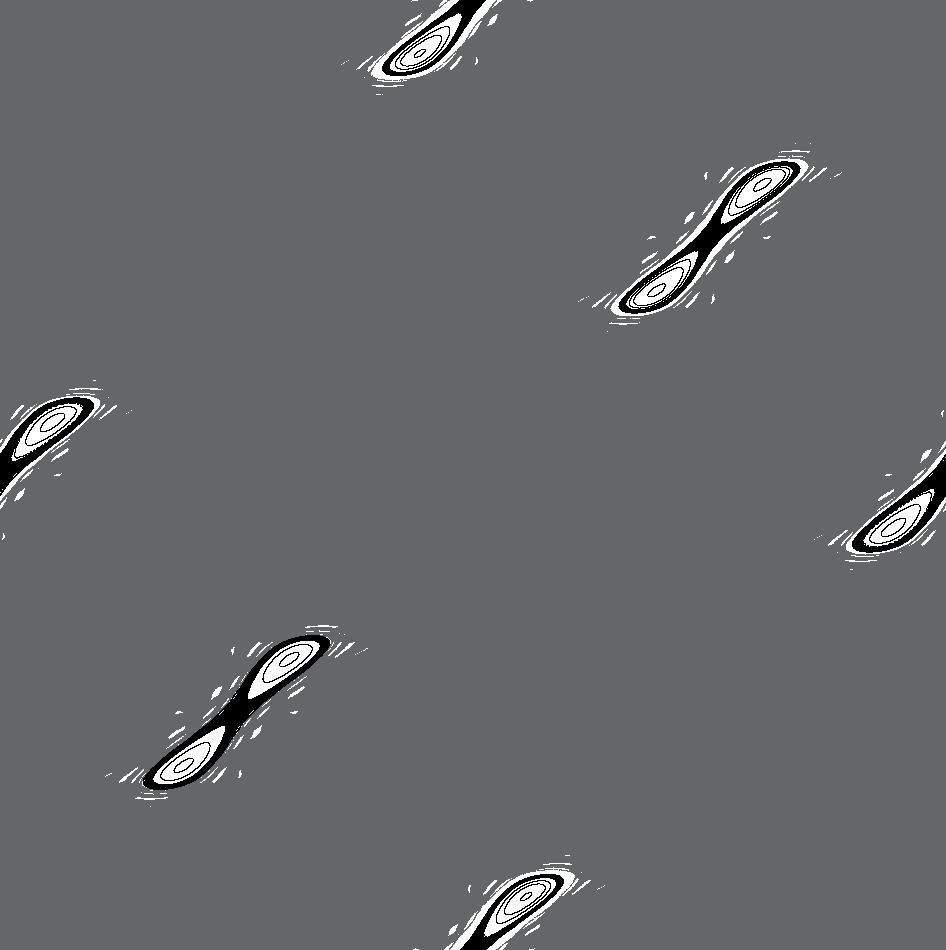}{Phase portraits of the map \Eq{TStepSparse} for $T=2$, $3$, and $4$ with $a_2 = -0.1732$. For $T \ge 6$ (not shown), the dynamics appears to be nearly uniformly chaotic, except for small period-six island chains along lines of slope $-1$.}{TStepSparsePhase}{2in}

For comparison, \Fig{ChirikovSparseBad} shows the sparse amplitude vectors, $\va$, that maximize the final mix-norm. Note that these have $a_2 >0$, the opposite of the optimal sparse case. These protocols vary more with $d$ than the optimal protocols did, but the difference in the final mix-norms is small. For the worst $d=1$ case $\Sob{\rho^T} = 0.2041$, which is less than that of the full RW case shown in \Fig{ChirikovHist}. A sparse amplitude vector is not as effective in inhibiting mixing as one that is more nearly constant.

\InsertFigFourB{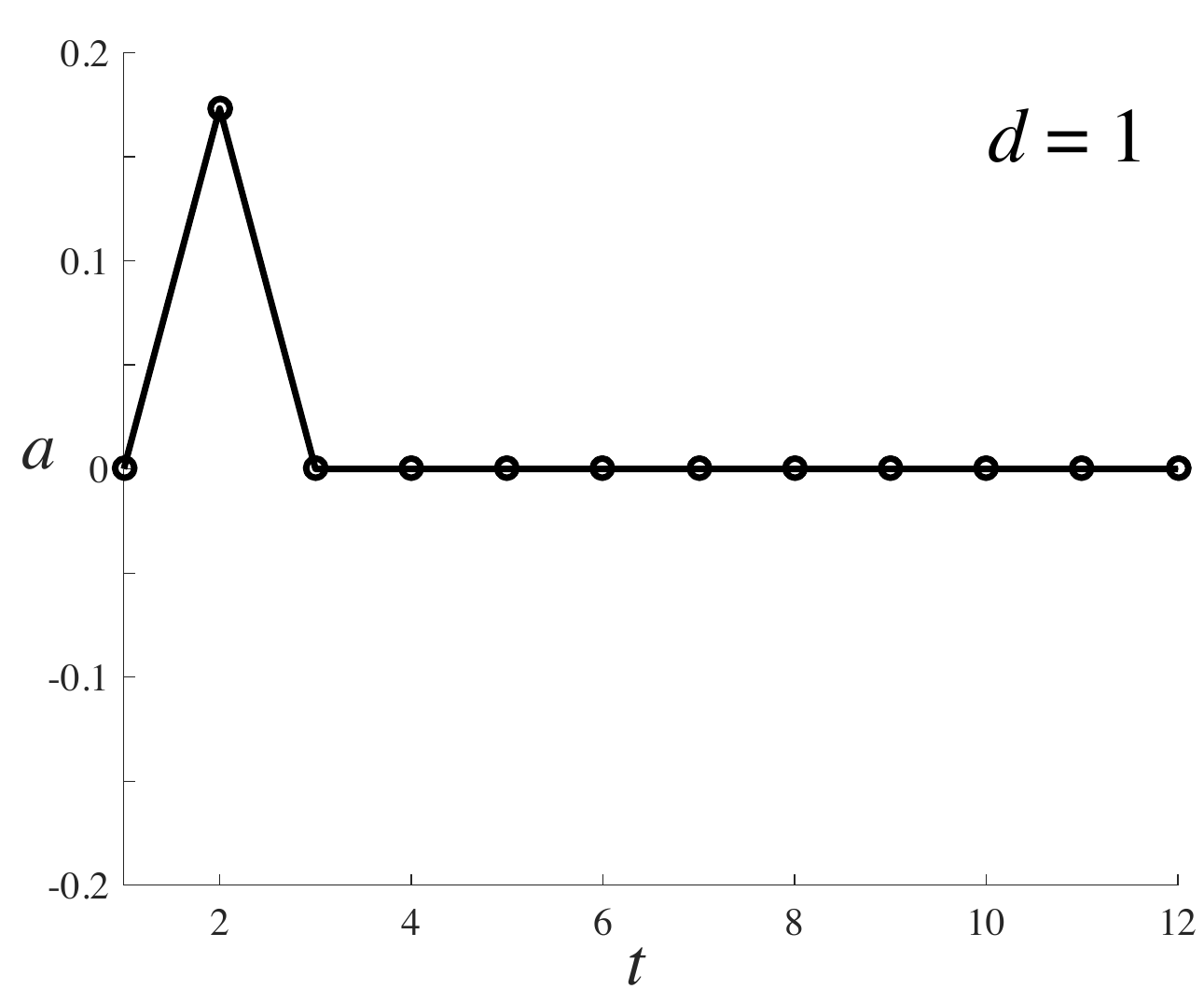}{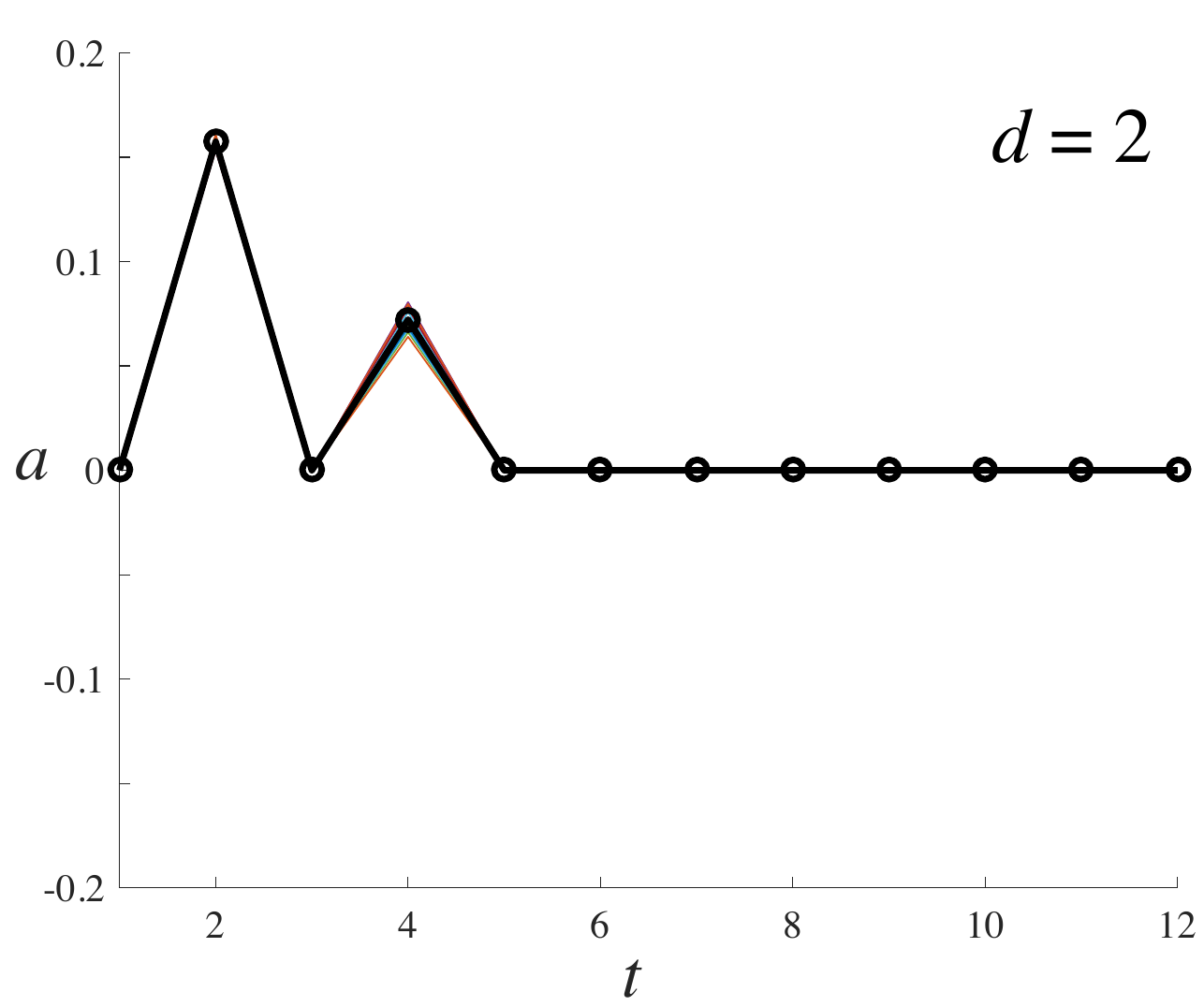}{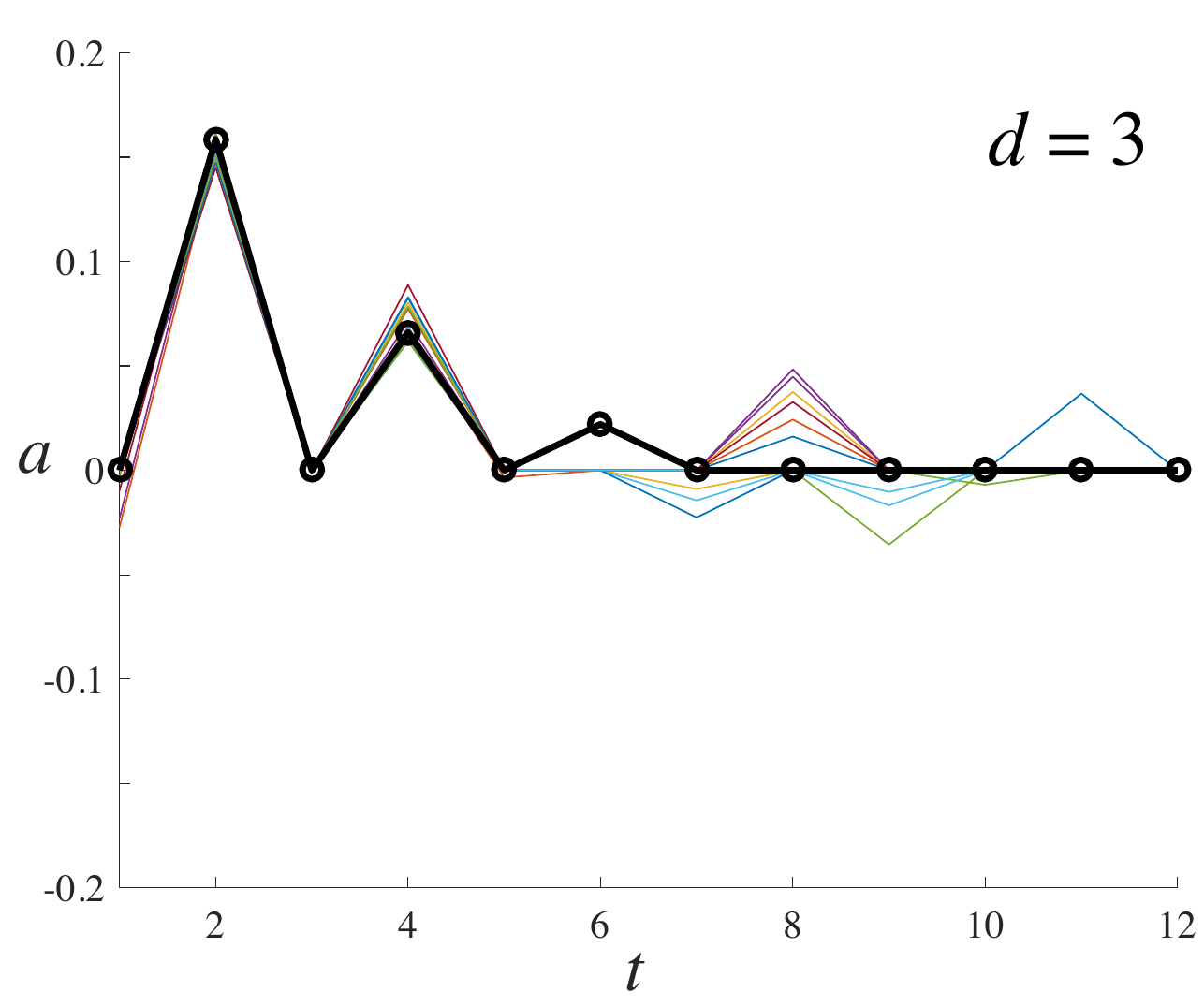}{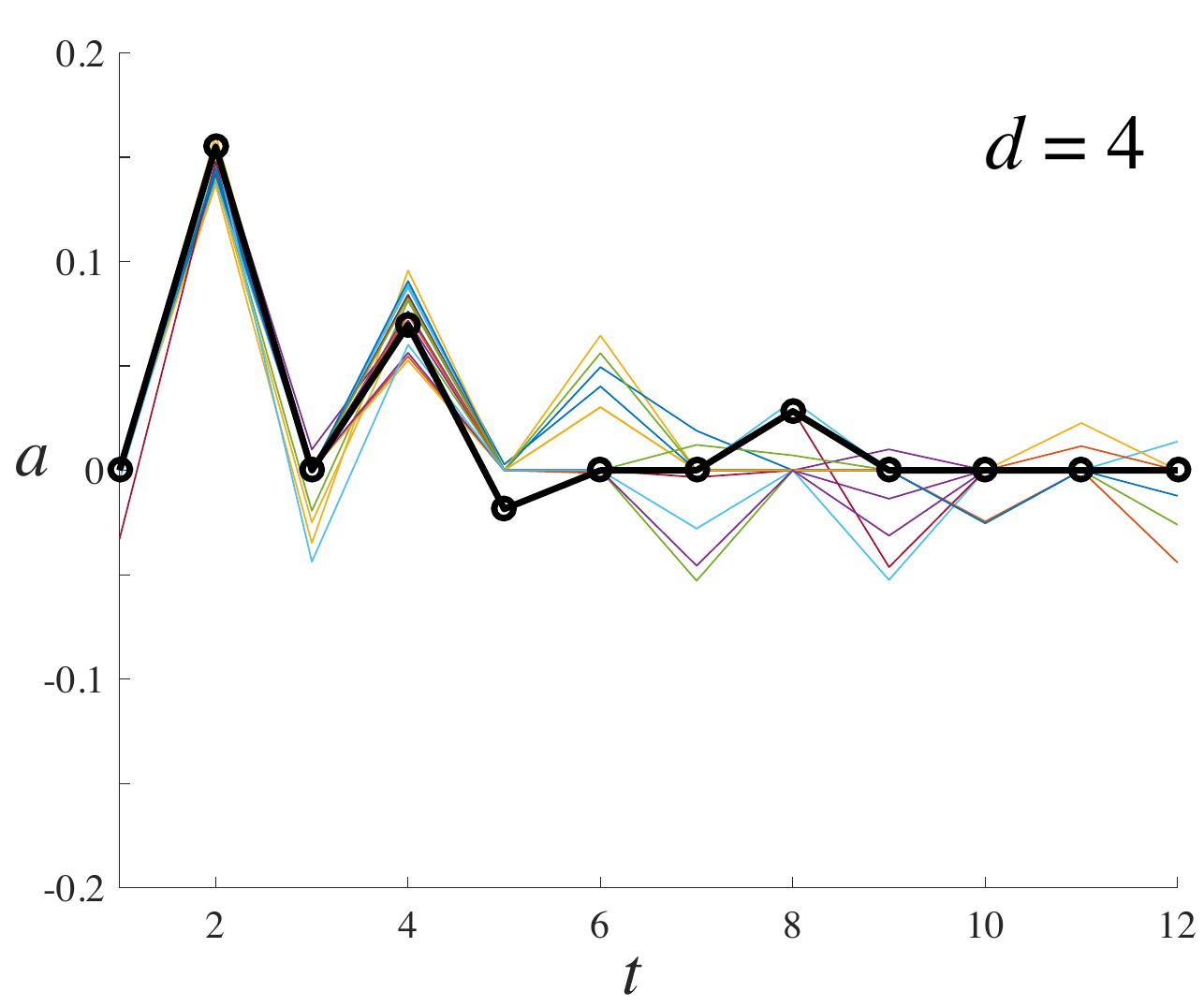}{Amplitude vectors for the worst ($d=1$) and the worst twenty ($d>1$) stirring protocols up to $d=4$. The open circles/black line indicates the worst protocol for each case.}{ChirikovSparseBad}{1.5in}

To further understand stirring in the $d=1$ sparse case, we can observe the effects of the horizontal shear in Fourier space. When $a = 0$, \Eq{StdFourierUpdate} gives
\[
	\hat{\rho}_m^t=\sum_{k\in\bZ}\hat{\rho}_{k,m_1+m_2}^{t-1}J_{m_1-k}(0)=\hat{\rho}_{m_1,m_1+m_2}^{t-1}.
\]
This implies that each Fourier coefficient has its $y$-wavenumber shifted by $m_1$ and, in particular, that the coefficients $\hat{\rho}_{0,m_2}^t$ remain unchanged. The effect of applying the map with $\va=0$ is that the Fourier amplitudes with $m_1 > 0$ are shifted up in  $y$-wavenumber; thus, for a smooth initial condition, the amplitudes of the low modes become asymptotically small. Of course the $m_1=0$ amplitudes are unaffected, so an optimal protocol must use the nonzero $a_t$ to minimize the magnitude of these coefficients. Conversely, a poor protocol must accomplish the opposite.

\begin{figure}[h!t]
	\centerline{
		\renewcommand{\arraystretch}{0.01}
         	\begin{tabular}{cccc}
         	\includegraphics[width=1.5in,height=1.5in]{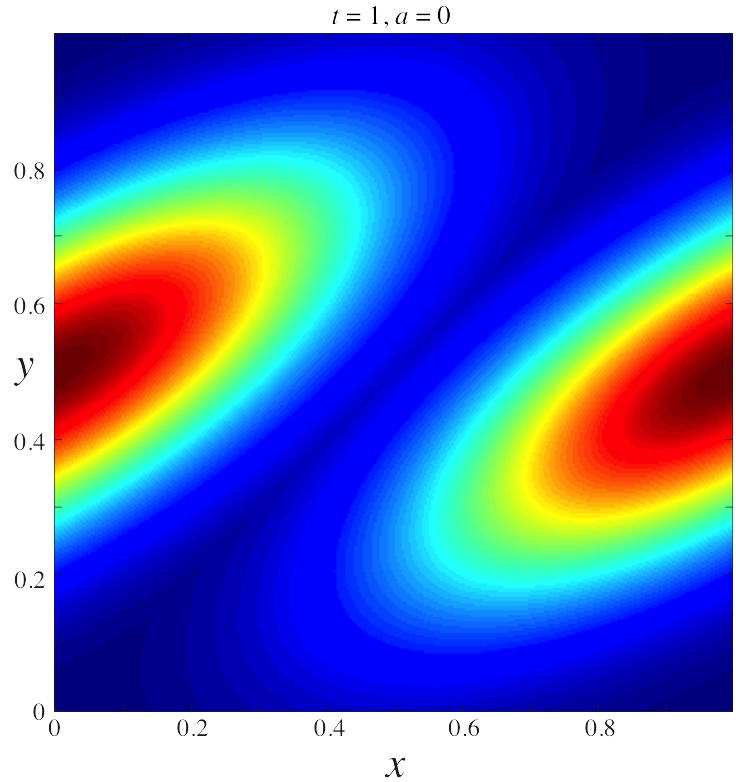} &
         	\includegraphics[width=1.5in,height=1.5in]{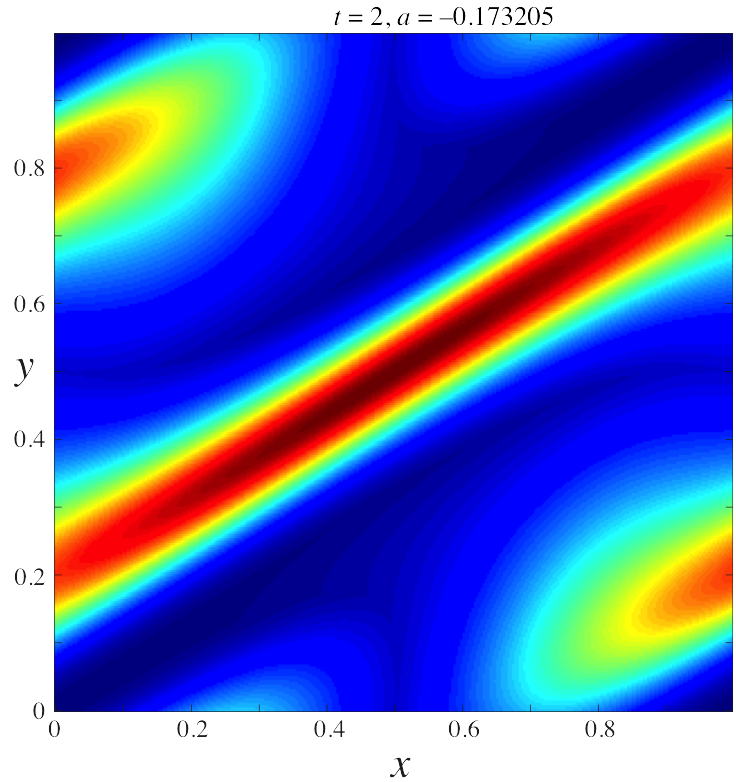} &
         	\includegraphics[width=1.5in,height=1.5in]{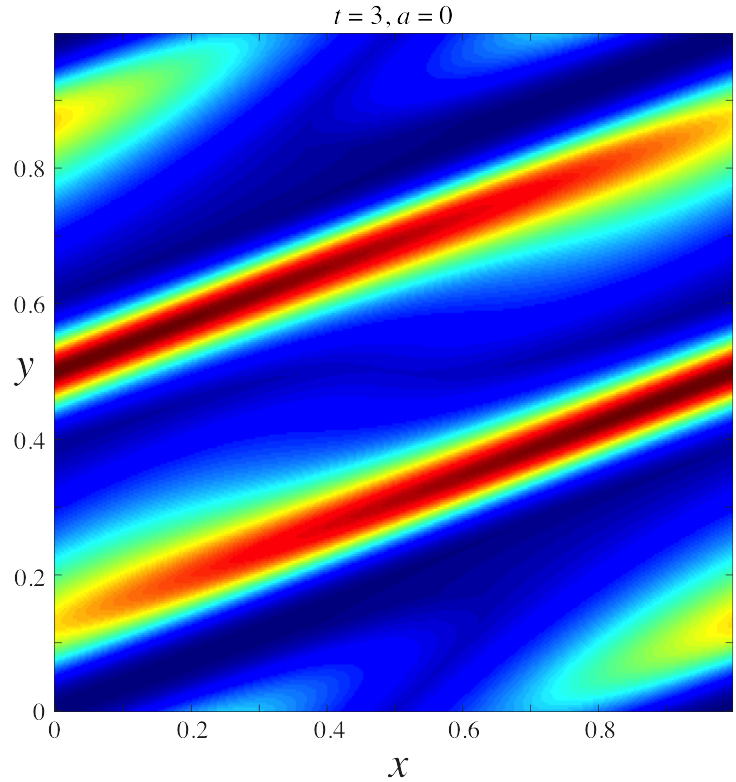} &
         	\includegraphics[width=1.5in,height=1.5in]{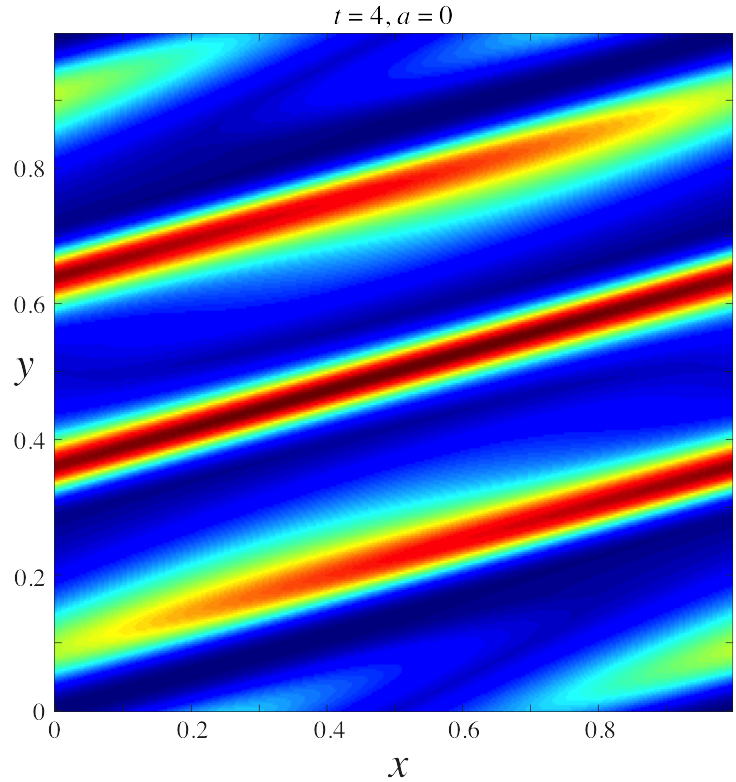}\\
         	\includegraphics[width=1.45in,height=1.45in]{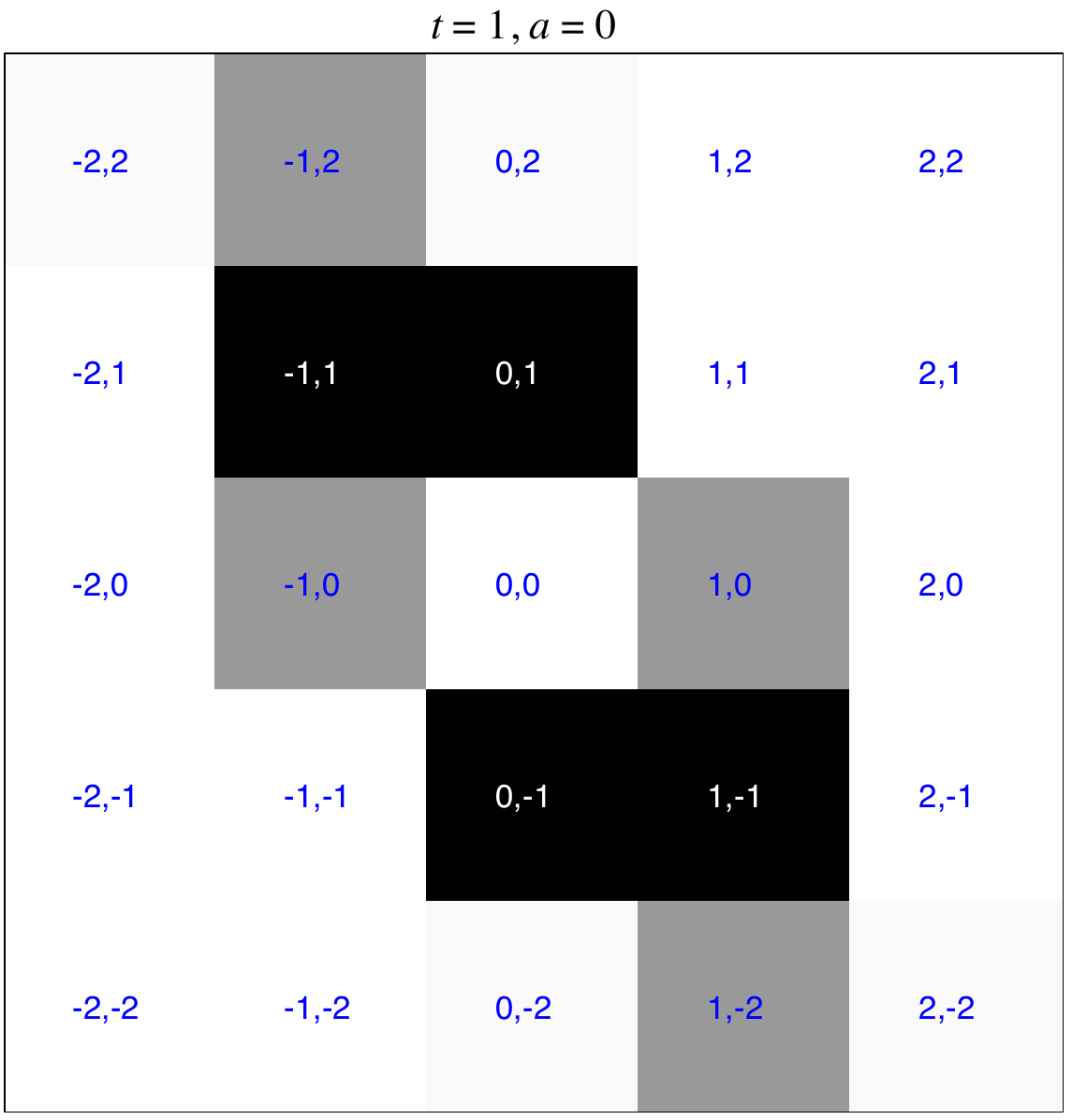} &
         	\includegraphics[width=1.45in,height=1.45in]{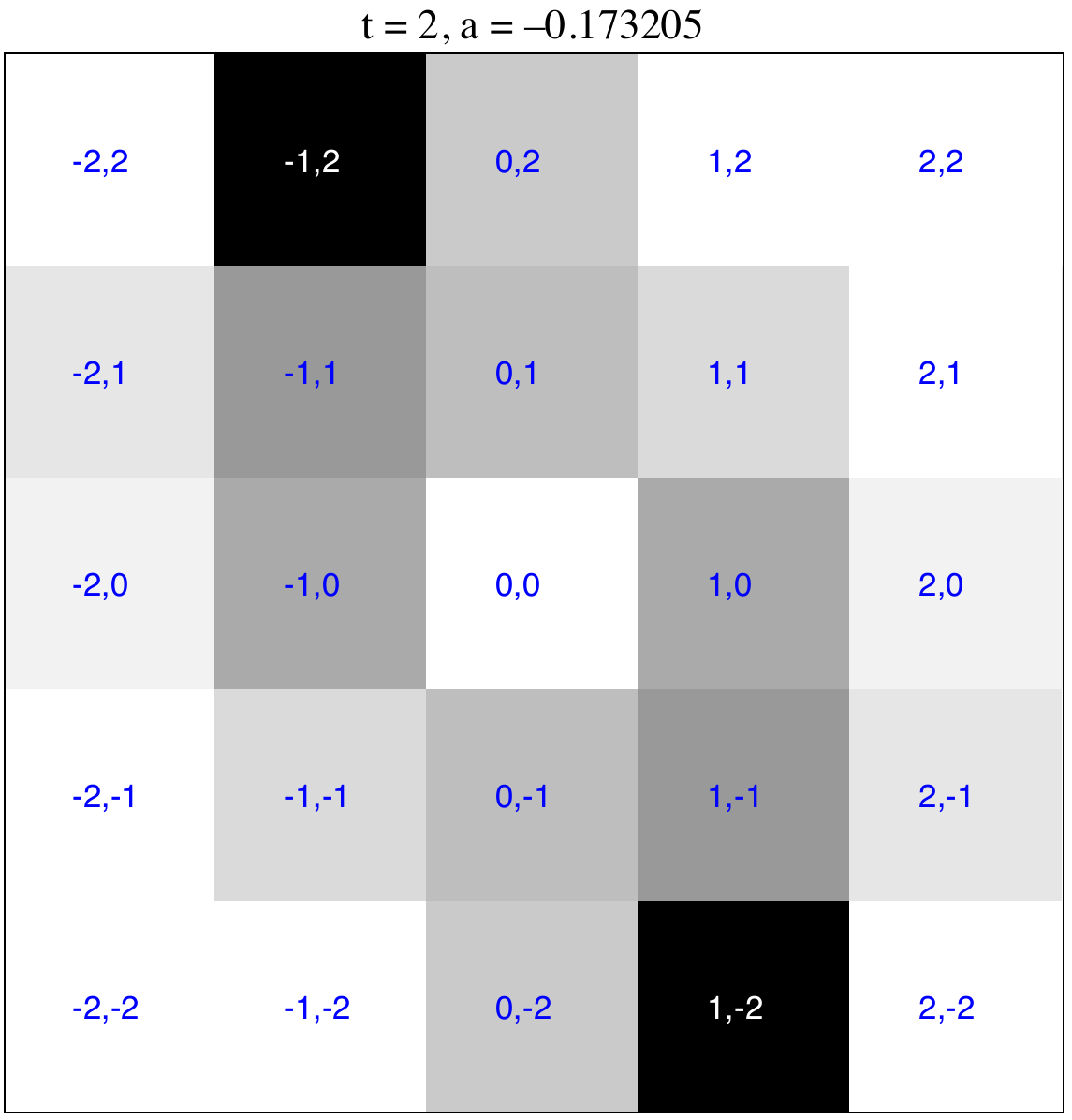} &
         	\includegraphics[width=1.45in,height=1.45in]{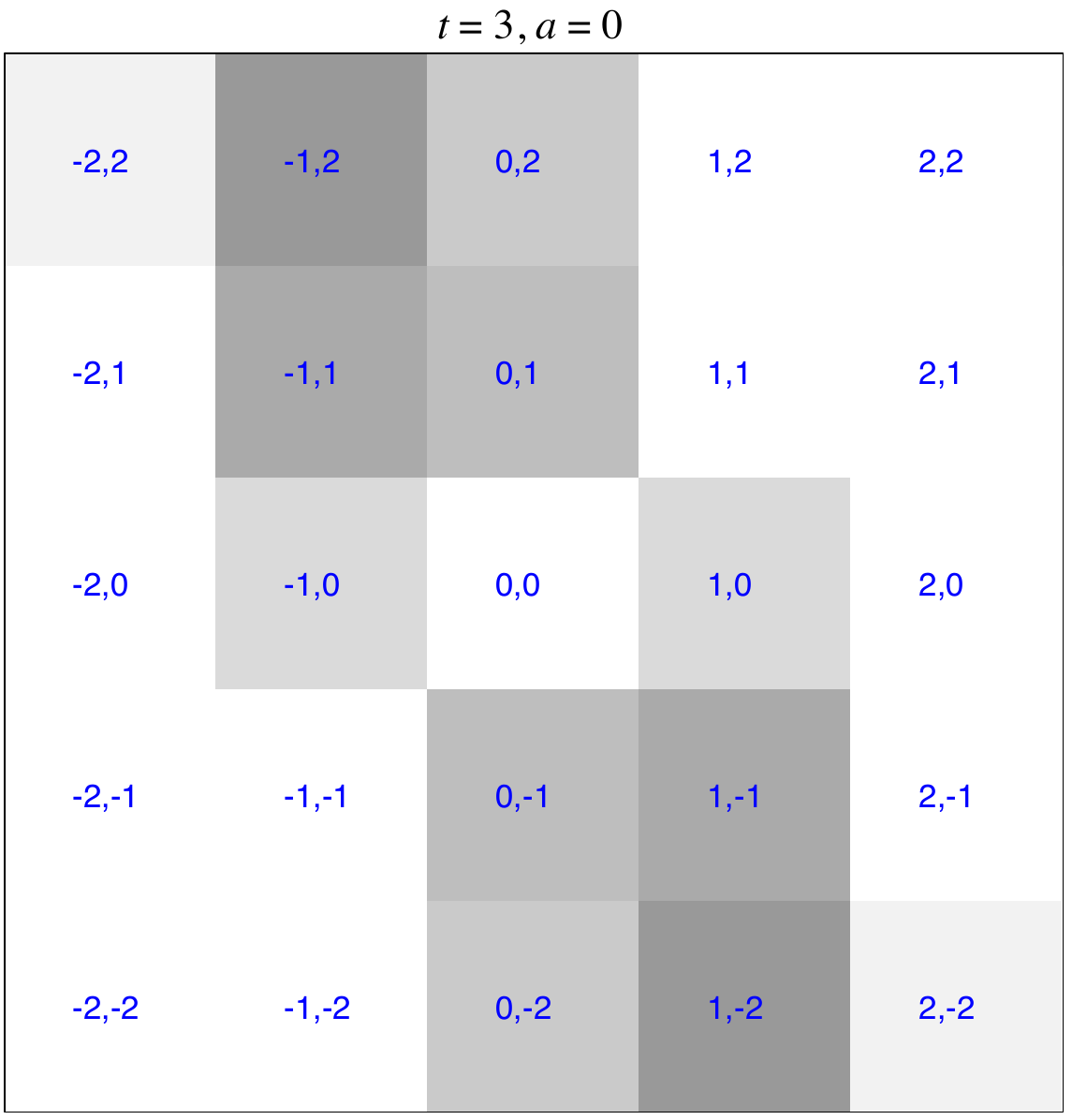} &
         	\includegraphics[width=1.45in,height=1.45in]{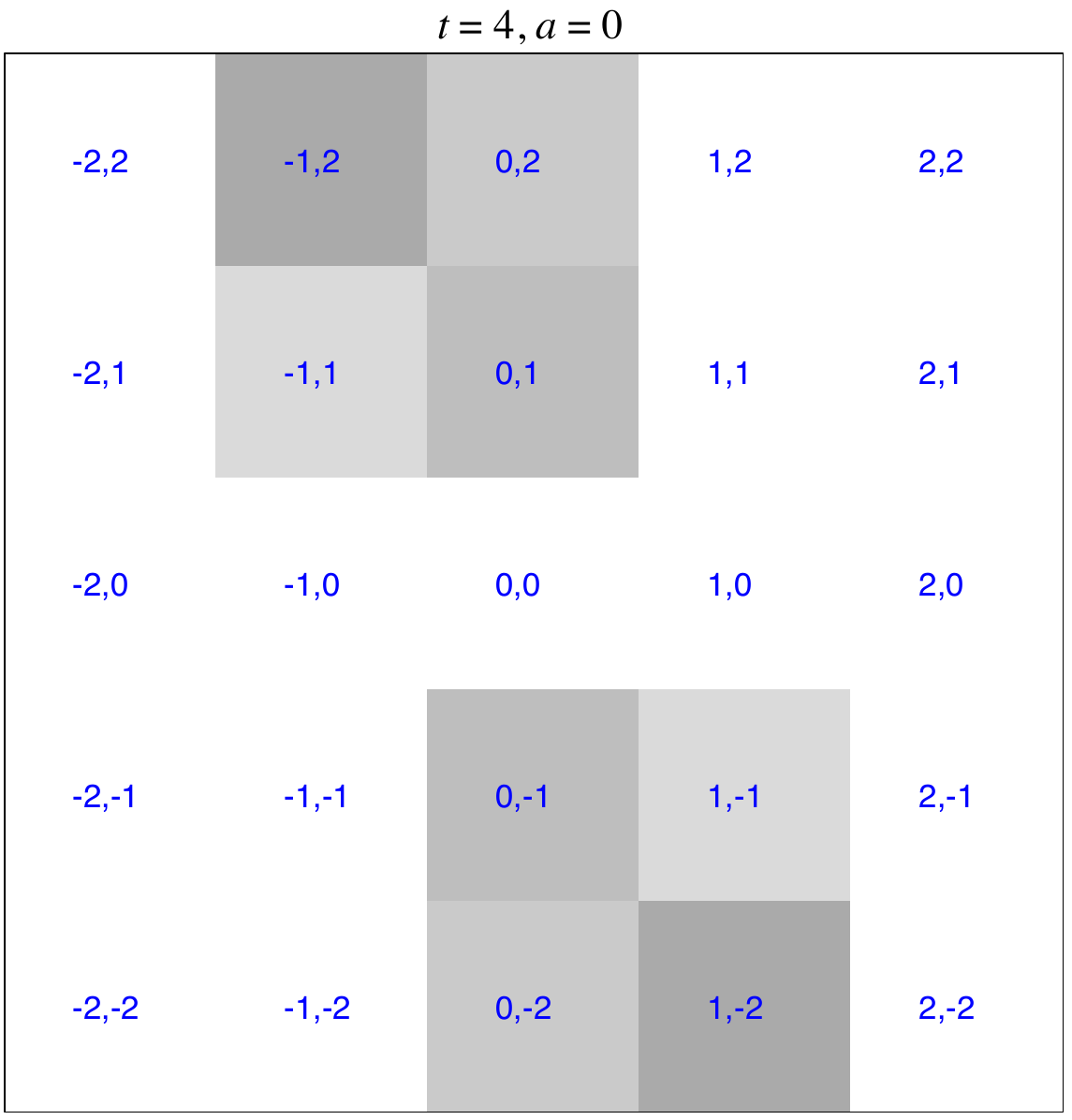}
			\end{tabular}
	}
	\caption{\footnotesize Evolution of $\rho^t(x,y)$ for the $d=1$ sparse RO protocol of \Fig{ChirikovSparse}. Top row shows the evolution in physical space. Bottom row shows the evolution of magnitudes of Fourier coefficients with mode numbers up to 2; the value of the invariant $(0,0)$ mode, $\hat{\rho}_{0,0}$ set to zero. The numbered pairs indicate the wave numbers and the greyscale indicates the coefficient magnitude, black being the maximum and white being zero. For a movie of all $12$ steps, see \url{http:/amath.colorado.edu/faculty/jdm/movies/ChirikovSparse.mp4}.}
	\label{fig:ChirikovMin}
\end{figure}

Let us now consider the effect of a single nonzero $a_t$ on the amplitudes with zero $y$ wave number. Since the horizontal shear has no effect on these modes, we may write the updated coefficients in terms of the initial vertically shearing step. Using 
\Eq{StdFourierUpdate}
the coefficients $\hat{\rho}_{0,m_2}$ are given by
\[
	\hat{\rho}_{0,m_2}^t=\sum_{k\in\bZ}\hat{\rho}_{k,m_2}^{t-1}J_k(-2\pi m_2a_t).
\]
After one step with $a_0=0$, the largest Fourier coefficients from the initial condition \Eq{InitialGaussian} with $m_1 = 0$ are $\hat{\rho}_{0,\pm 1} \approx -0.1204$. 
For a nonzero $a_2$, the $0,1$ mode becomes
\beq{TruncatedStep}
	\hat{\rho}_{0, 1}^2 \approx \hat{\rho}_{-2,1}^{1}J_{-2}(-2\pi a_2)+\hat{\rho}_{-1,1}^{1}J_{-1}(-2\pi a_2)+\hat{\rho}_{0,1}^{1}J_0(-2\pi a_2)+\hat{\rho}_{1,1}^{1}J_1(-2\pi a_2)+\hat{\rho}_{2,1}^{1}J_2(-2\pi a_2),
\eeq
where we truncate at the second mode number.
For the optimal case, $a_2=-0.1732$, \Eq{TruncatedStep} gives
\begin{align*}
	\hat{\rho}_{0,1}^2
	&\approx
	(-0.0008)(0.1340)+(-0.1204)(-0.4675)+(-0.1204)(0.7251)+ \\
		 & \quad \quad \quad (-0.0008)(0.4675)+(0.0000)(0.1340)\\
	&\approx-0.0315 .
\end{align*}
For this negative $a_2$, the Bessel functions $J_{-1}$ and $J_0$ have opposite signs. This makes the largest terms nearly cancel, causing the value $\hat{\rho}_{0,1}^2$ to decrease by a factor of $4$. This can be seen in \Fig{ChirikovMin}, which shows the progression of the density in both physical space (top row) and Fourier space (bottom row). The larger Fourier amplitudes in the figure are shown as darker squares, and moving from the $t=1$ column, to the $t=2$ column, shows that the $(0,\pm1)$ mode amplitudes decrease significantly. The following steps of shear then push the other large coefficients, the $(\pm1,\mp2)$ modes in the figure, to larger $y$-wavenumbers, so that the magnitude of the low Fourier modes diminishes significantly after just the four time steps shown. 

\begin{figure}[h!t]
	\centerline{
		\renewcommand{\arraystretch}{0.01}
         	\begin{tabular}{rrrr}
         	\includegraphics[width=1.5in]{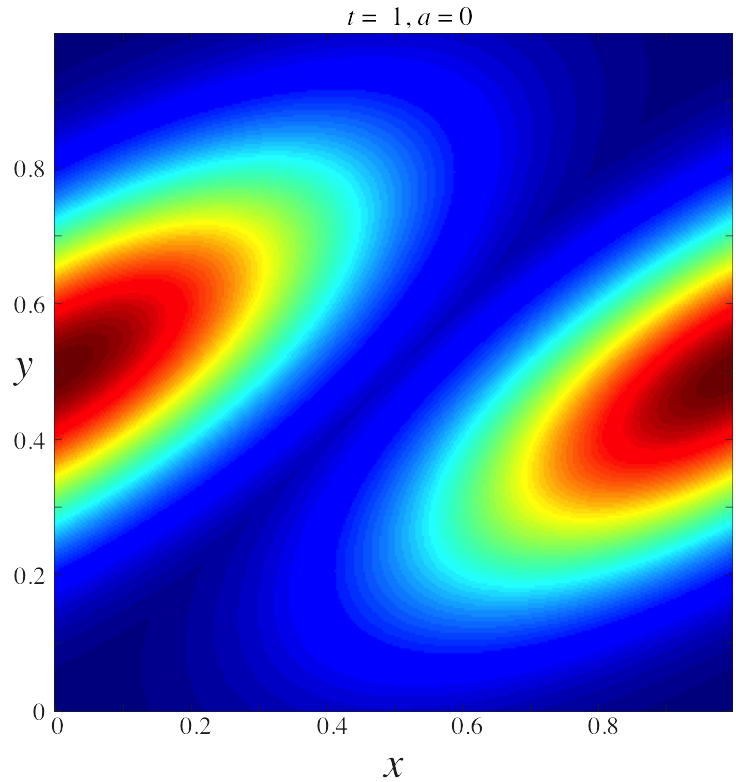} &
         	\includegraphics[width=1.5in]{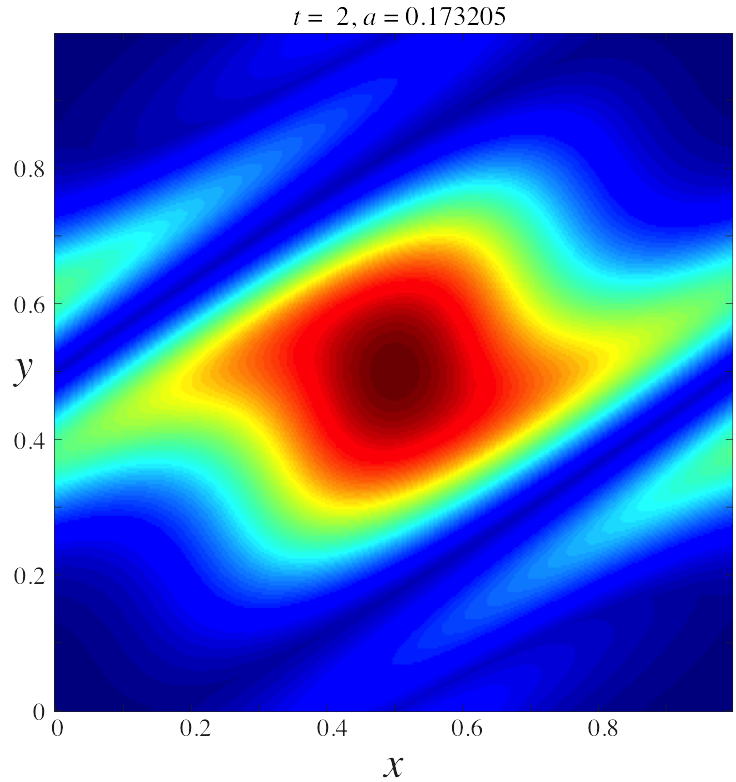} &
         	\includegraphics[width=1.5in]{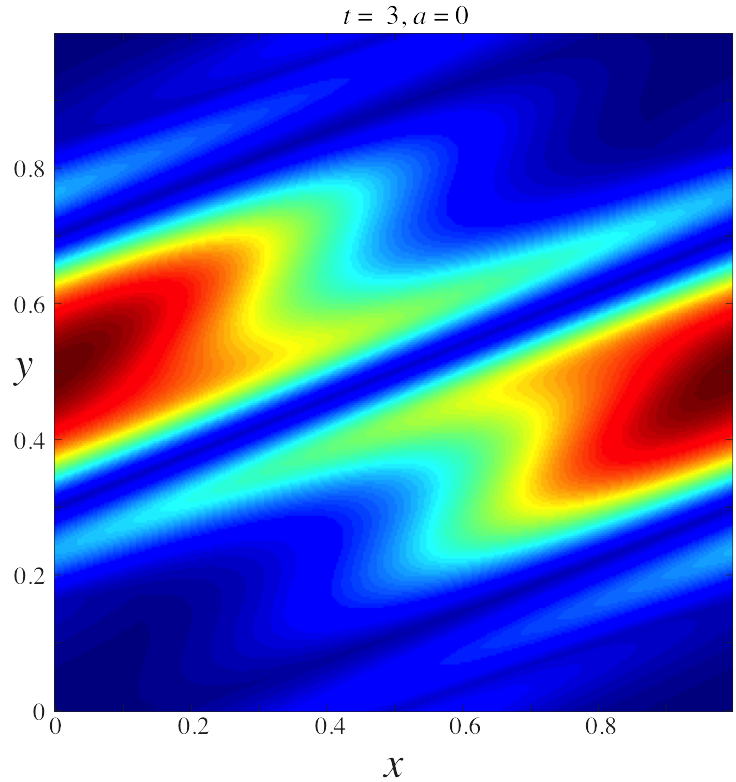} &
         	\includegraphics[width=1.5in]{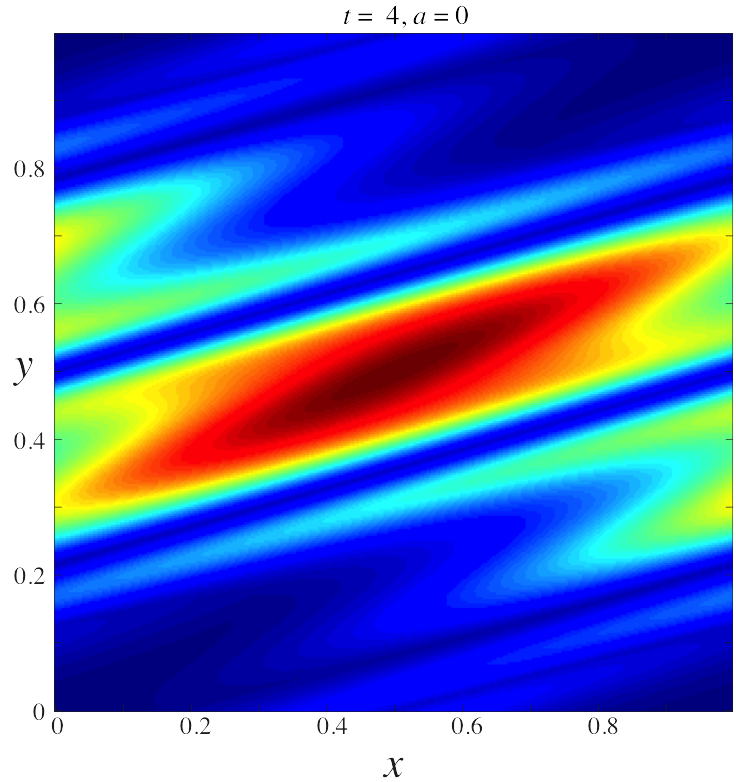}\\
         	\includegraphics[width=1.45in]{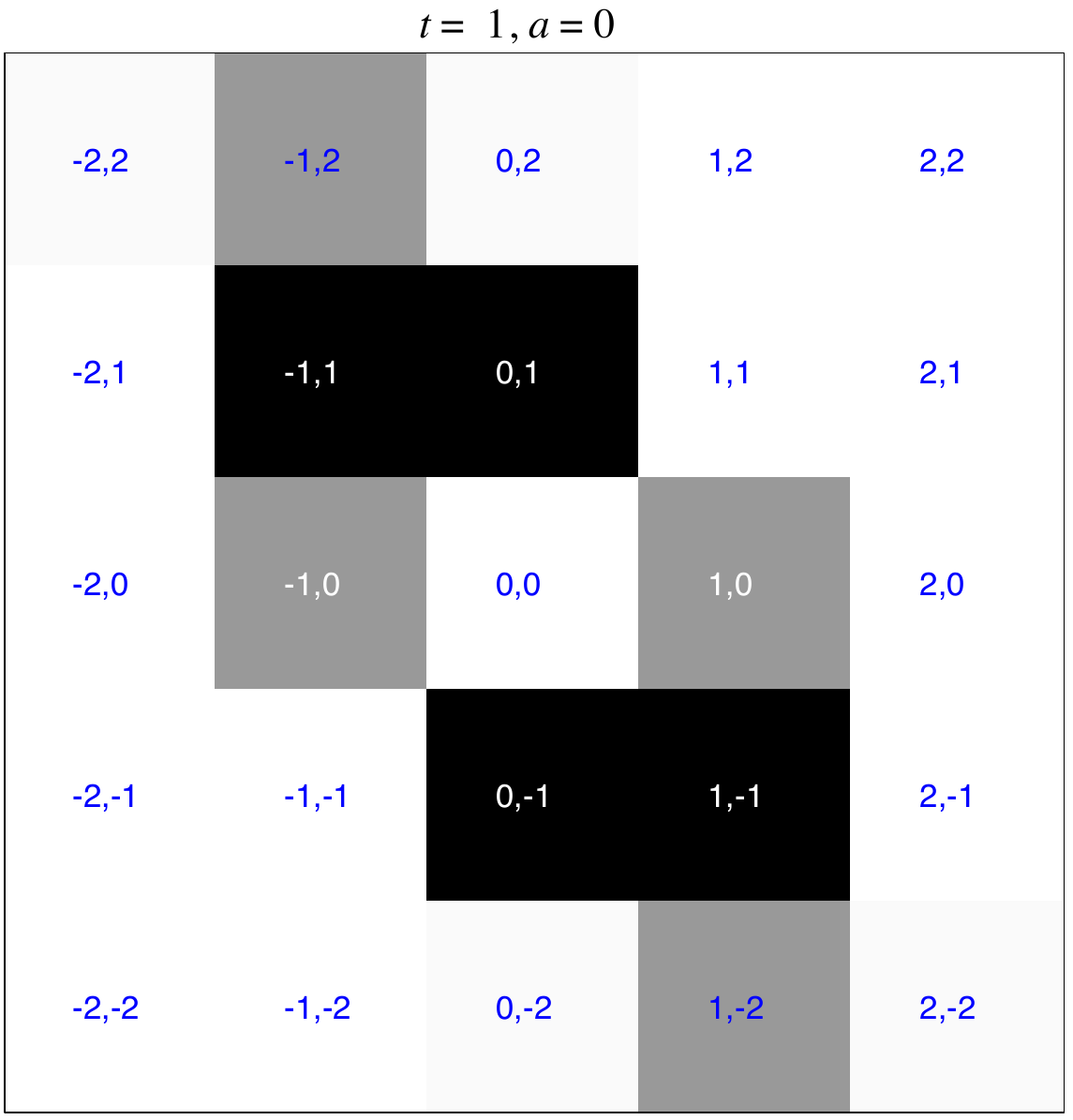} &
         	\includegraphics[width=1.45in]{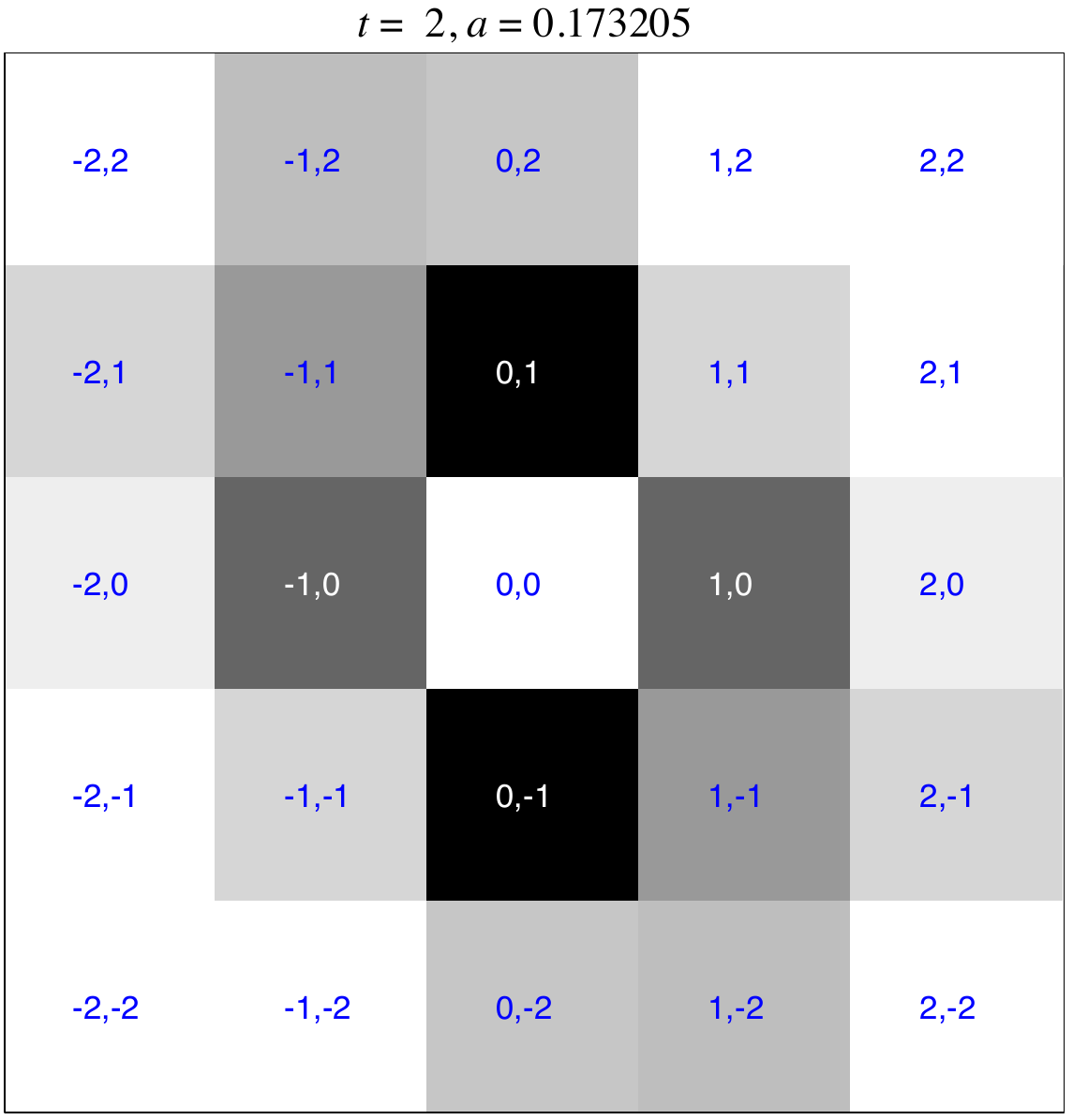} &
         	\includegraphics[width=1.45in]{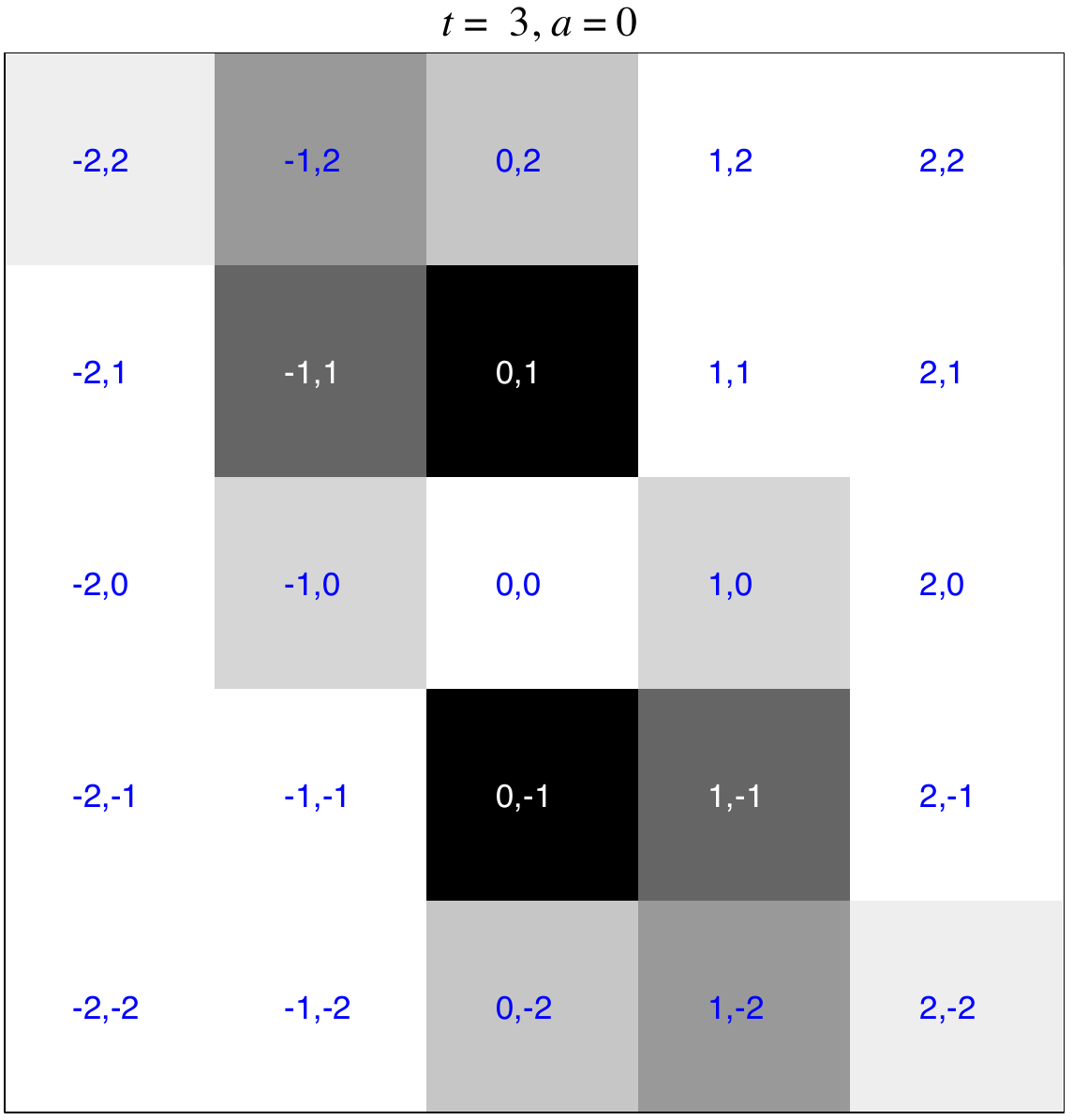} &
         	\includegraphics[width=1.45in]{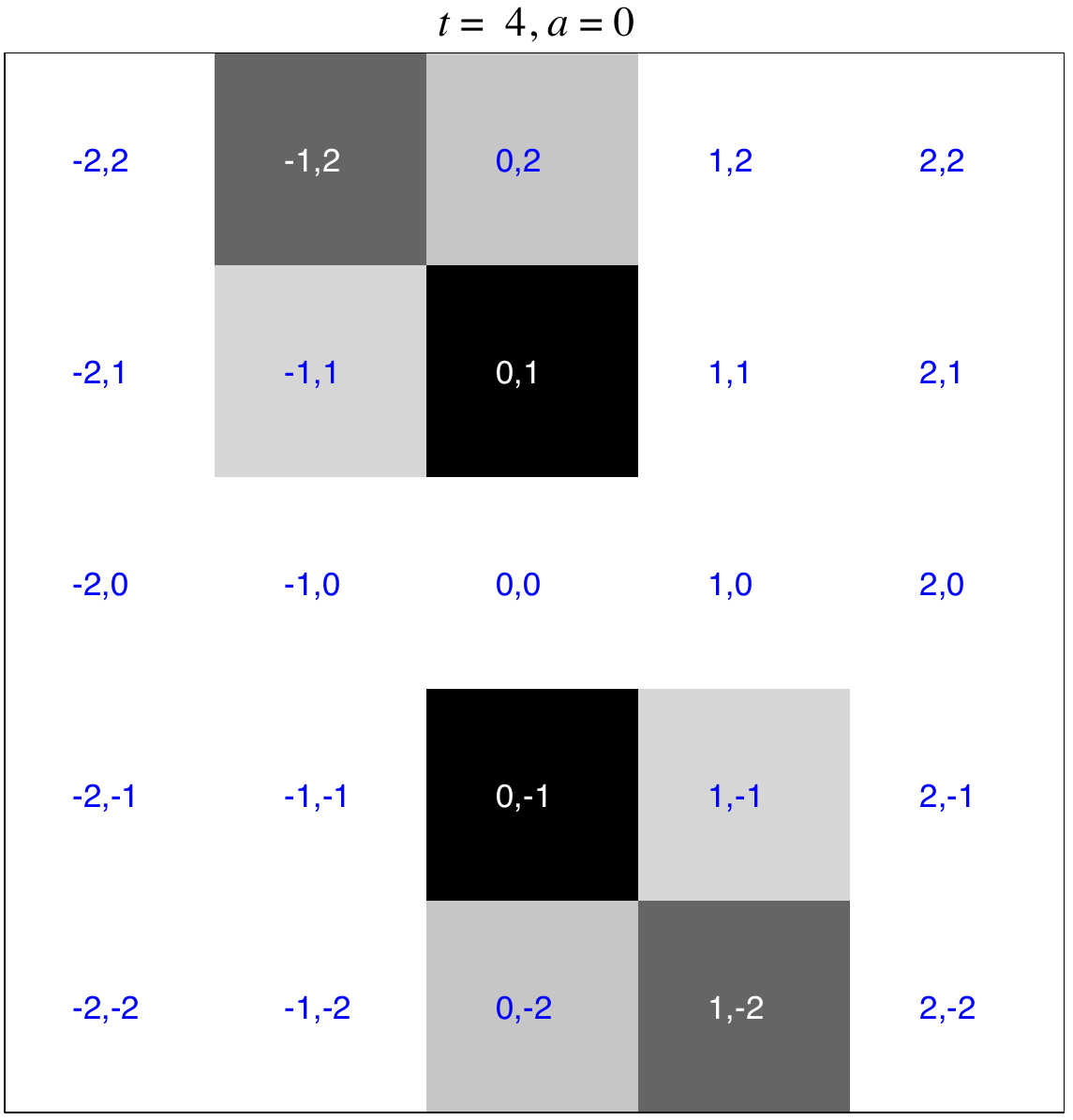}
			\end{tabular}
	}
	\caption{\footnotesize Evolution of $\rho^t(x,y)$ for the $d=1$ sparse RW protocol of \Fig{ChirikovSparseBad}.  The eight panels are otherwise equivalent to those in \Fig{ChirikovMin}.  For a movie of all $12$ steps, see \url{http:/amath.colorado.edu/faculty/jdm/movies/ChirikovSparseBad.mp4}.}
	\label{fig:ChirikovMax}
\end{figure}

By contrast, the $d=1$ RW protocol has the opposite effect. When $a_2= +0.1732$, \Eq{TruncatedStep} gives
\begin{align*}
\hat{\rho}_{0, 1}^2
	&\approx
	(-0.0008)(0.1340)+(-0.1204)(0.4675)+(-0.1204)(0.7251)+\\
	& \quad \quad \quad (-0.0008)(-0.4675)+(0.0000)(0.1340)\\
	&\approx -0.1433
\end{align*}
The sign change causes this mode amplitude to grow in magnitude, resulting in a clumping of density that can not be undone by the remaining steps of purely horizontal shear. This is also illustrated in \Fig{ChirikovMax}, which shows the first four steps of the sparse RW protocol in physical and Fourier space.


\subsection{Other Initial Conditions}\label{sec:StdOtherIC}

Given this intuition for optimizing mixing for the Gaussian initial density \Eq{InitialGaussian}, we now determine if the sparse protocol is still effective for other initial conditions. We study initial densities constructed by randomly superimposing periodic bump functions of the form 
\beq{ShearSignTest}
	f(x,y; C, \eps,u,v)= C e^{-\eps \sin^2[\pi(x-u)]  -\eps \sin^2[\pi(y-v)]}, 
\eeq
so that the center of the bump can range over the torus. 
Adding a set of sinusoidal perturbations, we construct the density by defining
\beq{TestConcentration}
	\rho^0(x,y)=\sum_{k=1}^K f(x,y;C_k,\eps_k,u_k,v_k)+
		\sum_{l=1}^L C_l \sin(2\pi n_l \cdot z - \phi_l).
\eeq
We randomly choose the centers $(u,v) \in [0,1)^2$, the widths $\eps \in [0.1,5]$, and $K$ and $L$ from $1$ to $5$. The wavenumbers $n_l$ ranged over $\{-10,10\}^2$ and the phases $\phi_l$ over $[0,2\pi)$. The amplitudes $C_l$ of the sine terms are kept smaller than $20\%$ of the maximum bump amplitude (to create a noisy effect), and each off the function \Eq{TestConcentration} is scaled so that $\Sob{\rho^0}=1$.


The question becomes: is a $d=1$ sparse protocol, in which we apply all the energy at a single step, still nearly optimal for varying initial conditions? If this is the case, than the optimal $d=2$ case, in which we divide the energy between two steps should resemble a $d=1$ protocol, in that the energy at one step should be much larger than that at the second. For each of $1200$ initial conditions, we found the best $d=2$ protocol from $10^4$ random trials with the energy \Eq{StdChirikovEnergy}. Figure~\ref{fig:Chirikov2DTest}(a) shows a histogram of step times $t_{max}$ at which the larger of the two nonzero amplitudes, $|a_{t_{max}}|$, was applied. The majority of these optimal protocols have $t_{max} \le 2$: as for \Eq{InitialGaussian} the largest amplitude is used early in the twelve applications of the map. Figure~\ref{fig:Chirikov2DTest}(b) shows the fraction of protocols for which $|a_{t_{max}}|^2 > r E$; that is, the fraction of trials for which the largest amplitude step contains more than a fraction $r$ of the available energy. Nearly three-quarters of protocols apply more than 70\% of the available energy in one step. This supports the conclusion that the scheme applying a large shear  early in the evolution is nearly optimal for the majority of initial conditions.

\InsertFigTwo{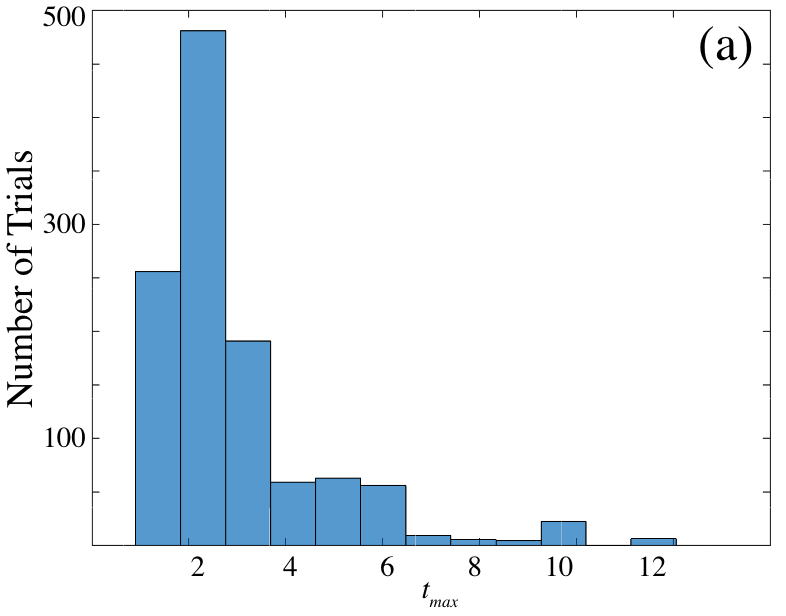}{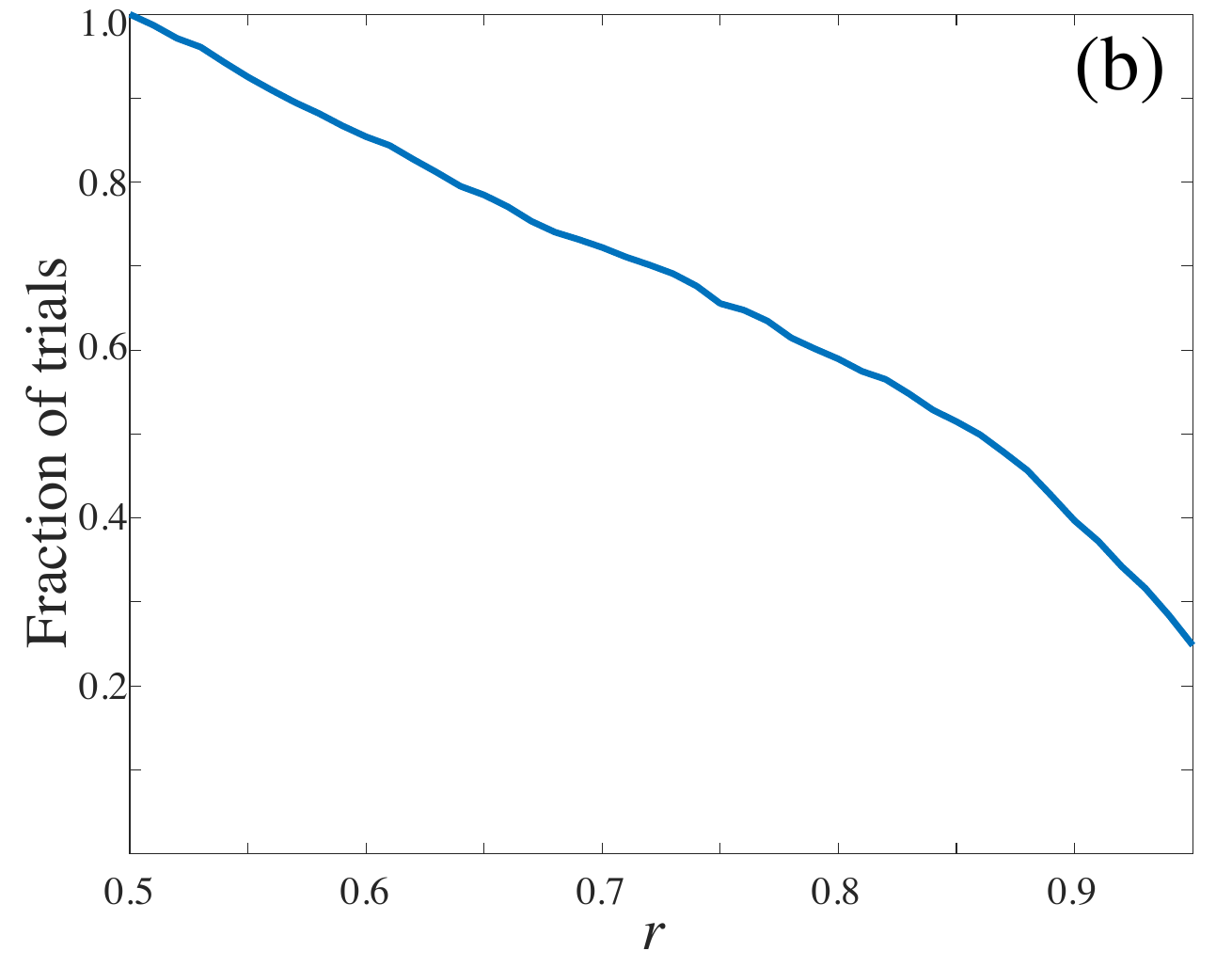}{(a) Histogram of time at which largest $|a_t|$ occurs for $d=2$ RO protocols. (b) The fraction of the RO protocols that satisfy $a_{t_{max}}^2>rE$ as a function of $r$. We perform $10^4$ trials to find the $d=2$ RO protocols for 1200 different initial conditions of the form \Eq{TestConcentration} on a grid of size $N=128$ with \Eq{StdChirikovEnergy}.}{Chirikov2DTest}{3in}

\section{Harper Map}\label{sec:HarperMap}

The Harper map \Eq{HarperMap} differs from the standard map in that the horizontal shear is nonlinear with the additional amplitude parameter $b$. The mixing protocol is now determined by two  parameter vectors $\va$ and $\vb$ so that the dimensionality of the optimization problem is doubled, allowing for more flexibility. Moreover, since $b_t$ can vary, the energy \Eq{EnergyHarper} in the horizontal shear is no longer fixed as it was for \Eq{EnergyChirikov}. In this section, we search for an optimal mixing protocol by means of the random optimization method and also develop other methods that require less computation.

For most of the results in this section, we again use the Gaussian density profile \Eq{InitialGaussian}, the time constraint $T=12$, and fix
\[
	E_H=\Vert{\va}\Vert^2+\Vert{\vb}\Vert^2= 0.5.
\]
Unless otherwise specified, we use a $301\times301$ grid, setting $M = 150$.

\subsection{Fixed Points}

Whenever $ab \neq 0$, the Harper map has exactly four fixed points, $(0,0)$, $(0,\tfrac12)$, $(\tfrac12,0)$ and $(\tfrac12,\tfrac12)$. For small $a$ and $b$ the map can be approximated by the system
\[ 
	(\dot{x},\dot{y}) = (\partial_y H, -\partial_x H)
\]
with Hamiltonian
\beq{HarperHam}
H(x,y)=-\frac{a}{2\pi}\cos(2\pi x)-\frac{b}{2\pi}\cos(2\pi y).
\eeq
Contours of this Hamiltonian are shown in \Fig{HarperHam} for various parameters.
For the flow, the stability of the fixed points depends only on the relative signs of $a$ and $b$. When $a$ and $b$ are of the same sign, the points $(0,0)$ and $(\tfrac12,\tfrac12)$ are hyperbolic and $(0,\tfrac12)$ and $(\tfrac12,0)$ are elliptic. When $a$ and $b$ are of opposite sign, the stabilities are are reversed. Note that $(\tfrac12,\tfrac12)$, the center of the domain, is elliptic when $ab>0$ and hyperbolic when $ab < 0$. When $|a| > |b|$ there is more vertical shearing; conversely, when $|b|> |a|$, the horizontal shear is dominant. 

The dynamics of the map \Eq{GenStdMap} with \Eq{HarperMap}---for fixed parameter values---is similar to the Hamiltonian flow when $|a|$, $|b| \ll 1$, except for chaotic layers near the separatrices; two examples are shown in \Fig{HarperPhasePortrait}. For the map, the elliptic fixed points undergo a period doubling bifurcation at $|ab| = \pi^{-2}$, becoming hyperbolic with reflection. However, for the total energy levels that we will choose, none of the shear amplitudes will be large enough for this to occur.

\InsertFigSix{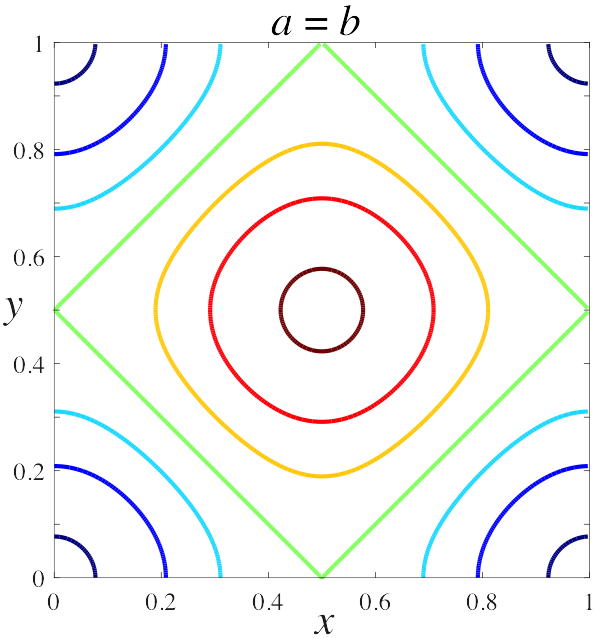}{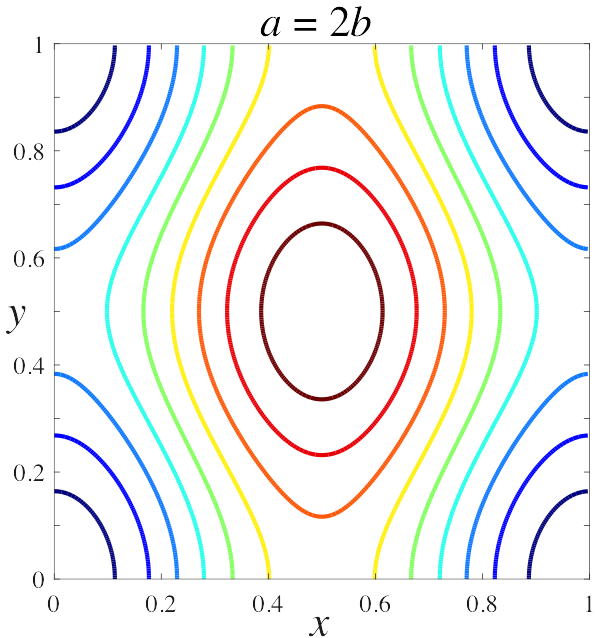}{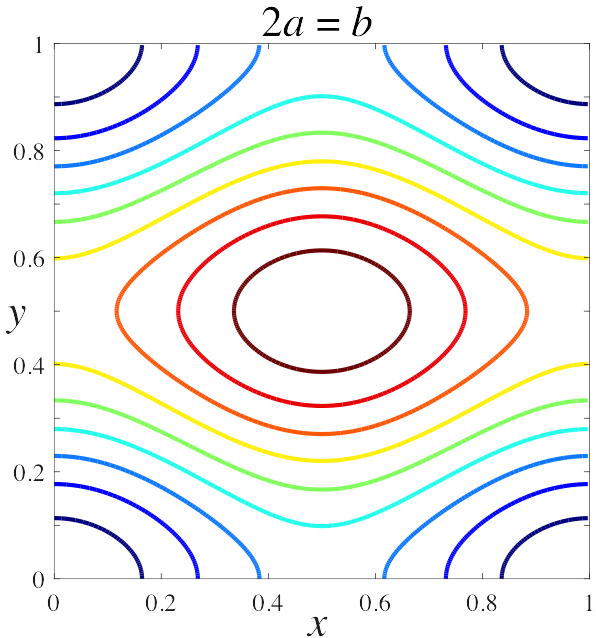}{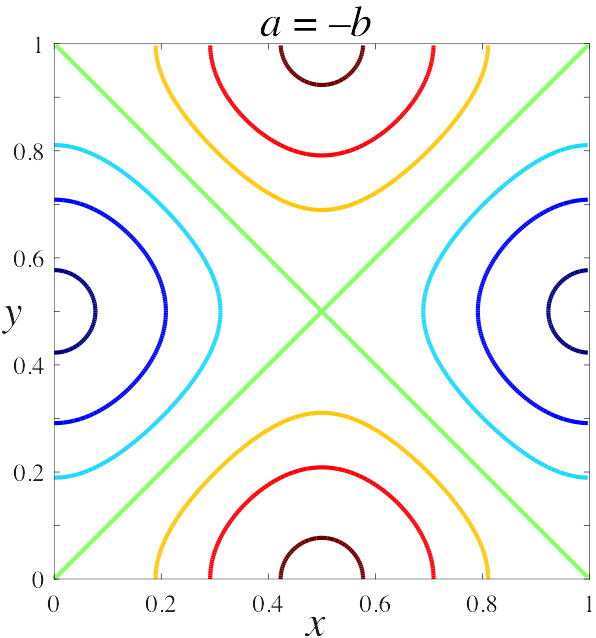}{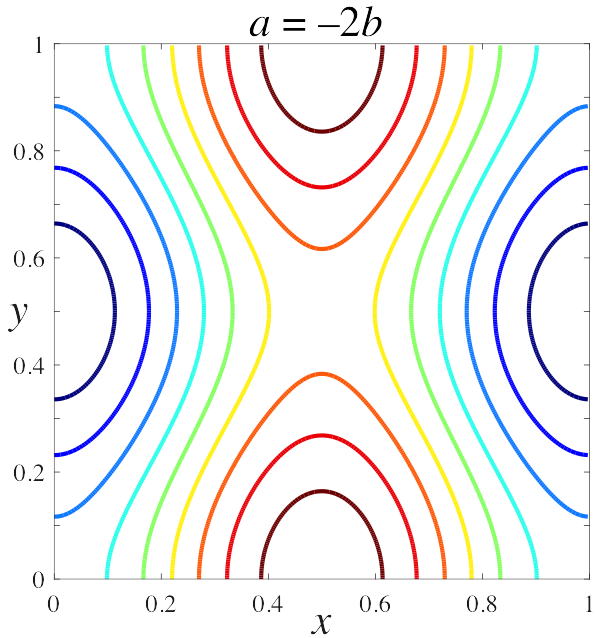}{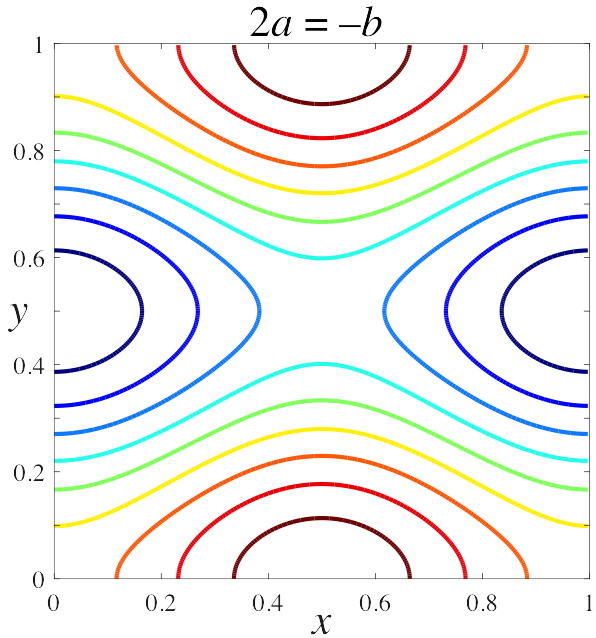}{Level sets of the Hamiltonian \Eq{HarperHam} with various relative values of $(a,b)$. }{HarperHam}{2in}

\InsertFigTwo{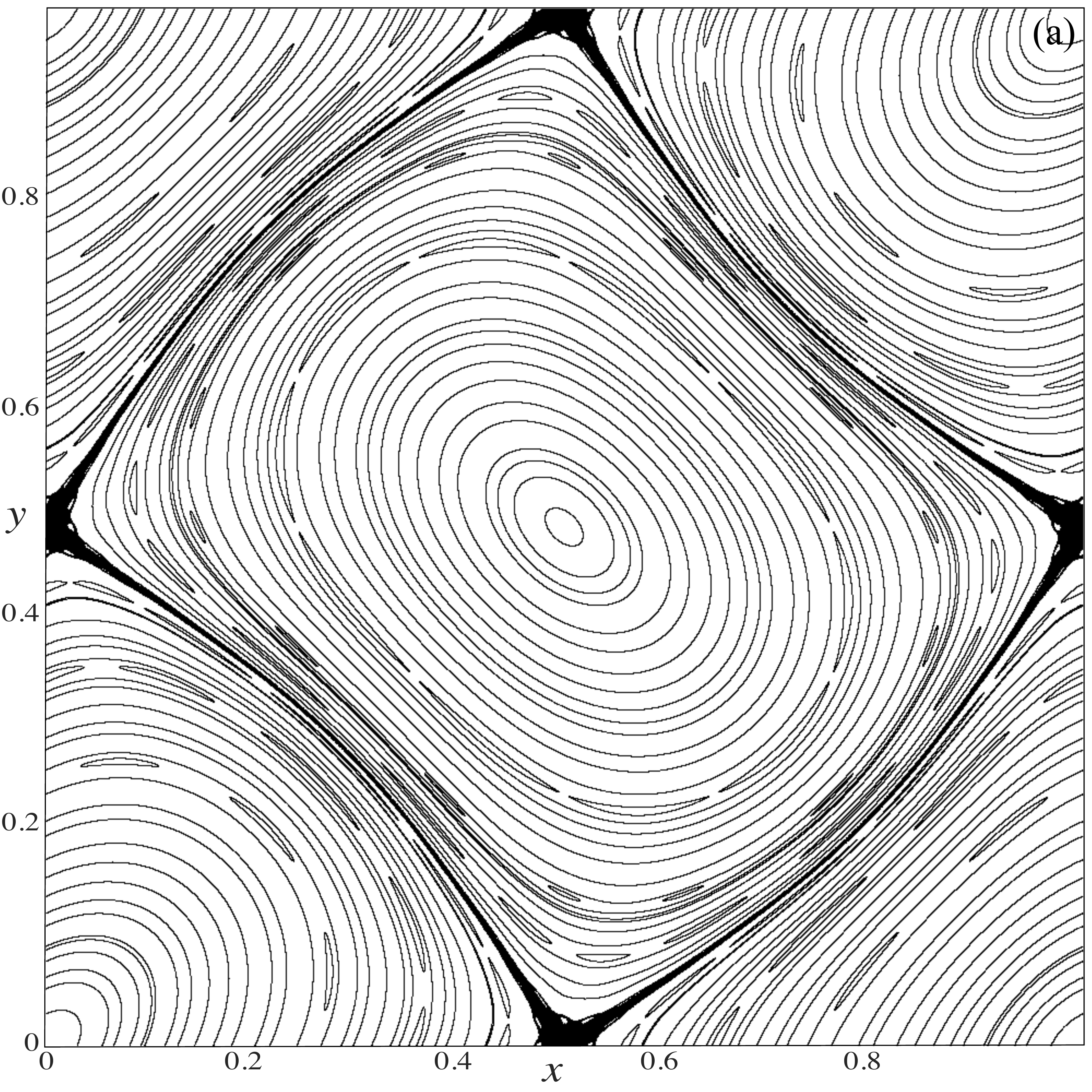}{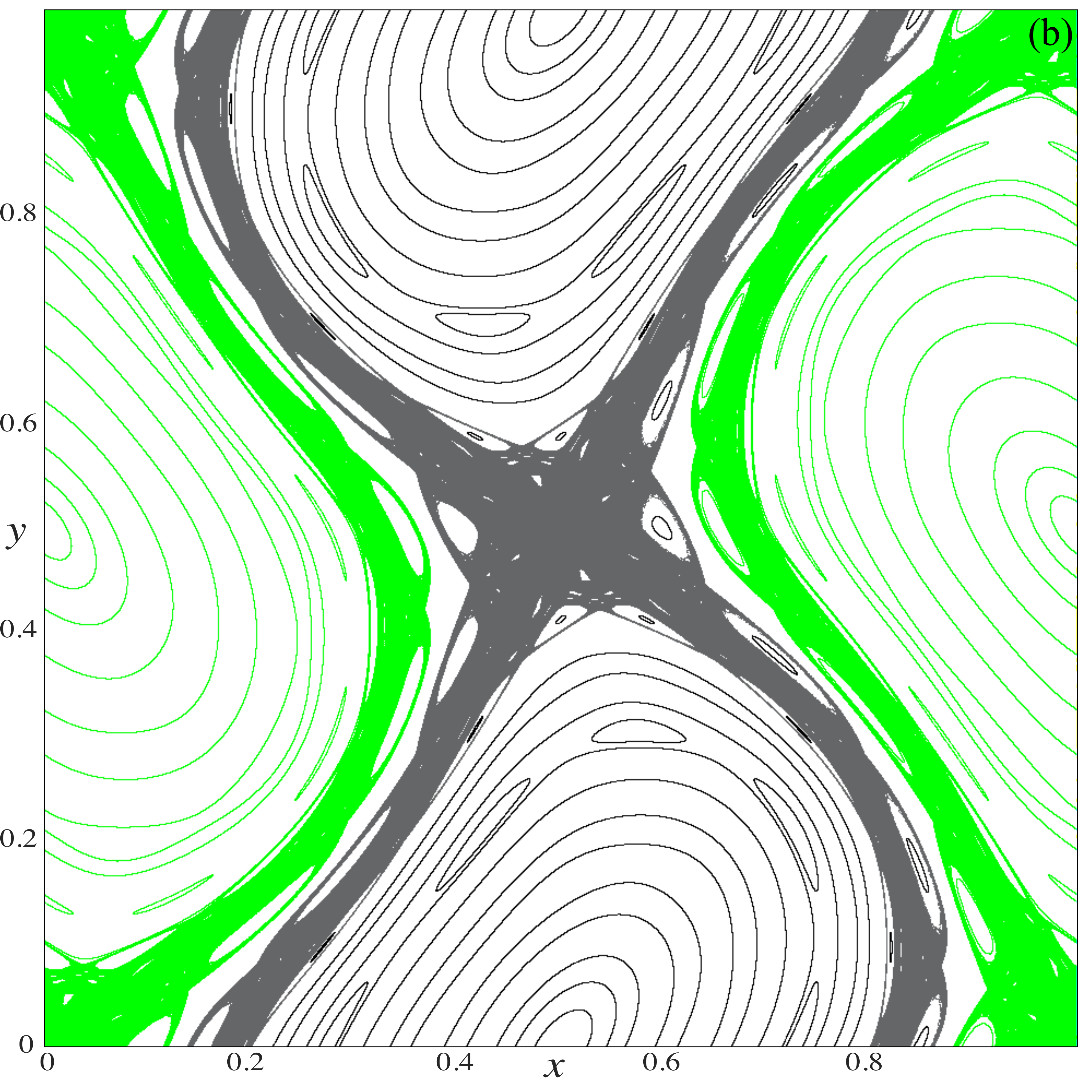}{Phase portraits of the Harper map \Eq{HarperMap} with constant amplitudes, (a) $a_t = b_t = 0.1443$ and (b) $(a_t,b_t) = (-0.211881,0.15726)$. With the relatively small amplitudes in (a) the evolution resembles the Hamiltonian flow shown in \Fig{HarperHam}, 
with a thin chaotic separatrix layer. For the larger amplitudes in (b) there is more chaos.}{HarperPhasePortrait}{2in}
\subsection{Constant Shear Amplitude}\label{sec:HarperConst}

\begin{figure}[h!t]
	\centerline{
		\renewcommand{\arraystretch}{0.01}
         	\begin{tabular}{cccc}
         	\includegraphics[width=1.5in]{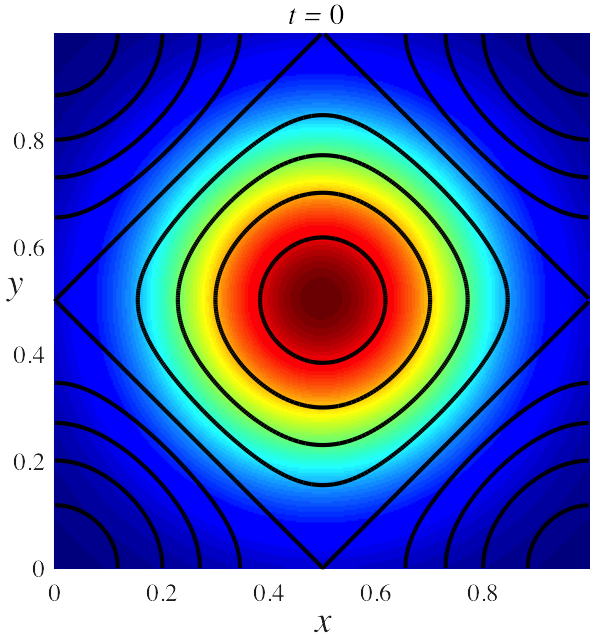} &
         	\includegraphics[width=1.5in]{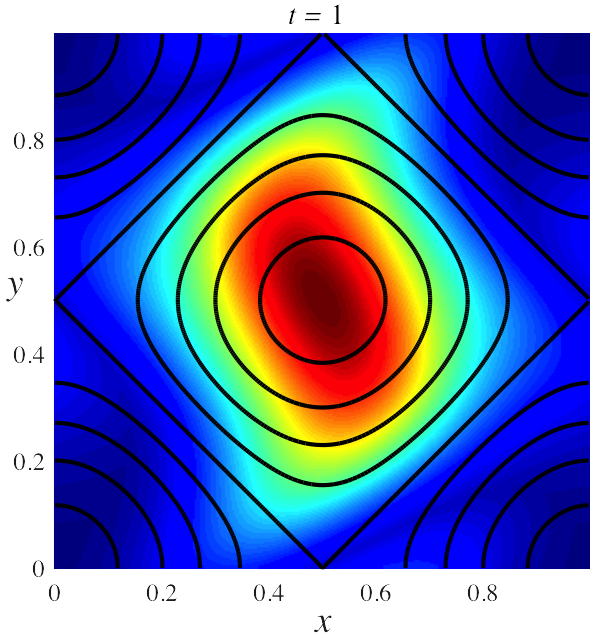} &
         	\includegraphics[width=1.5in]{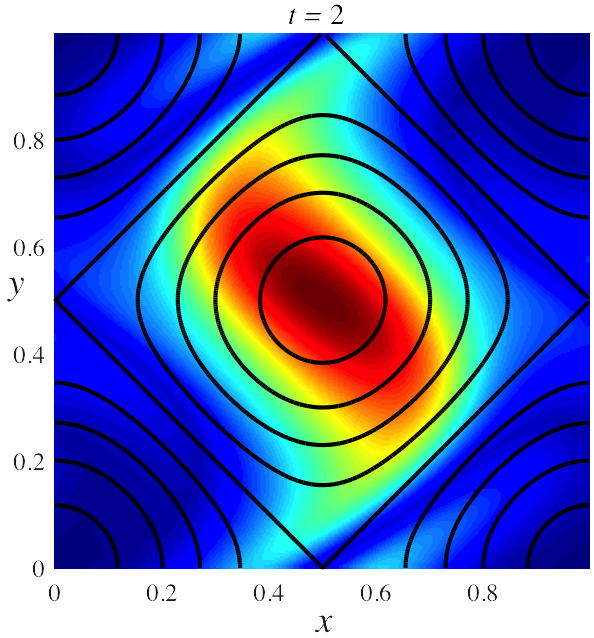} &
         	\includegraphics[width=1.5in]{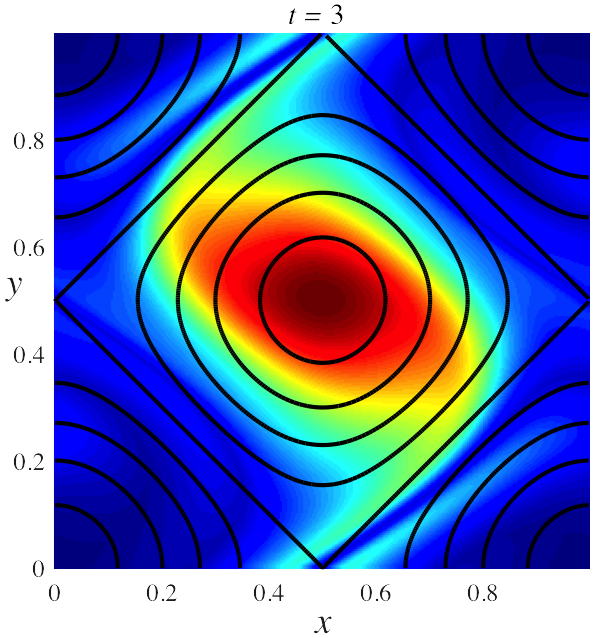}\\
         	\includegraphics[width=1.5in]{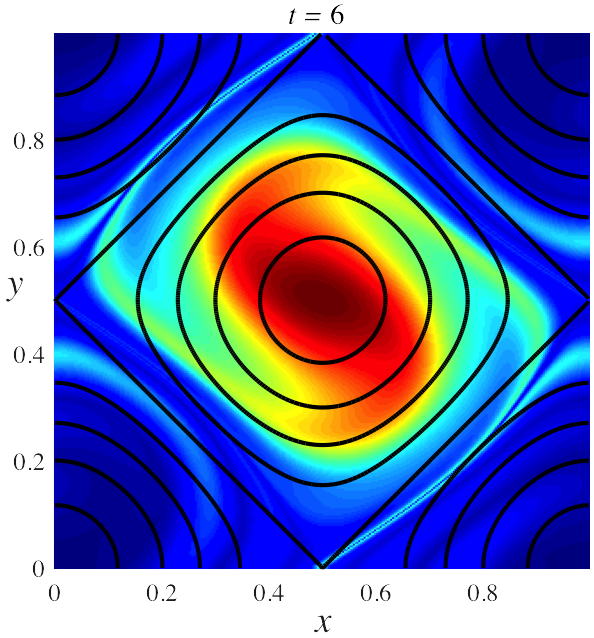} &
         	\includegraphics[width=1.5in]{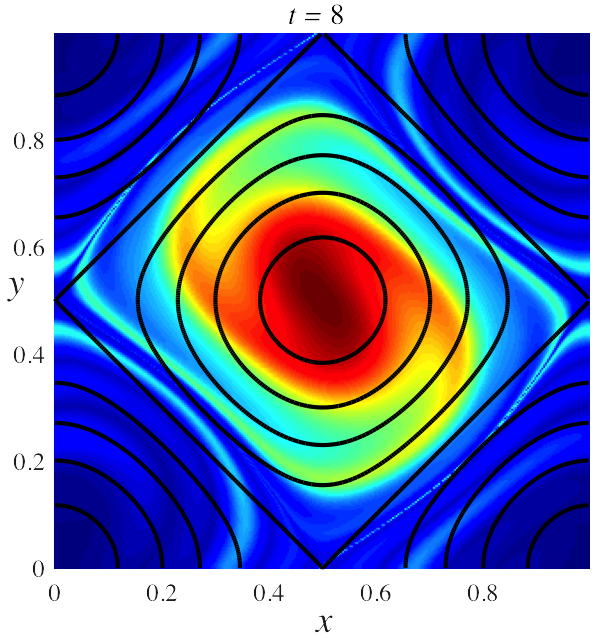} &
         	\includegraphics[width=1.5in]{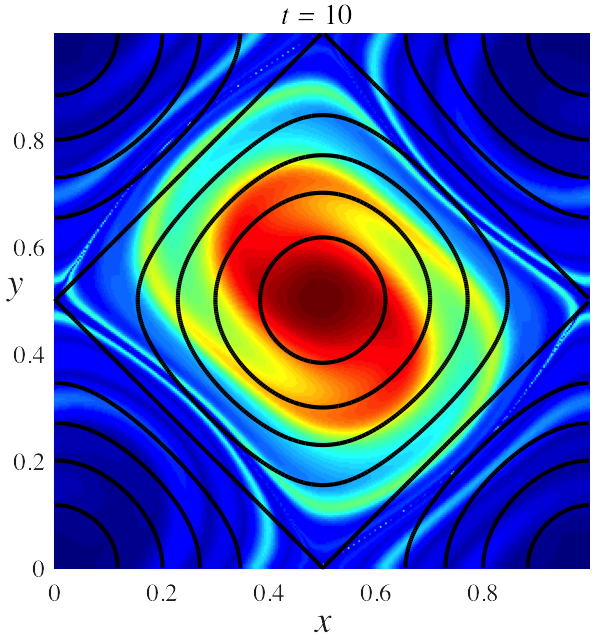} &
         	\includegraphics[width=1.5in]{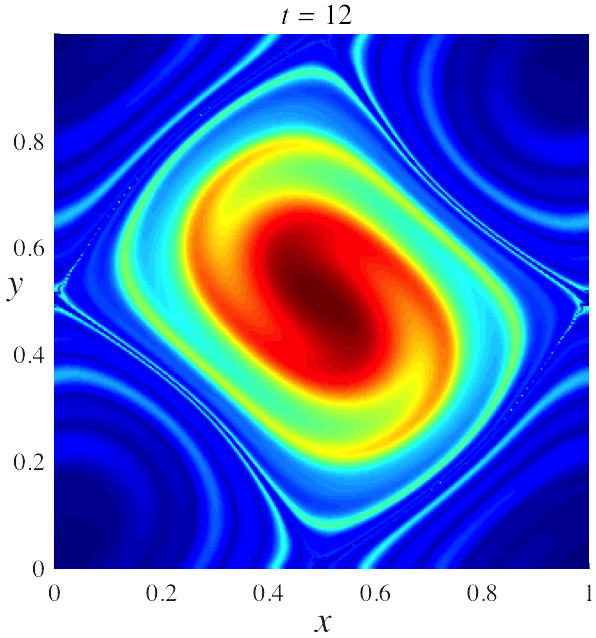}
			\end{tabular}
	}
	\caption{\footnotesize Evolution of $\rho^t(x,y)$ for the Harper map \Eq{HarperMap} with initial condition \Eq{InitialGaussian} for $t=0,1,2,3$ (top row) and $t=6,8,10,12$ (bottom row) with constant strengths $a_t=b_t=0.1443$.
Contours correspond to the Hamiltonian \Eq{HarperHam}.  For a movie of all $12$ steps, see \url{http:/amath.colorado.edu/faculty/jdm/movies/HarperConst.mp4}. }
	\label{fig:HarperConst}
\end{figure}

\begin{figure}[h!t]
	\centerline{
		\renewcommand{\arraystretch}{0.01}
         	\begin{tabular}{cccc}
         	\includegraphics[width=1.5in]{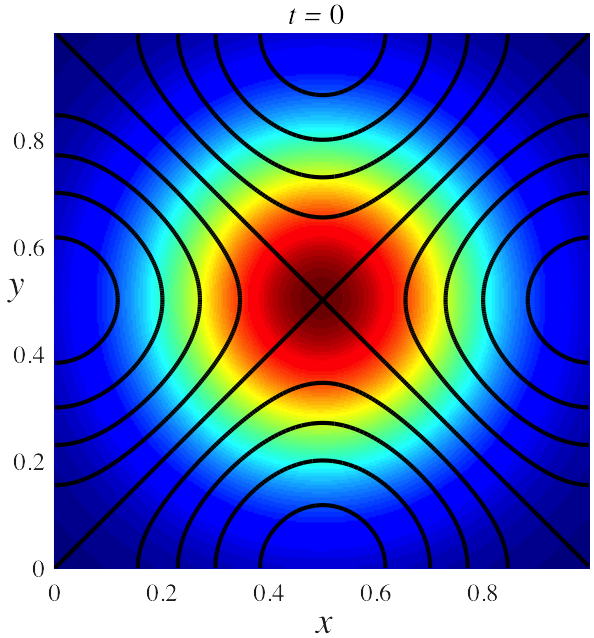} &
         	\includegraphics[width=1.5in]{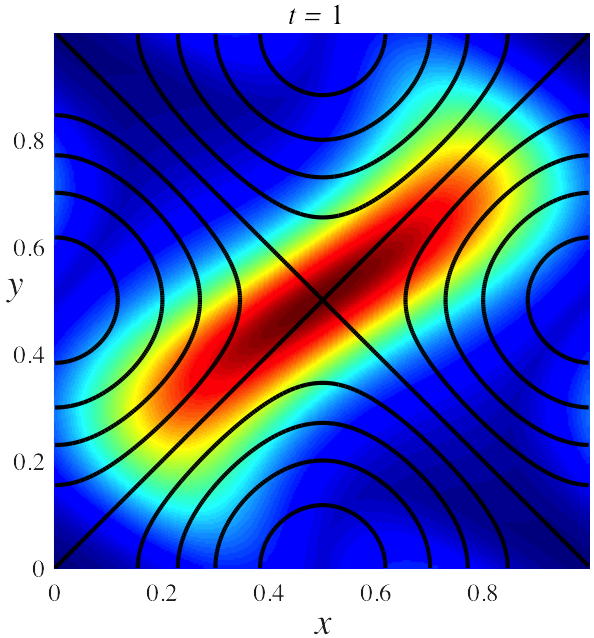} &
         	\includegraphics[width=1.5in]{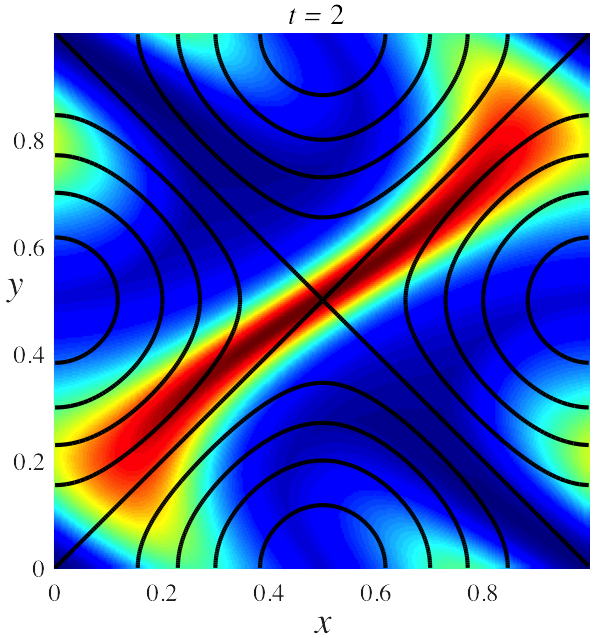} &
         	\includegraphics[width=1.5in]{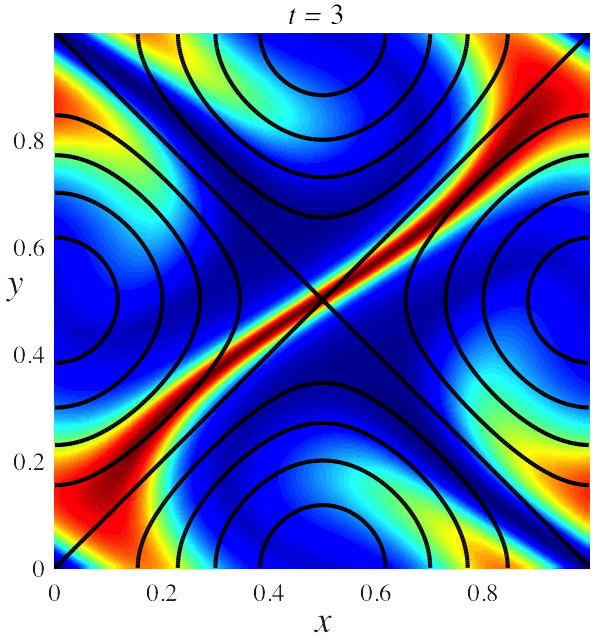}\\
         	\includegraphics[width=1.5in]{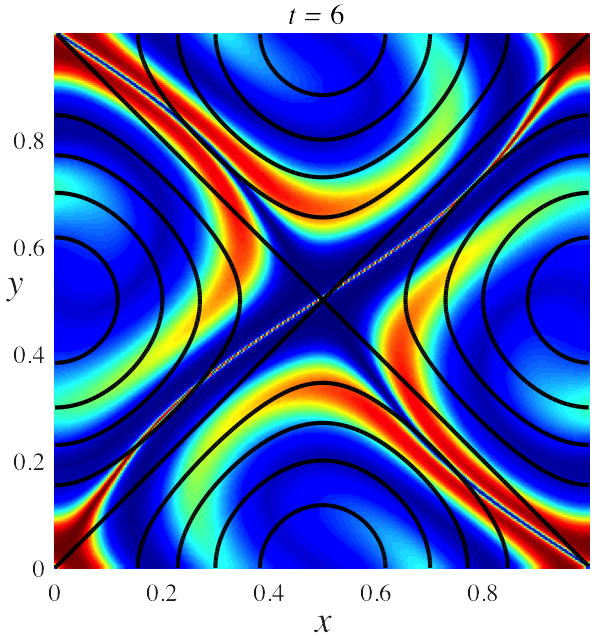} &
         	\includegraphics[width=1.5in]{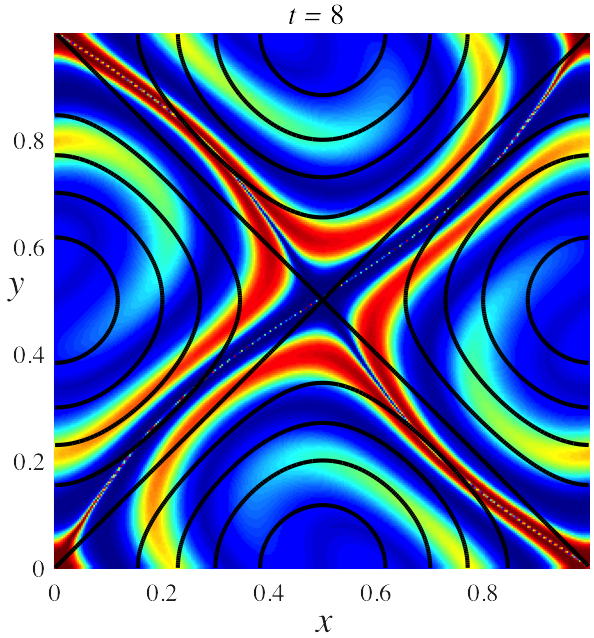} &
         	\includegraphics[width=1.5in]{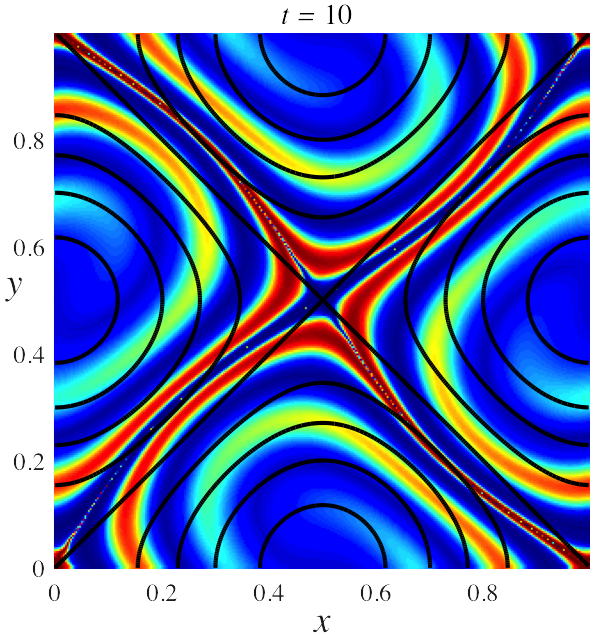} &
         	\includegraphics[width=1.5in]{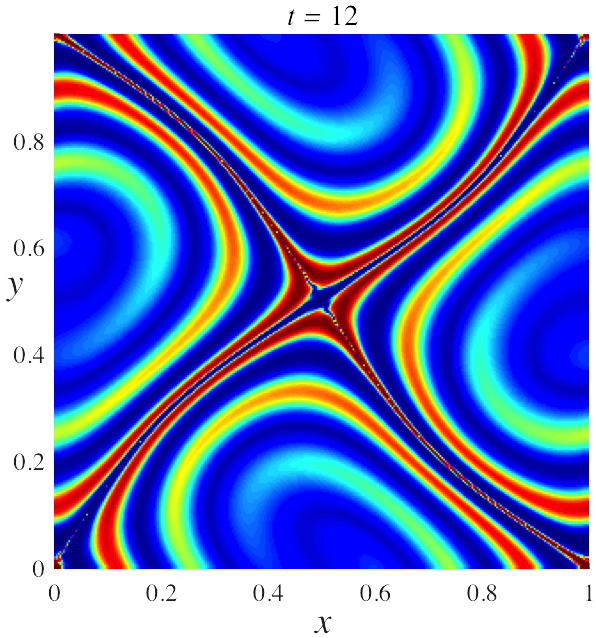}
			\end{tabular}
	}
	\caption{\footnotesize Evolution of $\rho^t(x,y)$ for the Harper map \Eq{HarperMap} with initial condition \Eq{InitialGaussian} for $t=0,1,2,3$ (top row) and $t=6,8,10,12$ (bottom row) with constant strengths $a_t=-b_t=0.1443$. Contours correspond to the Hamiltonian \Eq{HarperHam}.  For a movie of all $12$ steps, see \url{http:/amath.colorado.edu/faculty/jdm/movies/HarperConstNeg.mp4}.}
	\label{fig:HarperConstNeg}
\end{figure}

If the shears have constant amplitudes $|a_t| = |b_t| = \gamma_o$, then the energy constraint \Eq{EnergyHarper} implies that $\gamma_o=\sqrt{\frac{E_H}{2T}}$. It would seem to be counterproductive to have successive signs of the horizontal (or vertical) shears oscillate, so we will consider here two characteristic cases, $a_t=b_t=\gamma_o$, shown in \Fig{HarperConst}, and $a_t = -b_t = \gamma_o$, shown in \Fig{HarperConstNeg}. The contours of the Hamiltonian \Eq{HarperHam} are also shown in these figures to indicate the approximate dynamics of the next map to be applied---indeed as \Fig{HarperPhasePortrait}(a) shows, these contours reasonably approximate the dynamics since the amplitude, $\gamma_o = 0.1443$, is small when $E_H = 0.5$. 

Since the initial density is centered at $(\tfrac12,\tfrac12)$, which is elliptic when $a$ and $b$ have the same sign, and the standard deviation of the Gaussian profile \Eq{InitialGaussian} is $0.2236$, which is not too large, the dominant dynamics seen in \Fig{HarperConst} is a sheared rotation about the center which results in very little mixing. By contrast, when $a$ and $b$ have opposite signs, the point $(\tfrac12,\tfrac12)$ is a saddle, so the density in \Fig{HarperConstNeg} is stretched along the unstable manifold until it reaches the saddle at $(0,0) \equiv (1,1)$, where it divides and then moves back towards $(\tfrac12, \tfrac12)$ along its stable manifolds. 

Figure~\ref{fig:HarperConstSob} shows the decay of the mix-norm for these two cases and various values of the maximum mode number, $M$. As with the standard map, the decrease in the mix-norm slows as $t$ increases. As the grid size increases, the values of $\Sob{\rho^T}$ converge---little difference is seen when $M > 25$, well below our standard resolution.

\InsertFigTwo{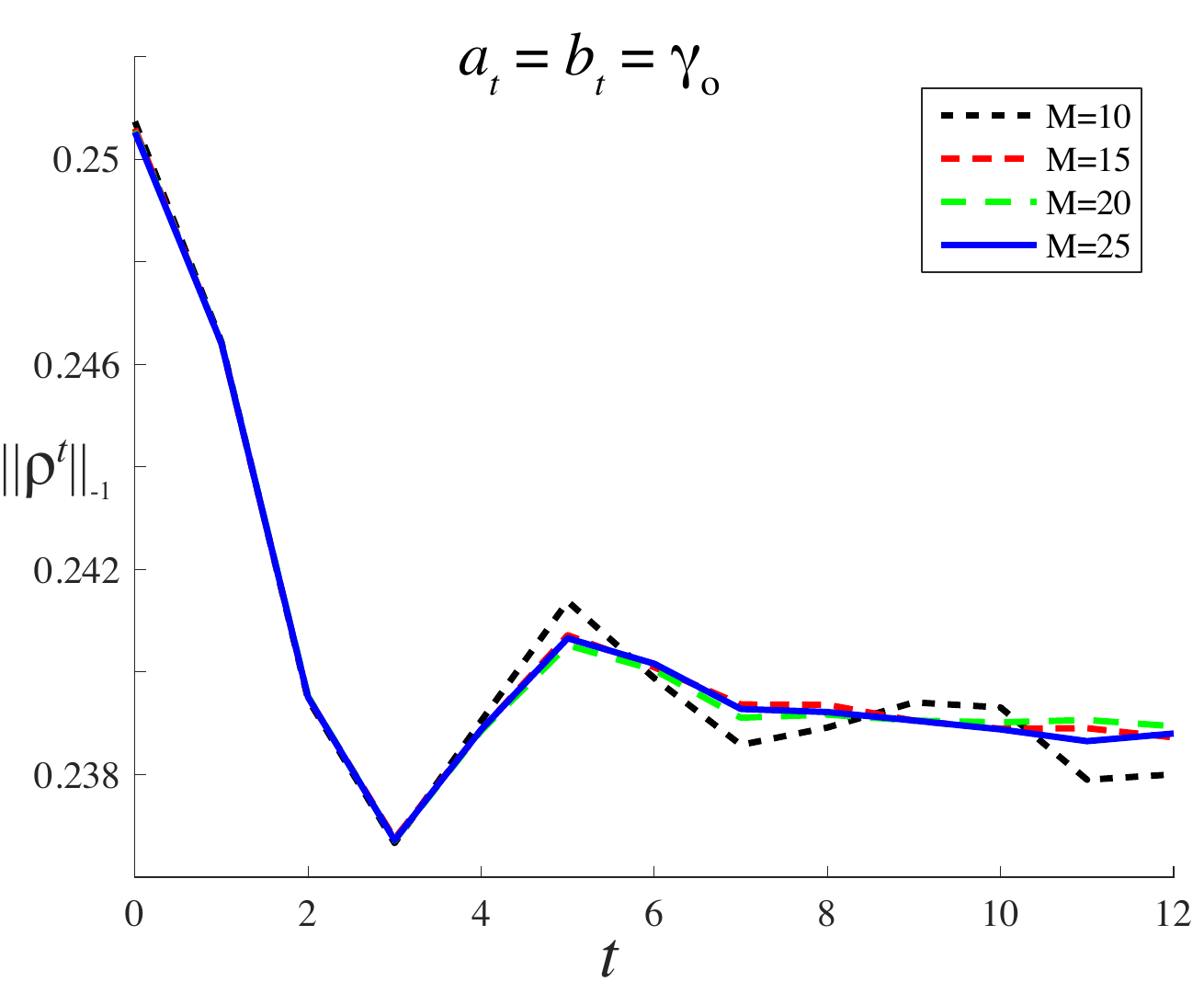}{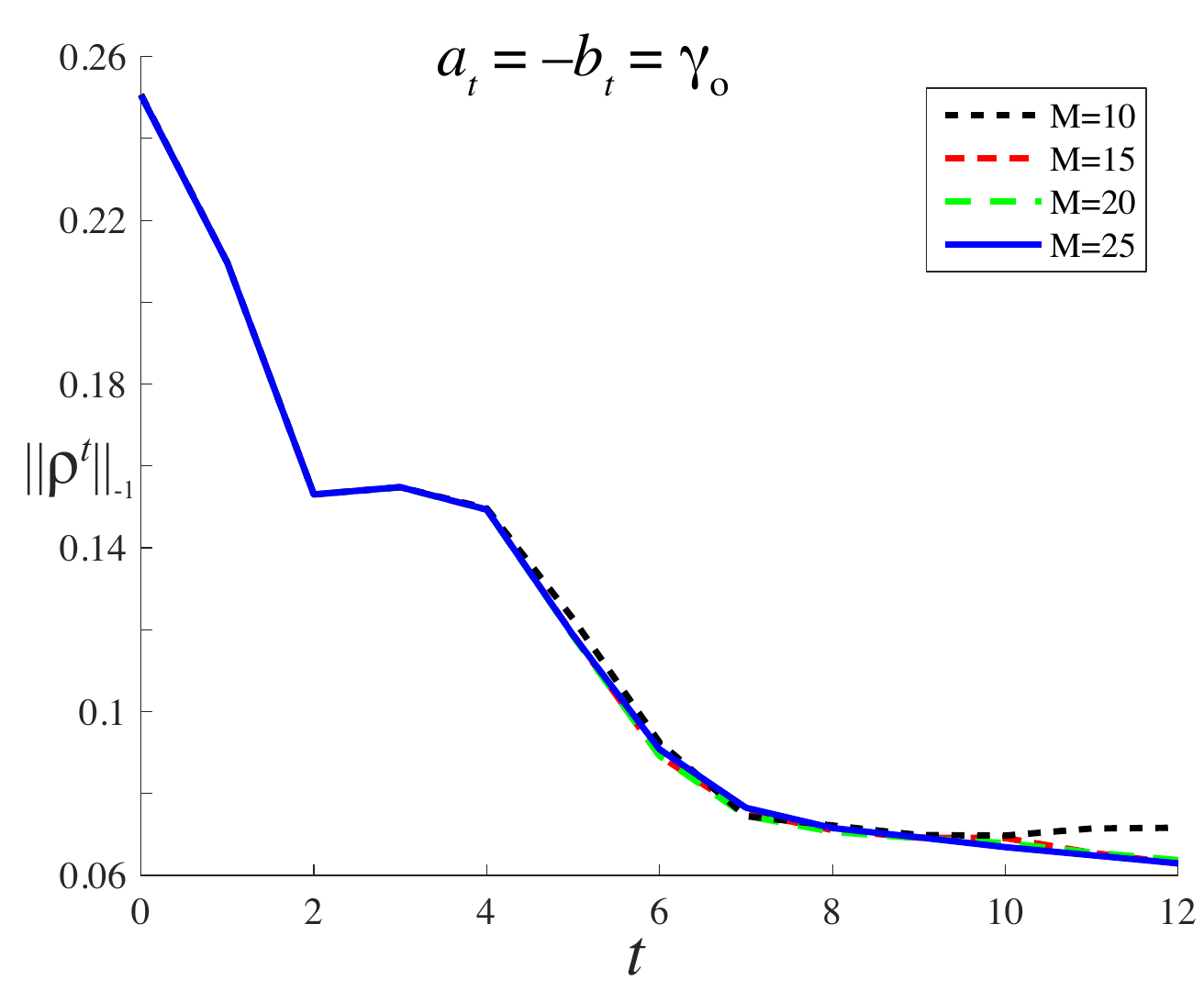}{Decay of $\Sob{\rho^t}$ for the Harper map \Eq{HarperMap} with initial condition \Eq{InitialGaussian} for twelve iterates with constant strength $a_t=b_t=0.1443$ (left) and $a_t=-b_t=0.1443$ (right). The values in the legend denote the maximum mode number, $M$, with grid size $N = 2M+1$.}{HarperConstSob}{3in}

Among the four possible constant amplitude cases, $(a_t = \pm 0.1443, b_t = \pm 0.1443)$ for $E_H=0.5$, the optimal mixing protocol for the centered Gaussian \Eq{InitialGaussian} corresponds to the opposite sign case depicted in \Fig{HarperConstNeg}, achieving $\Sob{\rho^T} = 0.0636$. Of course, since the initial condition is symmetric about the fixed point, flipping both signs results in the same Sobolev norm. This raises the question: will the opposing amplitude stirring scheme be optimal for different energy budgets? The mix-norm as function of constant amplitudes $(a_t,b_t) = (a, b)$ is shown in \Fig{HarperConstEn}. It is interesting that though the optimal protocols for both large and small energy have $a \approx -b$, there is an intermediate regime where this scheme is not optimal.

\InsertFig{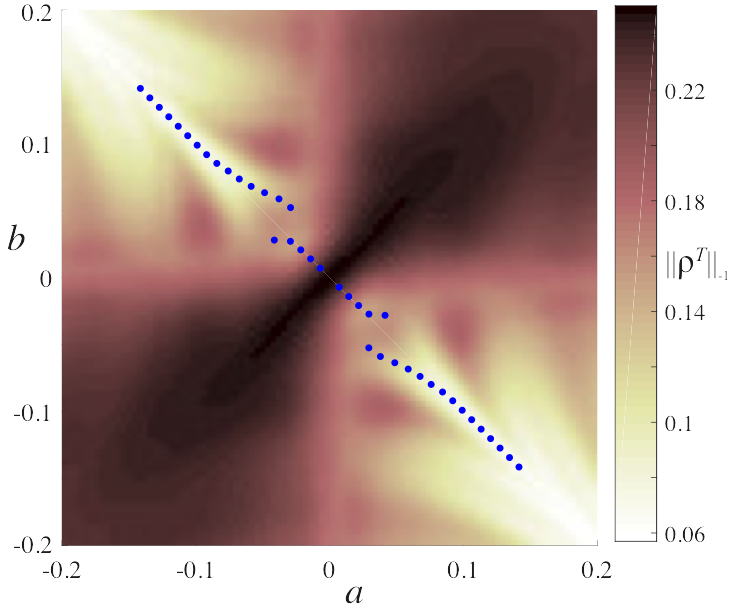}{Level sets of constant mix-norm for $(a_t,b_t)=(a,b)$ with $T=12$ and initial condition \Eq{InitialGaussian} for a grid of amplitudes in the square $-0.2 \le a, b \le 0.2$. The color bar shows the value of the mix-norm: dark colors correspond to less mixing, and the optimal case is white. The blue points indicate the optimal protocol as a function of energy $E_H = T(a^2+b^2)$.}{HarperConstEn}{3in}

\subsection{Random Optimal Protocol}\label{sec:HarperRO}

Computing the optimal mixing protocol for the Harper map, requires choosing the $n = 2T$ dimensional vector $(\va,\vb)$ on the energy sphere
\[
	\|\va\|^2+\|\vb\|^2=E_H
\]
that has the smallest $\Sob{\rho^T}$. 
Following the same procedure as \Sec{StdRO}, \Fig{HarperHist} shows a histogram of the mix-norm for our standard Gaussian initial density. Note that the two constant amplitude protocols (orange and red dots in the figure) correspond to near extremes of the range of $\Sob{\rho^T}$.

\InsertFig{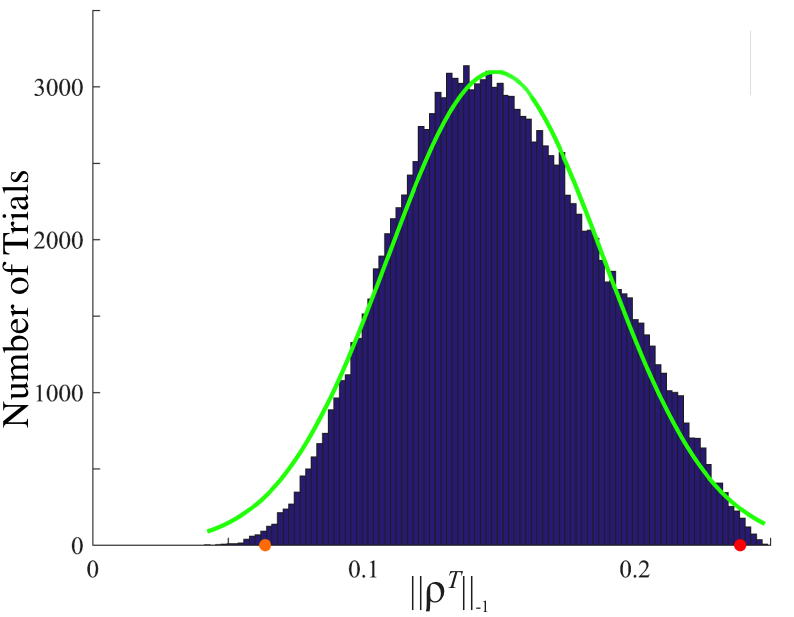}{Distribution of $\Sob{\rho^T}$ after $\sim 1.4\times 10^6$ trials of the random optimization method applied to the Harper map \Eq{HarperMap} with $T=12$ and initial condition \Eq{InitialGaussian}. The green curve indicates the best fit normal distribution with mean $0.1486$ and standard deviation $0.05636$. The red and orange dots indicate $\Sob{\rho^T}$ for the constant amplitude elliptic, \Fig{HarperConst}, and hyperbolic, 
\Fig{HarperConstNeg}, cases respectively.}{HarperHist}{3in}

\begin{figure}[h!t]
	\centerline{
		\renewcommand{\arraystretch}{0.01}
         	\begin{tabular}{cccc}
         	\includegraphics[width=1.5in]{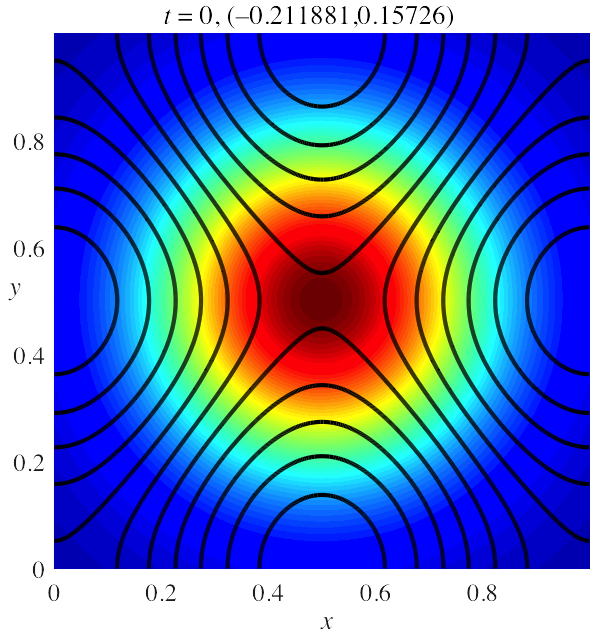} &
         	\includegraphics[width=1.5in]{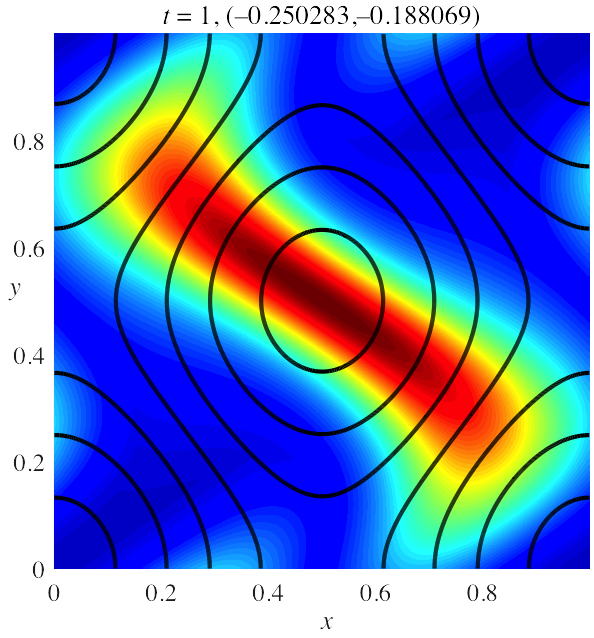} &
         	\includegraphics[width=1.5in]{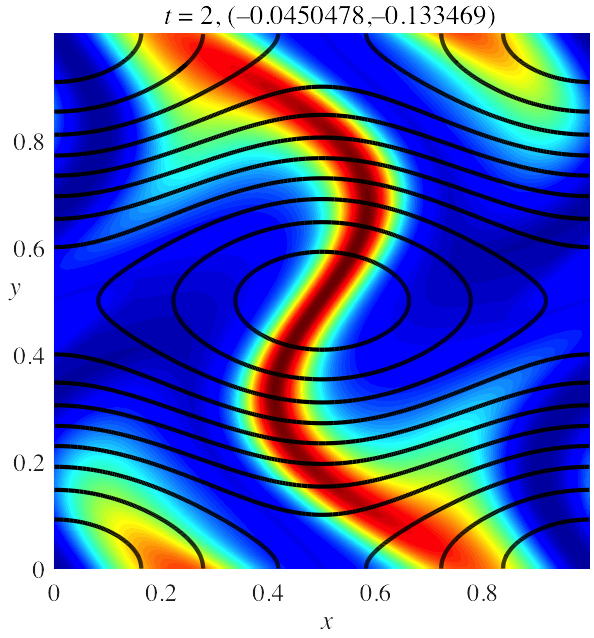} &
         	\includegraphics[width=1.5in]{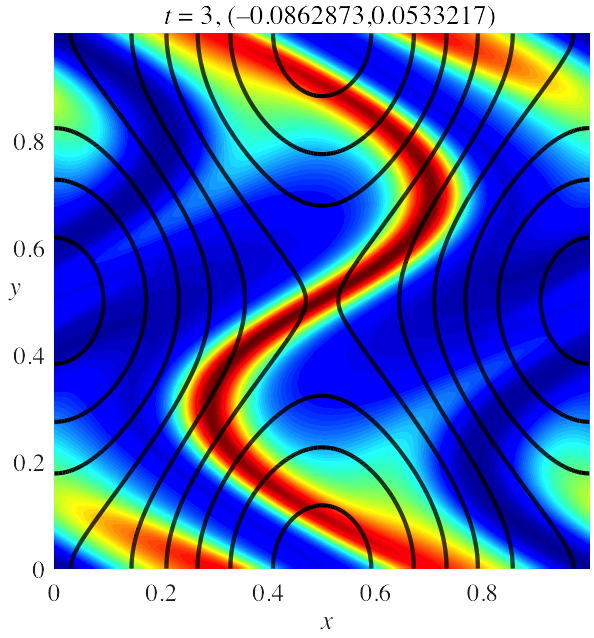}\\
         	\includegraphics[width=1.5in]{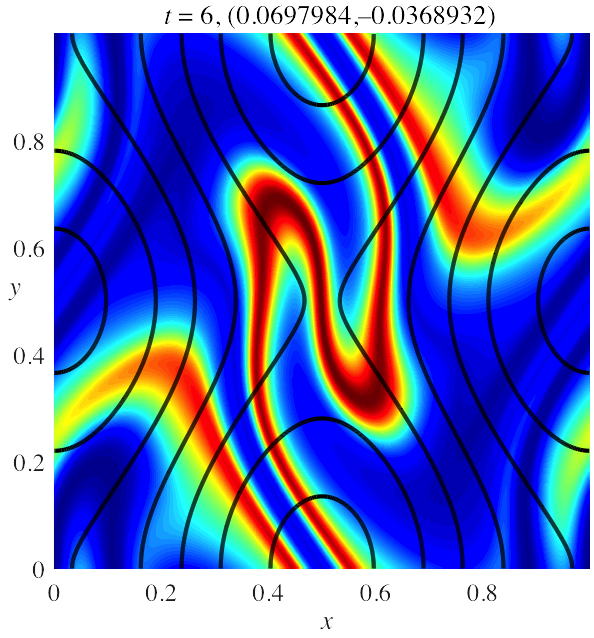} &
     		 \includegraphics[width=1.5in]{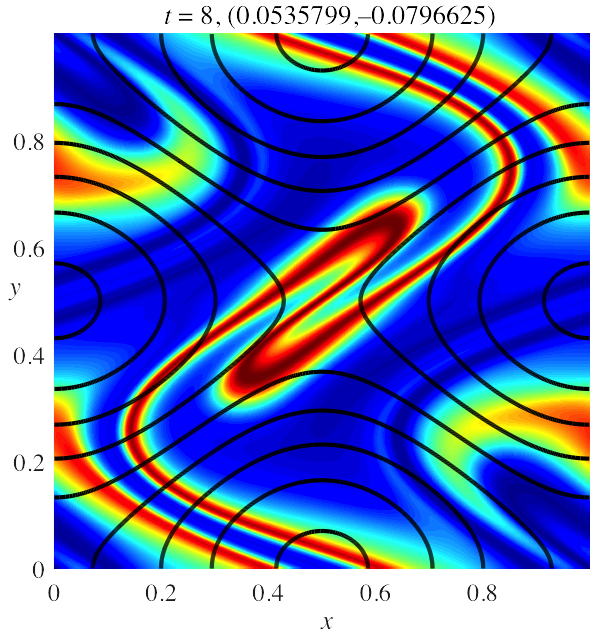} &
         	\includegraphics[width=1.5in]{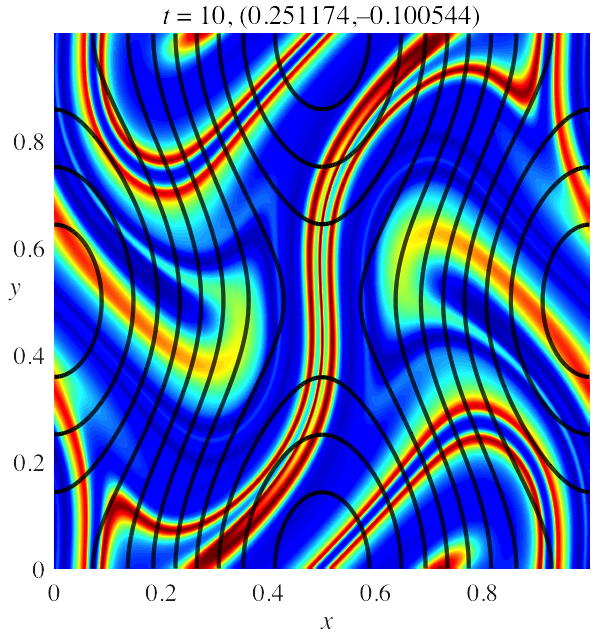} &
         	\includegraphics[width=1.5in]{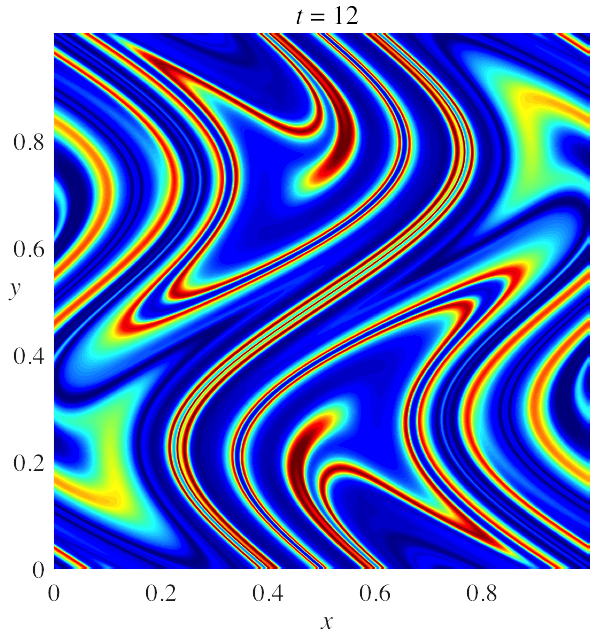}
			\end{tabular}
	}
	\caption{\footnotesize Evolution of $\rho^t(x,y)$ for the Harper map \Eq{HarperMap} with initial condition \Eq{InitialGaussian} for $t=0,1,2,3,6,8,10,12$ with best protocol of 
$\sim1.7\times 10^7$ random trials. Overlaid on each frame, except the last, are the contours of the Hamiltonian \Eq{HarperHam} for the \textit{upcoming} map step. For a movie of all $12$ steps, see \url{http:/amath.colorado.edu/faculty/jdm/movies/HarperMC.mp4}.}
	\label{fig:HarperMC}
\end{figure}

The evolution of the density for the best protocol from $1.7\times 10^6$ trials is shown in \Fig{HarperMC}; again we call this the RO protocol. Note that for this protocol $a_1$ and $b_1$ have opposite signs, causing the initial density to be stretched along the unstable manifold of the point $(\tfrac12,\tfrac12)$; this is as expected from our earlier discussion of the constant case. It is interesting that the density evolution in \Fig{HarperMC} exhibits not only the stretching seen in the constant amplitude hyperbolic case of \Fig{HarperConstNeg}, but also considerable folding, the second hallmark of two-dimensional chaos.

It is also important that the amplitudes of the RO protocol do not have equal magnitudes. Indeed, for several of the steps ($t = 5,8,10,$ and $12$) one of the amplitudes is at least three times smaller than the other, so that the dynamics is nearly a horizontal or vertical shear (Recall that the Hamiltonian contours for these steps are overlaid on the previous density.). In \Sec{HarperShear}, we exploit this by examining stirring protocols that allow only one of the shear amplitudes to be nonzero for each step. 

In addition, the energy of the shears in the RO protocol varies considerably with time. Defining the one-step energy by
\beq{StepEnergy}
	E^{RO}_t = a_t^2 + b_t^2, 
\eeq
we see in \Fig{HarperROEnergy}, most of the energy is expended in steps $t=1$, $2$, $10$ and $11$.

\InsertFig{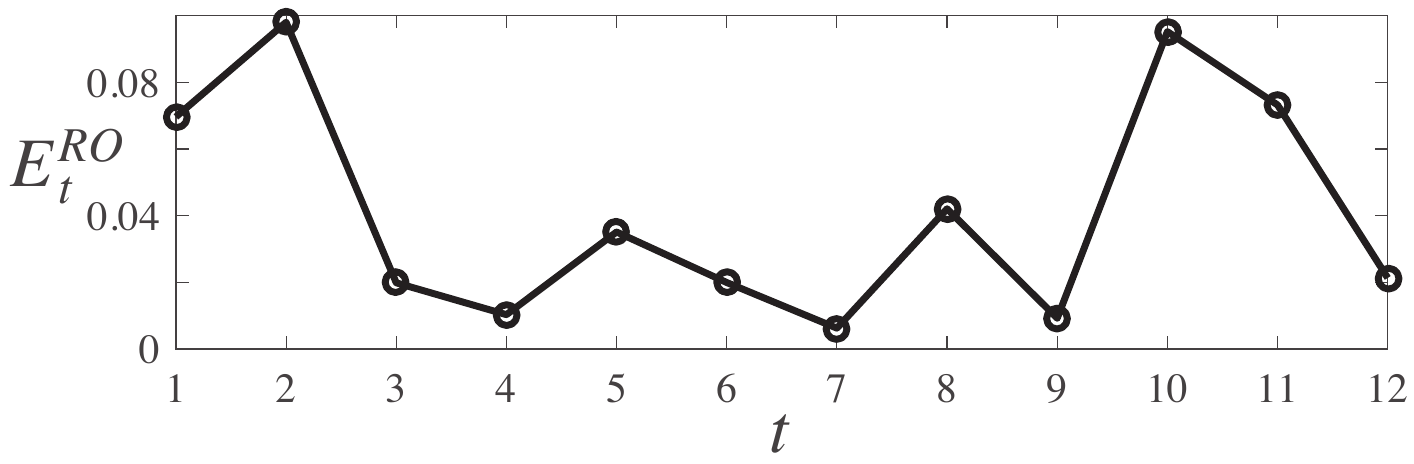}{Energy profile of the RO solution for initial density \Eq{InitialGaussian}, $E_H=0.5$, and $T=12$.}{HarperROEnergy}{3.5in}

The evolution of $\Sob{\rho^t}$ for the RO protocol is shown in \Fig{HarperSobMC}(a); this figure also includes the two constant amplitude cases from \Sec{HarperConst}. As seen in \Tbl{HarperTable}, at $t=12$, the RO protocol just outperforms the hyperbolic constant case, showing that it is possible to do better than just stretching the density along the manifolds of the central fixed point. It is interesting, however, that the constant amplitude case has lower mix-norm when $4<t<9$. During this period, as can be seen in \Fig{HarperMC}, the RO protocol is setting up improved stirring for the last three steps by folding the density contours; this is accomplished using disparate amplitudes of the two shears.

\begin{table}[h!t]
\centerline{
\begin{tabular}{l|c|c}
 & $\Sob{\rho^T}$ & \% of Worse Protocols\\
\hline
Elliptic Constant & 0.2388 & 0.26\\
Hyperbolic Constant & 0.0636 & 99.81\\
RO & 0.0364 & 100\\
SW	& 0.0401	 & 100\\
Shear RO & 0.0334 & 100\\
CS SW & 0.0693 & 99.54\\
CS FS & 0.0680 & 99.63\\
CS Full & 0.0543 & 99.98\\
\end{tabular}
}
\caption{{\footnotesize  Mix-norms for the Harper map with initial condition \Eq{InitialGaussian}, energy $E_H = 0.5$, and total steps $T=12$. The initial condition has $\Sob{\rho^0} = 0.2505$. The second column gives the percentage of random protocols from \Fig{HarperHist} with a larger mix-norm. }
       \label{tbl:HarperTable}}
\end{table}

As the energy varies, so does the achievable minimum mix-norm. The optimal protocol as a function of energy gives a Pareto frontier, shown in \Fig{HarperSobMC}(b). This figure was constructed using $6.5\times 10^5$ trials for the initial condition \Eq{InitialGaussian} spread over 20 values of $E_H \in [0,2]$. Also shown, for each energy, are the mix-norms for the worst trial and for the constant amplitude, hyperbolic protocol. Note that the latter protocol appears to be optimal when $E_H <0.3$; however, for larger energies the RO protocol beats the constant case.
 	
\InsertFigTwo{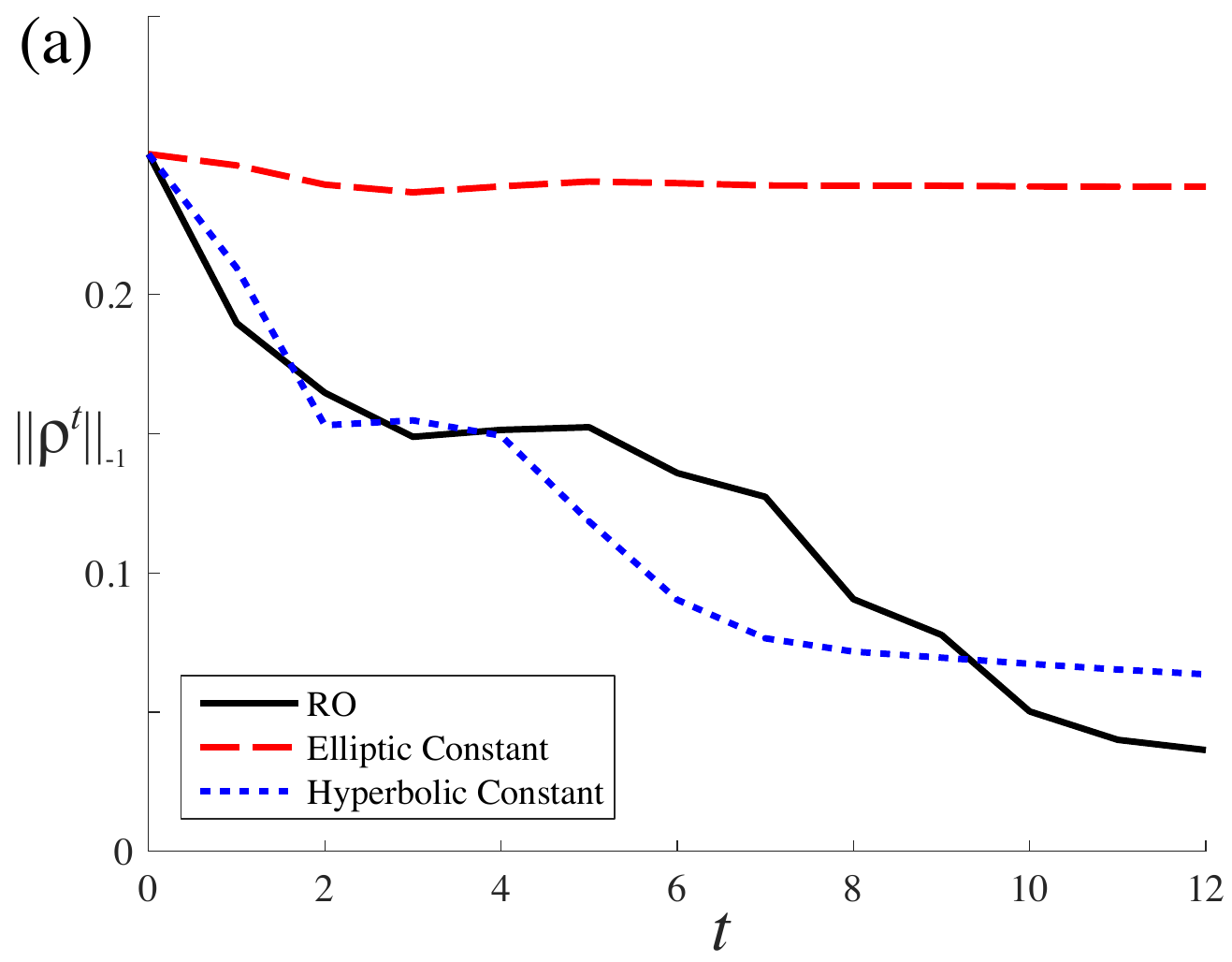}{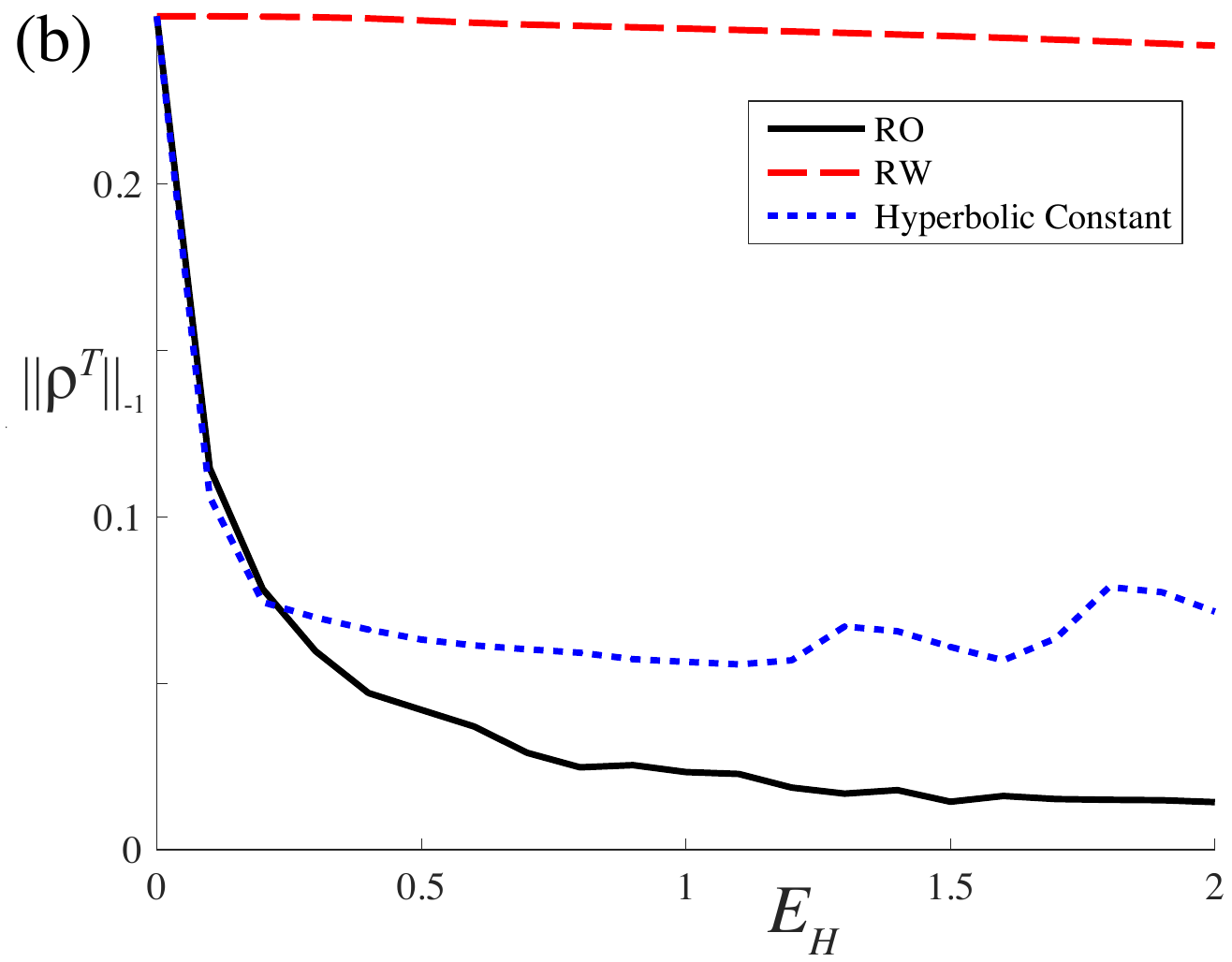}{(a) Mix-norm as a function of time for the RO protocol for the Harper map \Eq{HarperMap} with $E_H = 0.5$, solid (black) curve. Also shown are the elliptic constant, $a_t=b_t = \gamma_o$, (dashed red curve) and the hyperbolic constant, $a_t=-b_t = \gamma_o$ (dotted blue curve) protocols of \Sec{HarperConst}. (b) Pareto frontier, showing the attainable mix-norm as a function of energy for the Harper map, solid (black) curve. The dashed (red) curve shows the worst trial, RW, and the mix-norm achieved by the hyperbolic constant protocol is shown as the dotted (blue) curve.}{HarperSobMC}{3in}

\subsection{Stepwise Minimization}\label{sec:HarperSW}

The decay of the mix-norm for the RO protocol shown in \Fig{HarperSobMC}(a) is not monotone. Though $\Sob{\rho}$ decreases sharply during the first three steps, it appears to increase slightly during steps four and five, finally  decreasing for the remaining steps. This is odd, since one might expect the optimal twelve-step protocol to minimize the mix-norm at each step as much as possible. Instead of considering the time-history as a whole, could it be as effective to choose $a_t$ and $b_t$ independently to maximize the decrease in the mix-norm from step $t-1$ to step $t$?

For simplicity, we will look for a protocol that has same stepwise values of the energy as the RO protocol. Thus, if the amplitudes $(a_t,b_t)$ for the RO protocol at time $t$ have energy \Eq{StepEnergy}, then we will search on this circle for the optimal one-step protocol: for each time $t$, we apply \Eq{Perron} to find
\beq{SWProtocol}
	(a_t, b_t) = \mbox{argmin}\{ \Sob{\rho^{t-1} \circ f^{-1}_{a_t,b_t}} \;:\; a_t^2 + b_t^2 = E^{RO}_t \}. 
\eeq
Recall that the energy $E^{RO}_t$ was shown in \Fig{HarperROEnergy}. Since \Eq{SWProtocol} requires succession of a one-dimensional searches at each time, it is much quicker than the full random optimization. However, the restriction that to a known energy distribution means that we will not be looking for more general protocols. We call this new stepwise minimization protocol ``SW".

\begin{figure}[h!t]
\begin{center}
\includegraphics[height=.26\textwidth]{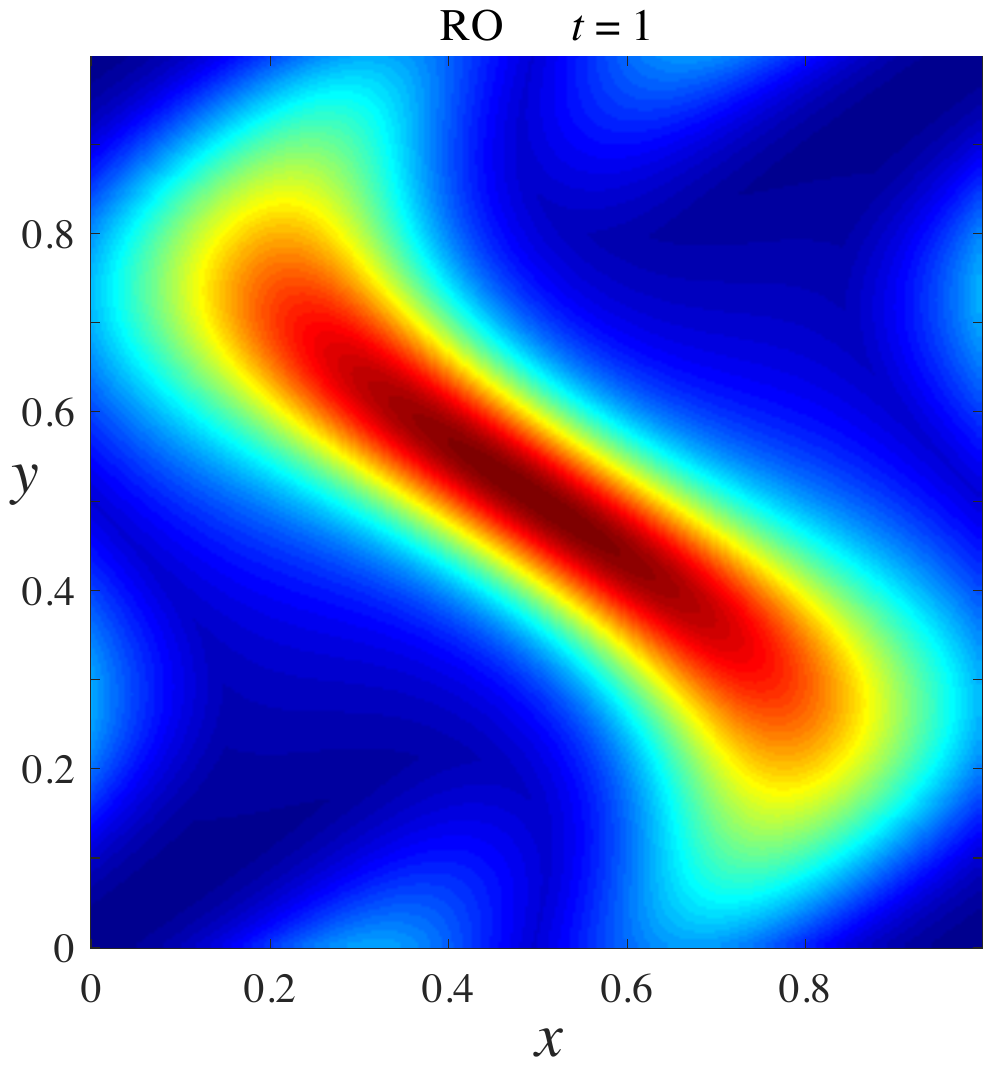}
 \hskip 0.3in
 \includegraphics[height=.26\textwidth]{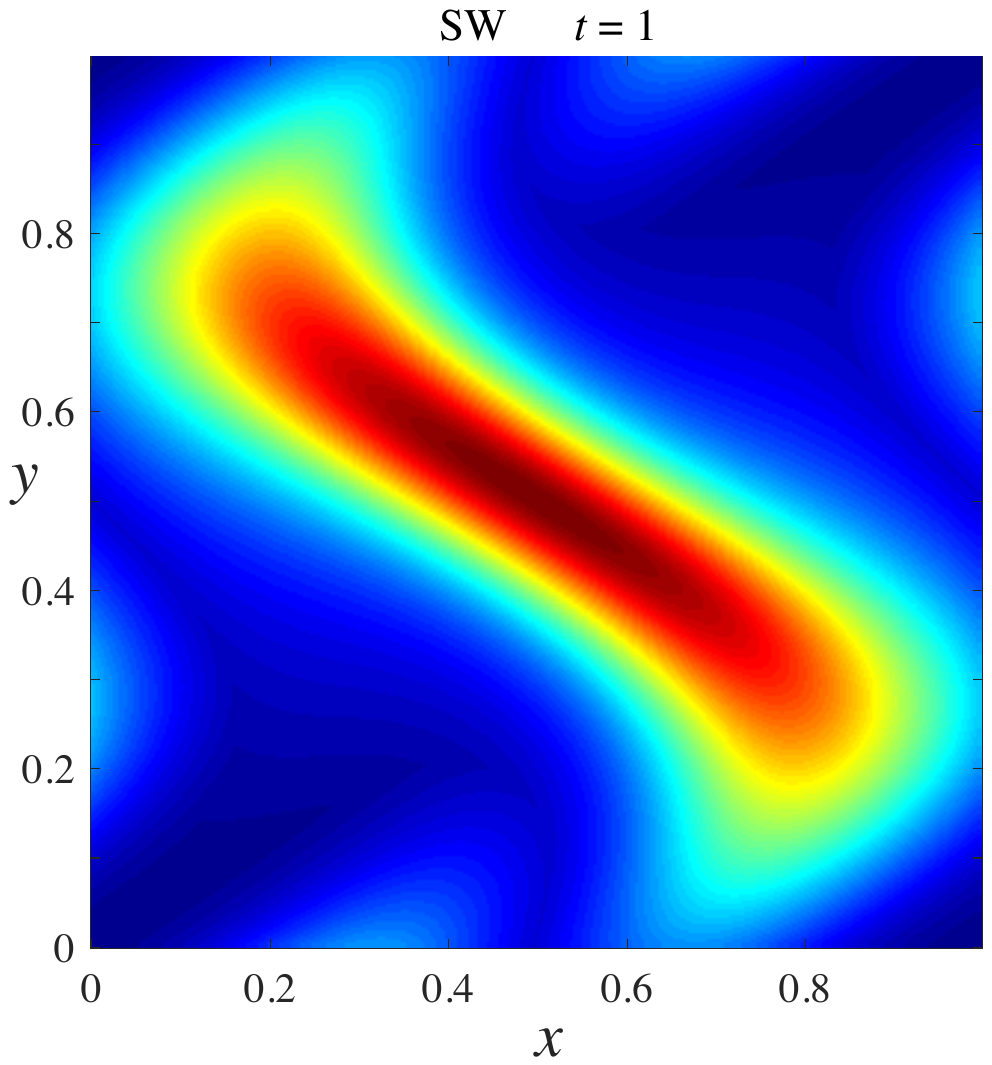}
 \hskip 0.3in
 \includegraphics[height=.26\textwidth]{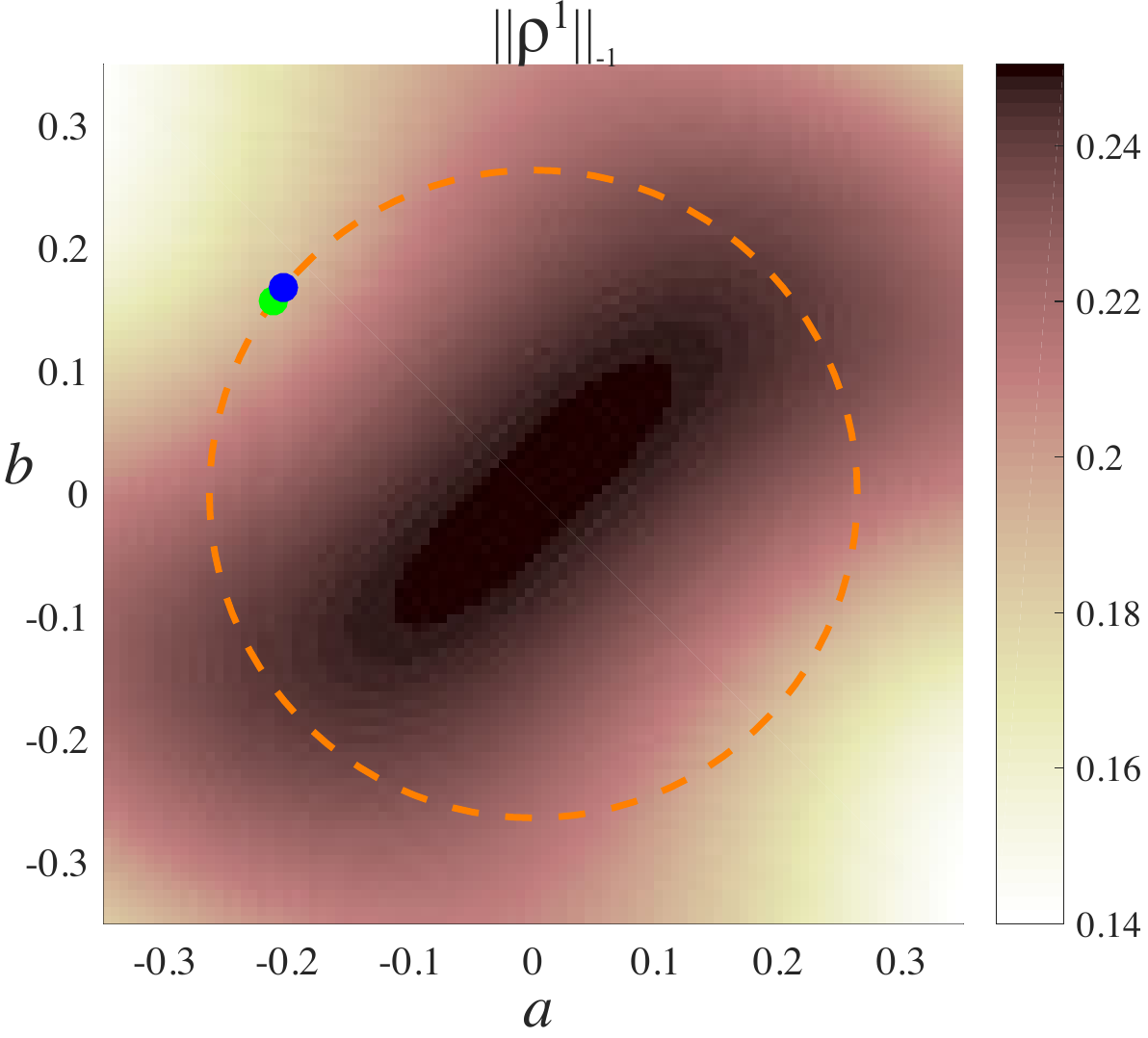}\\
 \includegraphics[height=.26\textwidth]{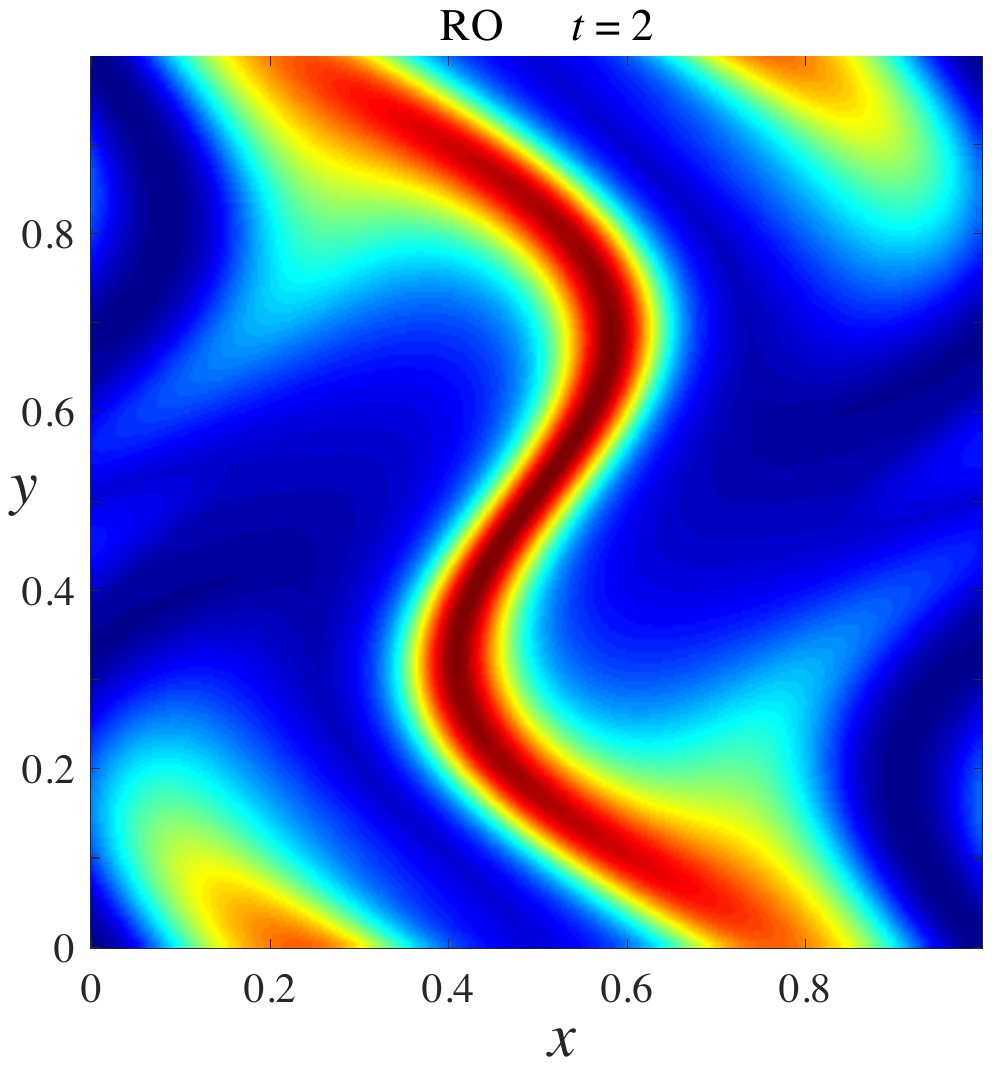}
 \hskip 0.3in
 \includegraphics[height=.26\textwidth]{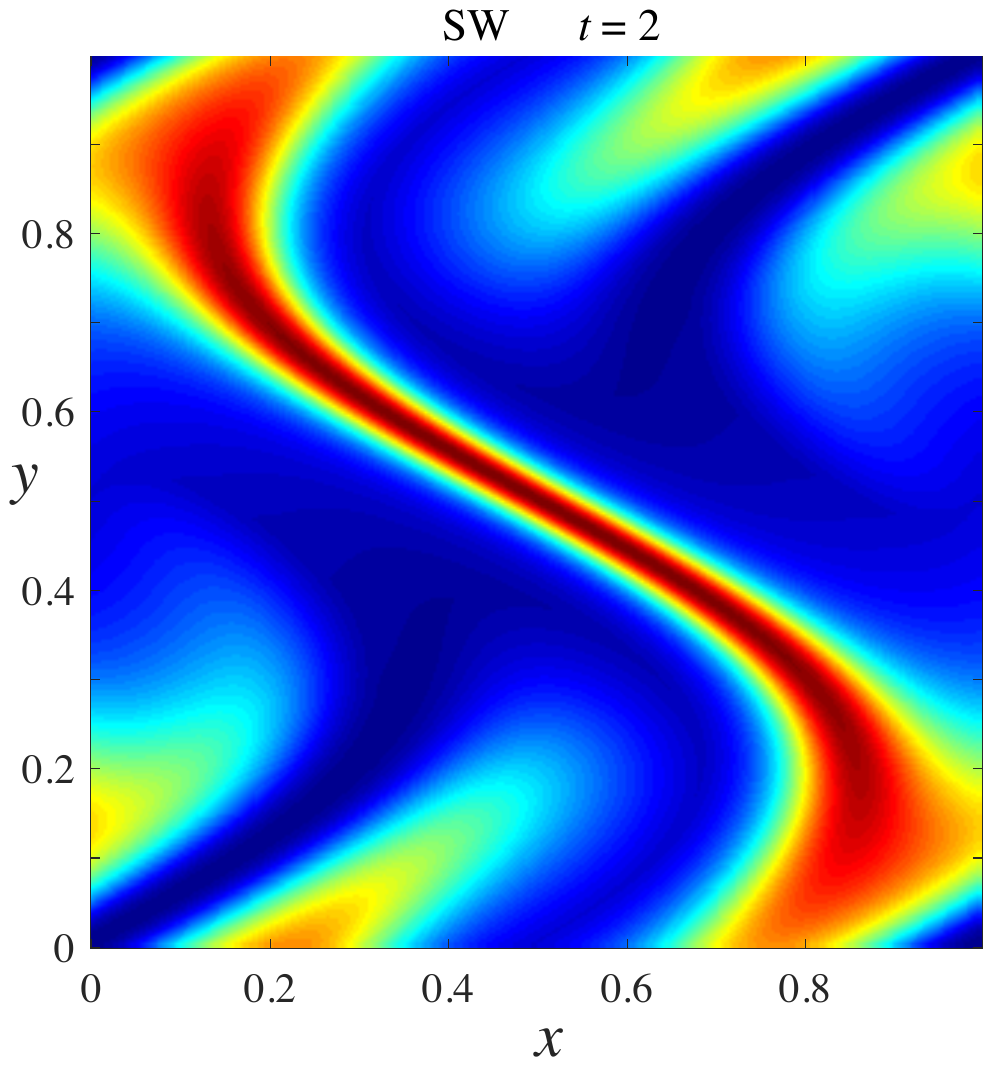}
 \hskip 0.3in
 \includegraphics[height=.26\textwidth]{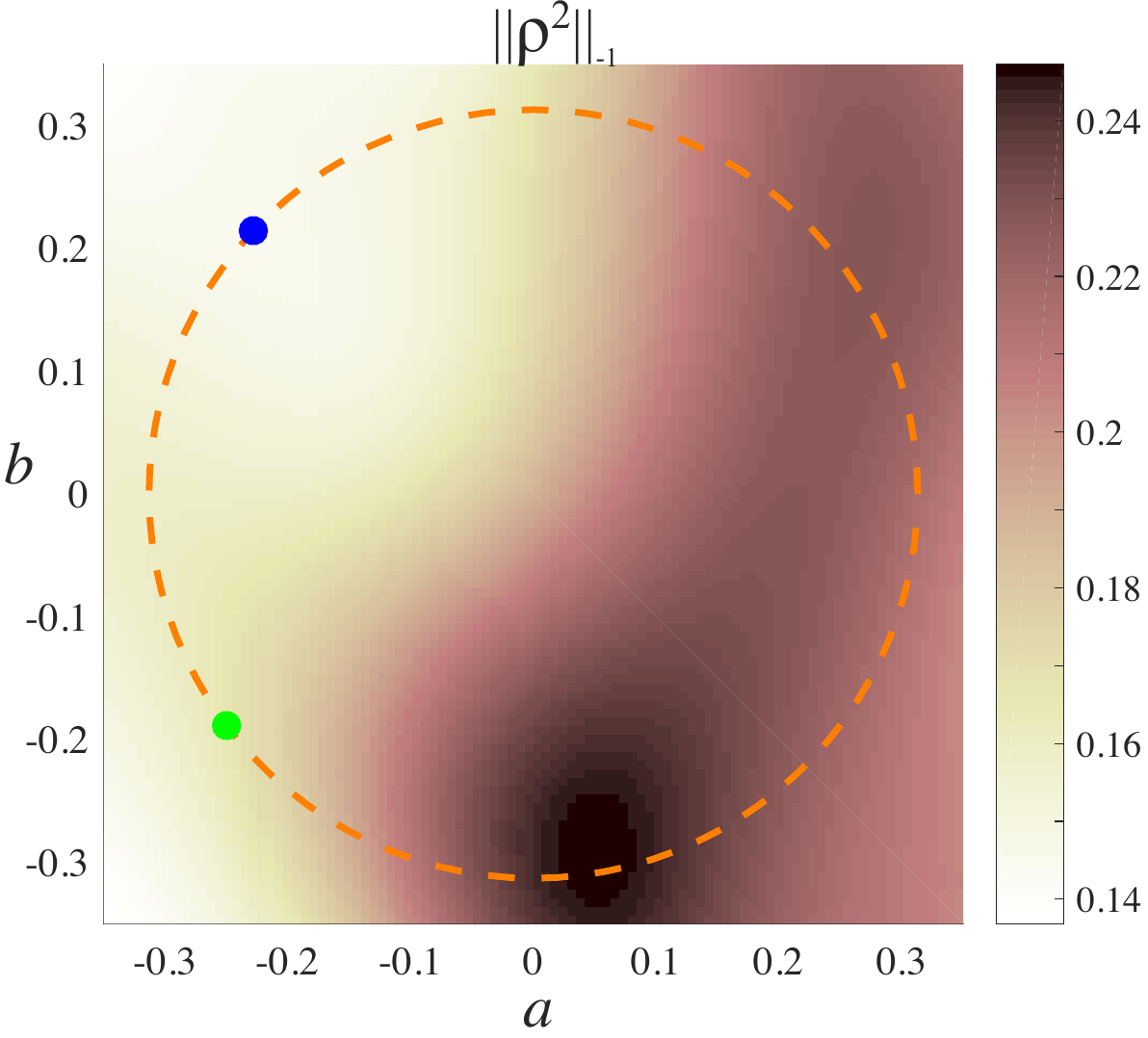}\\
 \includegraphics[height=.26\textwidth]{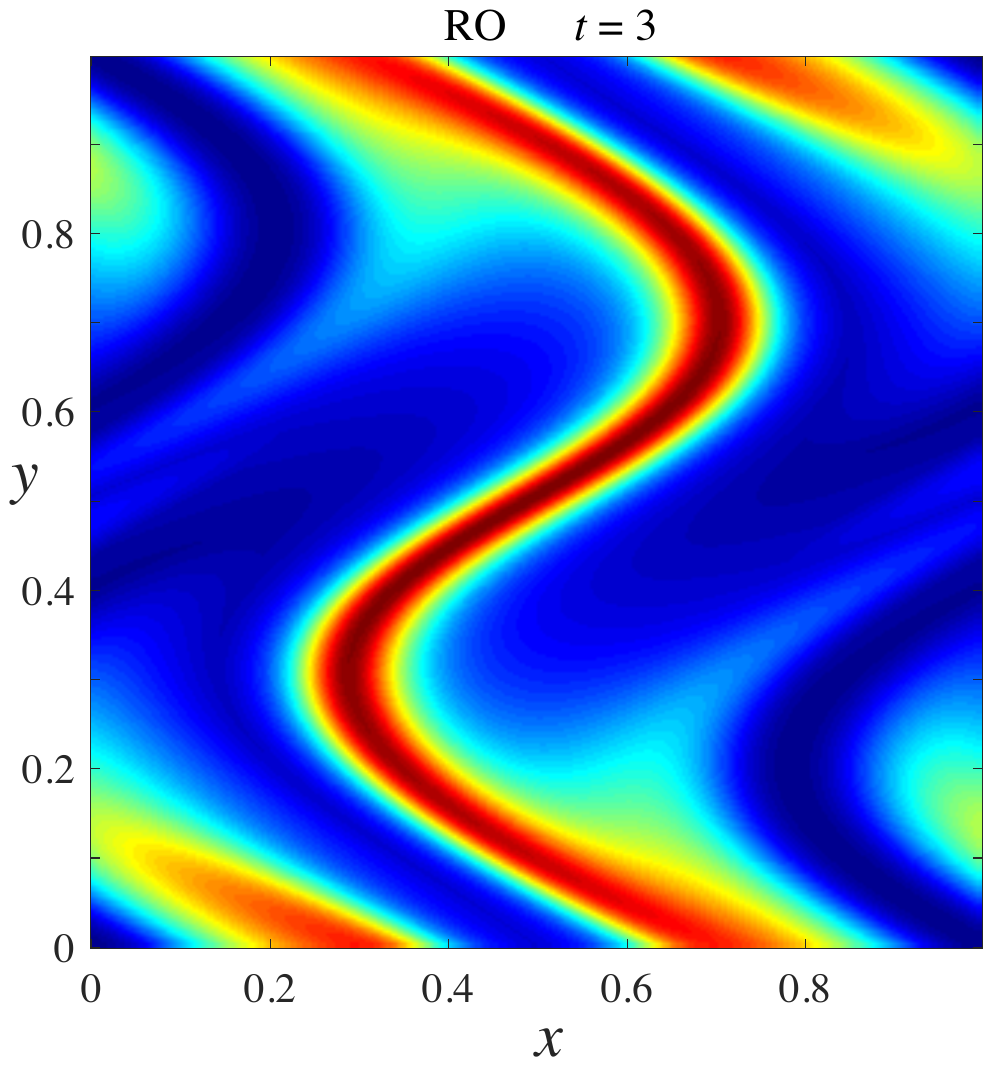}
 \hskip 0.3in
\includegraphics[height=.26\textwidth]{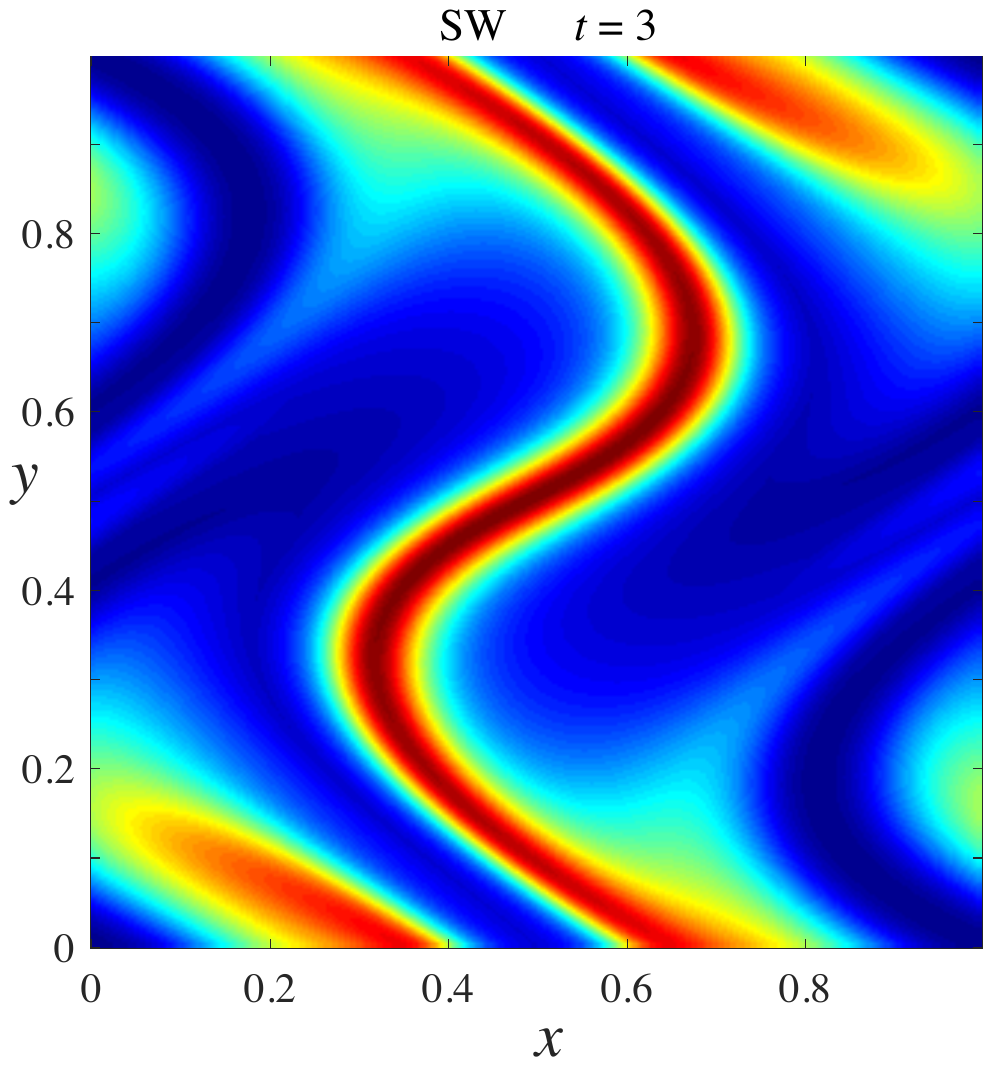}
 \hskip 0.3in
\includegraphics[height=.26\textwidth]{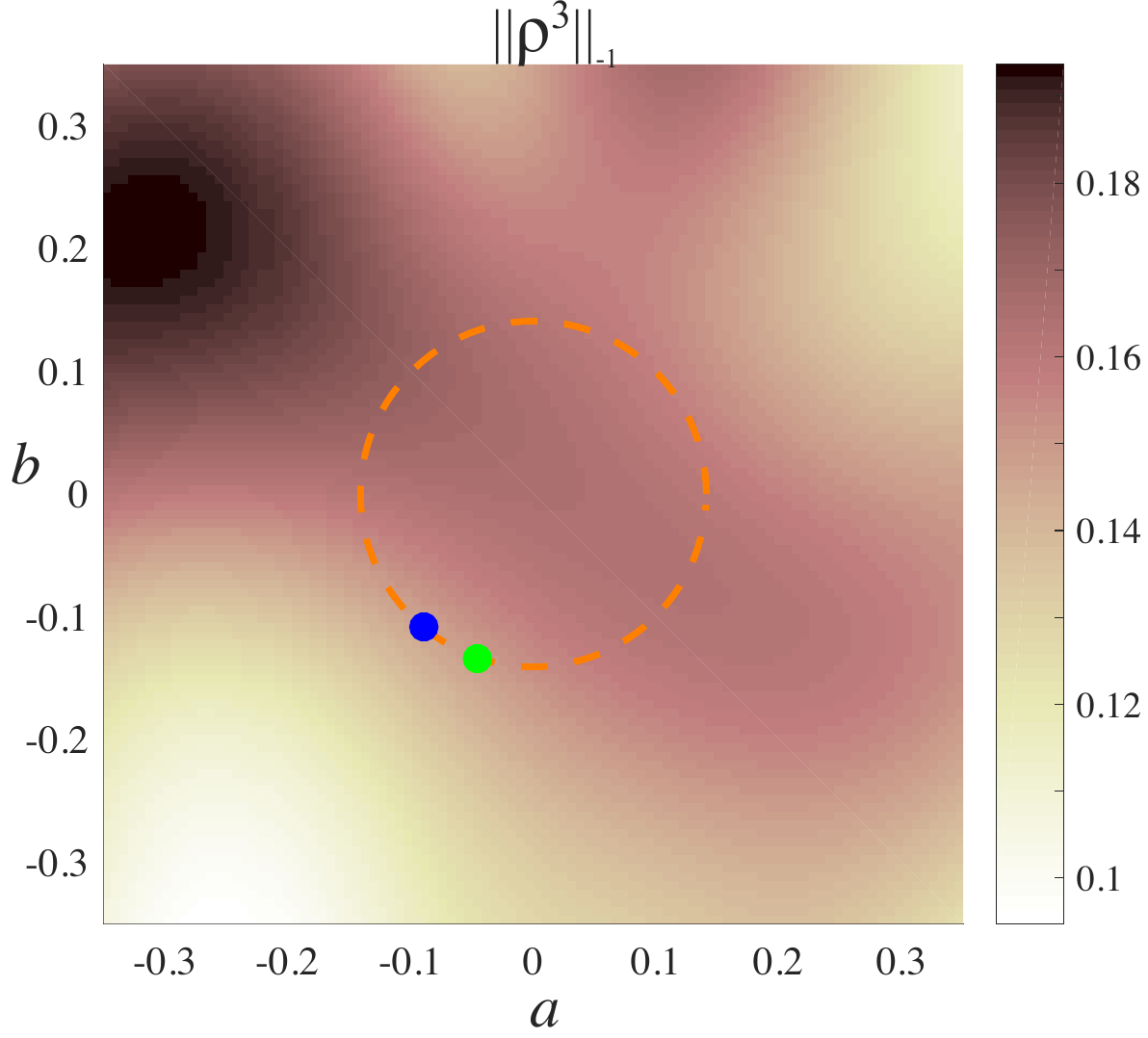}\\
\caption{\footnotesize Comparison of RO and SW protocols of the Harper map for $t=1,2,3$ using the initial condition \Eq{InitialGaussian}. The first column shows $\rho^t(x,y)$ for first three steps of the RO protocol, and the second  gives that of the stepwise protocol applied to the RO state $\rho^{t-1}$. The third column shows the mix-norm as a function of $(a,b)$ for step $t$. The dotted circle is \Eq{StepEnergy}, from which the parameters for the SW protocol are selected. The green point on this circle corresponds to the RO and the blue point to the SW parameters.}
\label{fig:HarperSW}
\end{center}
\end{figure}

Figure \ref{fig:HarperSW} illustrates the stepwise minimization by comparing the RO protocol of \Sec{HarperRO} to a SW protocol at three times. For this illustration, we used the density $\rho^{t-1}$ obtained by the RO protocol as the initial state for the SW minimization, shown in the second column. So to obtain the stepwise minimization at $t$, we perform $t-1$ steps of the RO protocol and then use \Eq{SWProtocol}. The amplitudes $(a_1, b_1)$ for the SW protocol are essentially the same as those for RO--this can be seen in the third column which shows $\Sob{\rho^t}$ as a function of $(a,b)$. The RO amplitudes, the green dot in this panel, and the selected SW amplitudes nearly overlap.

This is also true at $t=3$, though there is a larger difference, presumably because none of the selected random trials landed on the best point. However the difference at $t=2$ is nontrivial: the sign of $b_2$ flips between the RO and SW cases. Note that the central point $(\tfrac12,\tfrac12)$ is elliptic for the RO amplitudes, whereas it is a saddle for the SW amplitudes. This supports our hypothesis that the RO protocol, by looking ahead, does not minimize the mix-norm at every step: for this step it has emphasized folding instead of stretching.

Fully implementing \Eq{SWProtocol} (without reinitializing to the RO solution at every step) for the initial density \Eq{InitialGaussian} gives $\Sob{\rho^t}$  shown in 
\Fig{HarperSWSob}(a). This figure also shows the mix-norm of the RO protocol from \Fig{HarperSobMC} for comparison. Though the overall performance of the two protocols is about the same at $t=12$, the SW protocol reaching $\Sob{\rho^T} = 0.0401$, the RO protocol does slightly outperform the stepwise minimizer. It seems that an optimal mixing scheme cannot simply consider each step of the map independently, but must instead forgo a one-step improvement to setup the density for future steps of the map.

\InsertFigTwo{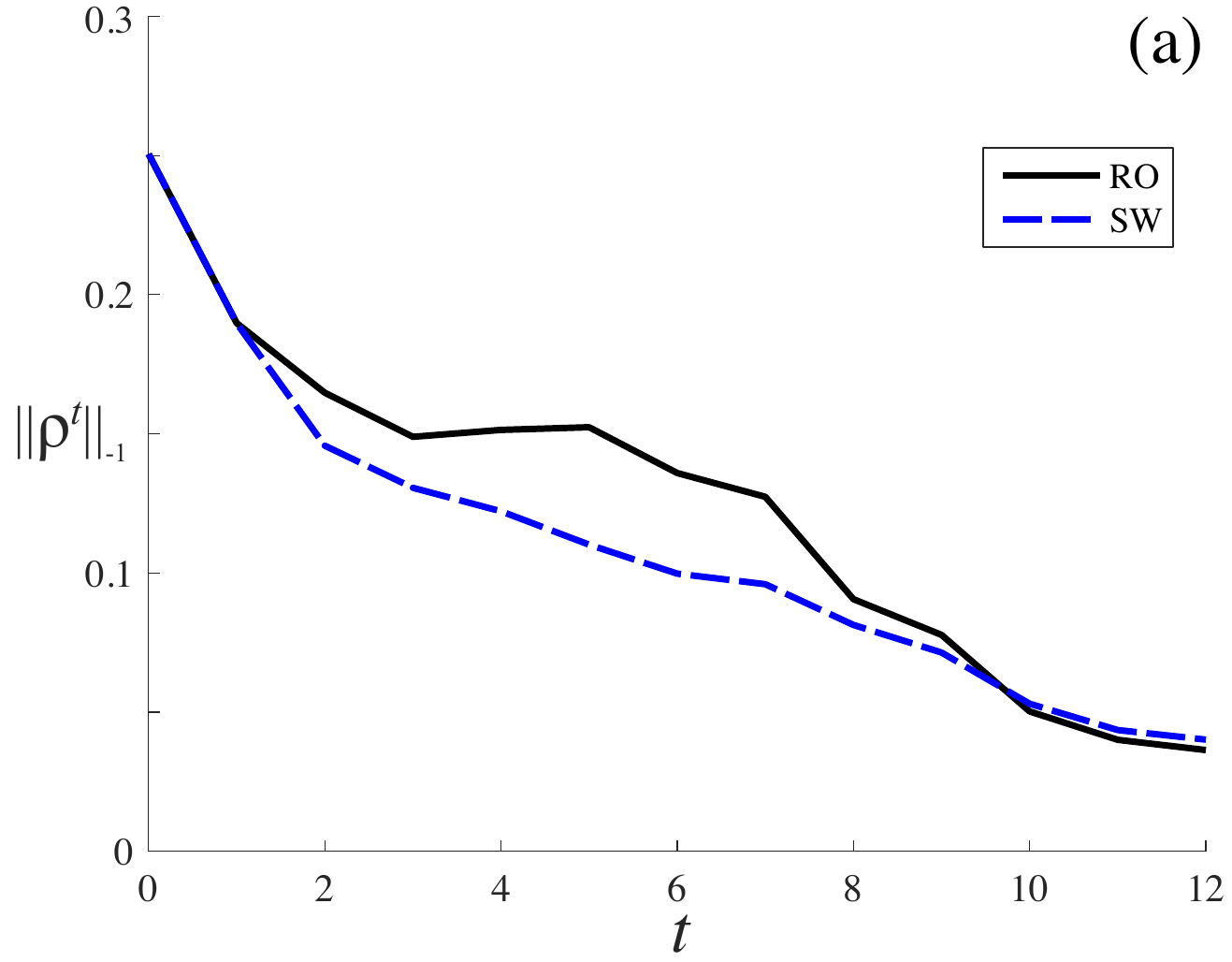}{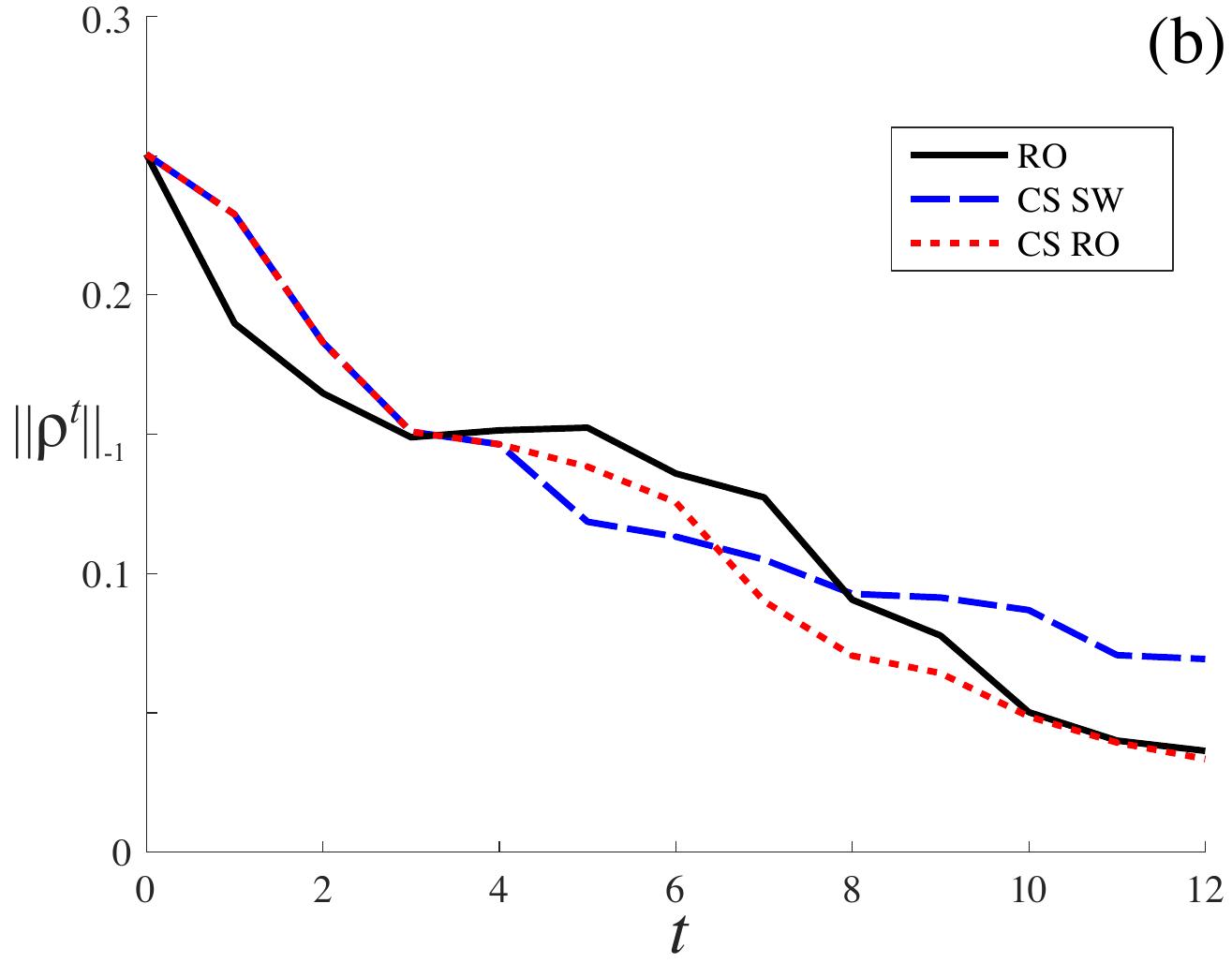}{(a) Progression of $\Sob{\rho^t}$ for the RO and SW protocols described in Figure \ref{fig:HarperSW}. For a movie of the SW protocol steps, see \url{http:/amath.colorado.edu/faculty/jdm/movies/HarperSW.mp4}. (b) Progression of $\Sob{\rho^t}$ for the RO, CS SW, and CS Full protocols. The latter used $\sim 4\times 10^5$ trials. The two shear protocols are identical until $t=3$. }{HarperSWSob}{3in}

\subsection{Horizontal or Vertical Shearing}\label{sec:HarperShear}

For many of the time steps of the RO protocol, as illustrated in \Fig{HarperMC}, one of the shear amplitudes is much smaller than the other, giving either primarily horizontal or vertical shearing. In this section we investigate protocols in which there is only one nonzero shear at each step. To simplify the problem even further, we assume that the energy per step is constant. Defining the amplitude $\gamma_o=\sqrt{E_H/T}$; there are now just four options for the parameter pair at a given step: $(a,b) \in \{(\gamma_o,0),(-\gamma_o,0),(0,\gamma_o),(0,-\gamma_o)\}$.

To find an optimal stirring protocol with the new constraints is easier than a full RO since there are only four choices for parameters at every step. In fact except for $t=1$, only three of these can decrease the mix-norm. For example if $(a_1,b_1) = (\gamma_o,0)$, then the choice $(a_2,b_2) = (-\gamma_o,0)$ simply undoes the previous step, so it cannot be an optimal choice. Therefore, only the three options $(\gamma_o,0),(0,\gamma_o),(0,-\gamma_o)$ need be considered at $t=2$. In this way, there are $4\times3^{T-1}$ distinct shear protocols to consider, i.e., $\sim 7\times 10^5$ when $T=12$. The scheme found from randomly sampling this discrete space is compared to the RO protocol in \Fig{HarperSWSob}(b). The optimal shear protocol (labeled ``Shear RO" in the figure) just beats the full RO protocol, as shown in \Tbl{HarperTable}.

An even simpler method is to do a stepwise minimization, as in \Sec{HarperSW}, now imposing the single-shear restriction. In this case there are only $4+(T-1)3$ choices, i.e., $40$ when $T=12$. The optimal result in this case gives the mix-norm curve labelled CS SW (Constant Shear SW) in \Fig{HarperSWSob}(b).  This simplified protocol achieves $\Sob{\rho^T} = 0.0693$, which is not comparable to the better performing protocols. Nevertheless, it still performs better than 99.6\% of the randomly sampled amplitude vectors shown in \Fig{HarperHist}. Note that, for this specific initial density profile, the hyperbolic constant scheme is slightly better (see \Tbl{HarperTable}); however, its effectiveness depends strongly on the localization of \Eq{InitialGaussian} near the saddle point. 

To gain some insight into how to choose the direction of shear based on the current density distribution, we can consider what the optimal one-step parameter choice would be when the density can be described by a single sinusoid.
Consider the initial condition
\beq{HarperOneModeTest}
	\rho^0(x,y)=2\cos(2\pi n \cdot z -\phi)
\eeq
for wave vector $n$, and phase shift $\phi$. The resulting Fourier expansion will have only two nonzero terms, both of modulus one. Using this initial condition and  fixing the amplitude $\gamma_o = 0.2041$, we compute the mix-norm $\Sob{\rho^1}$ as a function of mode numbers for $200$ randomly chosen phase shifts. Figure~\ref{fig:HarperOneMode} shows a density plot of the number of times choosing $a \neq 0$  results in better one-step mixing than choosing $b \neq 0$. According to \Fig{HarperOneMode}, we should choose $a \neq 0$ whenever $|n_1|<|n_2|$ and $b\neq 0$ if $|n_1|>|n_2|$. The lines $|n_1|=|n_2|$ are ambiguous.

\InsertFig{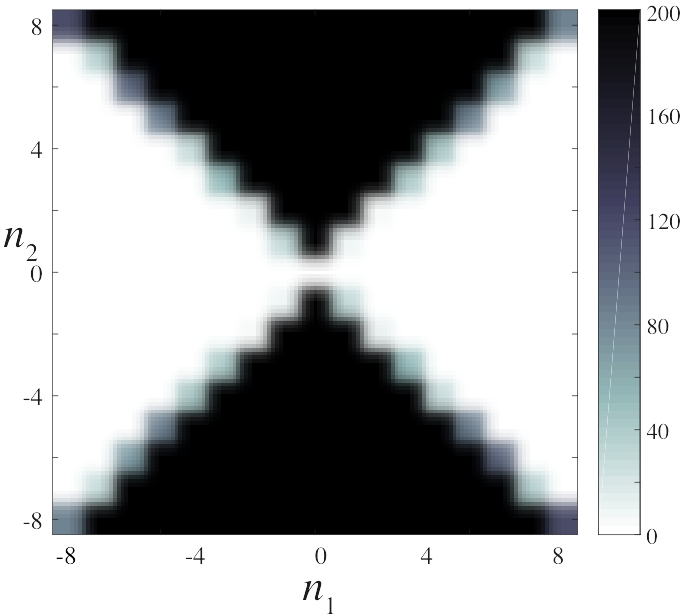}{Density plot of the number of times choosing $a \neq 0$ results in better one step mixing than choosing $b \neq 0$ with respect to mode numbers $(n_1,n_2)$ for initial condition \Eq{HarperOneModeTest} with $200$ random choices of phase shift $\phi$. Here the shear amplitudes, $|a|$ or $|b|$, are fixed to $\gamma_0 = 0.2041$, but varying $\gamma_0$ within reasonable bounds does not meaningfully change the results.}{HarperOneMode}{3in}

To understand why this is true, consider the simple case of \Eq{HarperOneModeTest} with $\phi = 0$, so that the Fourier expansion will contain the only nonzero terms $\hat{\rho}_{n}^{0}=\hat{\rho}_{-n}^{0}=1$. Recalling \Eq{HarperFourierUpdate}, the Fourier coefficients at $t=1$ will be
\[
	\hat{\rho}_{m_1,m_2}^1=J_{m_1+n_1}(-2\pi n_2 a)J_{-n_2-m_2}(2\pi m_1 b)+
			J_{m_1-n_1}(2\pi n_2 a)J_{n_2-m_2}(2\pi m_1 b).
\]
When $b=0$ and $a$ is nonzero, the only terms that survive are those that satisfy $m_2=\pm n_2$ so that
\[
	\hat{\rho}_{k,\pm n_2}^1=J_{m_1\mp n_1}(\pm2\pi n_2 a),  \quad  (b = 0),
\]
for $k \in \bZ$. On the other hand, if $a=0$ and $b$ is nonzero, we are left with the terms in which $m_1=\pm n_1$, giving
\[
	\hat{\rho}_{\pm n_1,k}^1=J_{\pm n_2-m_2}(\pm2\pi n_1 b), \quad (a = 0).
\]
Since the $H^{-1}$ norm \Eq{FourierSobolev} has the wavenumber in the denominator, the implication is that
we would then want to choose $b=0$ when $|n_1|<|n_2|$, since the resulting terms $\hat{\rho}_{k,\pm n_2}$ will have higher combined wave number than the terms $\hat{\rho}_{\pm n_1,k}$. The Bessel function terms will also tend to be smaller, at least when the shears have comparable amplitude and the mode numbers are large. This supports the more general results of \Fig{HarperOneMode}. 

To apply this insight to more general density profiles which have a full Fourier spectrum, we define the partial sums
\begin{align*}
&S_<=\sum_{|n_1|<|n_2|}\frac{|\hat{\rho}_{n_1,n_2}|^2}{n_1^2+n_2^2},
&S_>=\sum_{|n_1|>|n_2|}\frac{|\hat{\rho}_{n_1,n_2}|^2}{n_1^2+n_2^2}.
\end{align*}
The shear choice is then determined by the relative sizes of these partial norms: if $S_<$ is smaller we choose $(|a|,|b|)=(0,\gamma_o)$; otherwise, if $S_>$ is smaller we choose $(|a|,|b|)=(\gamma_o,0)$.

To make this a complete protocol, we must also select the signs of the shears.
Note that successive steps $(a_t,0)$ to $(0,b_{t+1})$ are equivalent to one step with amplitudes $(a_{t},b_{t+1})$, so that the choice of sign is equivalent to the selection of the location of elliptic and saddle points of the map.  The opposite switch, from $(0,b_t)$ to $(a_{t+1},0)$ does not give a single Harper map of the form \Eq{HarperMap}, but is instead equivalent to its inverse with the signs of both parameters switched. Thus the locations of the elliptic and saddle cases are still the same as those shown in \Fig{HarperHam}.

If the density is concentrated in a specific location, then it seems reasonable that a sign choice that makes the nearest fixed point a saddle will beat the sign choice that makes it elliptic.  Indeed, for the Gaussian centered at $(\frac{1}{2},\frac{1}{2})$, it was important to choose opposite signs to allow this fixed point to be hyperbolic, since making it elliptic would result in virtually no mixing.

To determine the relationship between the location of density peaks and the optimal sign choice, we can do a simple one-step experiment with various localized initial densities---we use the initial state \Eq{ShearSignTest}, varying the center $(u,v)$ of the profile.  Figure~\ref{fig:HarperShearSign} shows the best sign choice as a function of $(u,v)$. The color at a given point indicates the sign choice that most often resulted in the best one-step mixing
under randomly chosen values of the width parameter $\eps$. As expected, if the density is localized near $(\frac{1}{2},\frac{1}{2})$, we should choose $a$ and $b$ of opposite sign. This generalizes: if the density is localized near any fixed point, the relative signs should be chosen to make that fixed point hyperbolic. The regions in which we choose a given sign combination form nearly perfect squares around each fixed point. To understand why this is, consider the approximate eigenvalues and eigenvectors given by the Hamiltonian \Eq{HarperHam}. The fixed point $(\frac{1}{2},\frac{1}{2})$  has eigenvalue-eigenvector pairs
\[
	\lambda=\pm 2\pi \sqrt{-ab},\quad\textbf{v}=\left(\mp \frac{b}{\sqrt{-ab}},1\right).
\]
Now assume that we are using the constant shearing scheme such that $|a|=|b|= \gamma_o$. Also assume that the density is concentrated near $(\frac{1}{2},\frac{1}{2})$ so that we at least know we want $a = -b$ to make the point hyperbolic. This causes the eigen-pairs to reduce to
\[
	\lambda=\pm 2\pi \gamma_o,\quad\textbf{v}=(\mp\text{sign}(b),1).
\]
The vector $(\text{sign}(b),1)$ corresponds to the stable manifold and $(-\text{sign}(b),1)$ to the unstable. The closer the concentration of density is to a hyperbolic fixed point, the more it gets stretched along the unstable manifold of that point. Ideally the density should be localized to straddle the stable manifold so that it gets pushed towards the fixed point and then spread out along the two branches of the unstable manifold. This implies that if the concentration is located in the $1^{st}$ or $3^{rd}$ quadrants, relative to $(\frac{1}{2},\frac{1}{2})$, we should choose $b>0$ to make the local stable manifold have positive slope. By contrast, if the concentration lies in the $2^{nd}$ or $4^{th}$ quadrants we choose $b<0$ to make the local stable manifold have negative slope. Performing a similar analysis for each of the fixed points implies the results obtained experimentally in \Fig{HarperShearSign}. 

\InsertFig{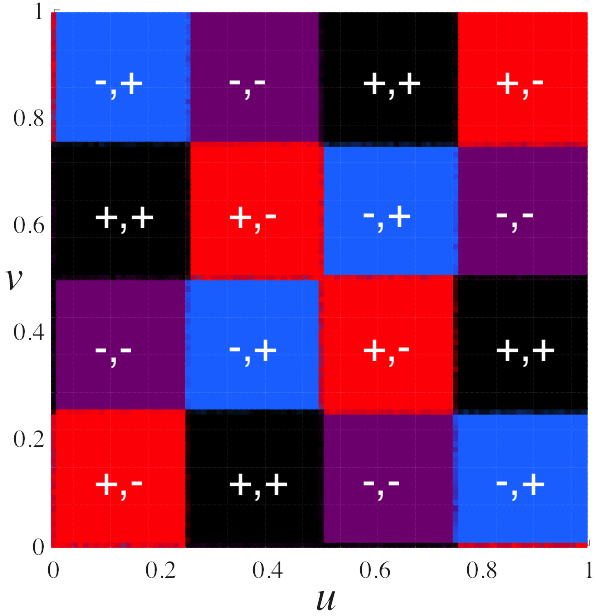}{The intensity of the color at a given point $(u,v)$ is proportional to the percentages of sign choices that yielded the best mixing over  different $\eps$ ranging from 0.1 to 5 using the initial density \Eq{ShearSignTest}.}{HarperShearSign}{3in}

From this we can extract an algorithm to determine effective sign choices. We denote the $16$ regions in \Fig{HarperShearSign} by $R_1, R_2,\ldots, R_{16}$, and define the indicator functions 
\[
	\chi_i(x,y)= \left\{\begin{array}{ll}
						1, & (x,y)\in R_i\\
						0, & (x,y)\notin R_i\end{array}\right..
\]
For each region we compute a partial mix-norm 
\[
	S_i=\Sob{\rho\cdot\chi_i}, \quad i = 1, 2, \ldots, 16.
\]
Under the hypothesis that region with the largest $S_i$ dominates the need for mixing,
we pick the signs of $(a,b)$  for this region according to \Fig{HarperShearSign}.

A combination of the Fourier and sign algorithms gives a full protocol choice for the multistep shear problem. We attempt two different schemes. The first allows for the signs of $(a,b)$ to change between steps of the map if the location of the density has changed. We refer to this as the ``CS Full" method. The second scheme chooses signs for $(a,b)$ based solely on the initial condition maintaining this sign choice throughout the protocol. We refer to this as the ``CS FS" (Fixed Signed) method. the CS Full method has the advantage of tracking the density and switching the fixed points accordingly. However, the CS FS method allows for more consistency in the direction of stirring. 

The CS Full algorithm performs better when we limit the instances in which a sign change can occur. Switching the signs, and thereby changing the direction of stirring, often causes a momentary increase in the mix-norm. Performing a sign change when it is not strictly necessary can negatively affect the results. However, forgoing a sign change can cause the mix-norm to level off. Recall that the RO solution outperformed the Hyperbolic Constant scheme because it sacrificed some steps to cause folding which will allow for more effective stretching later. Experimentally we find that CS Full performs the best when we only switch signs if the difference between the two largest $S_i$ values is more than 10\% of the total mix-norm.

\InsertFigThree{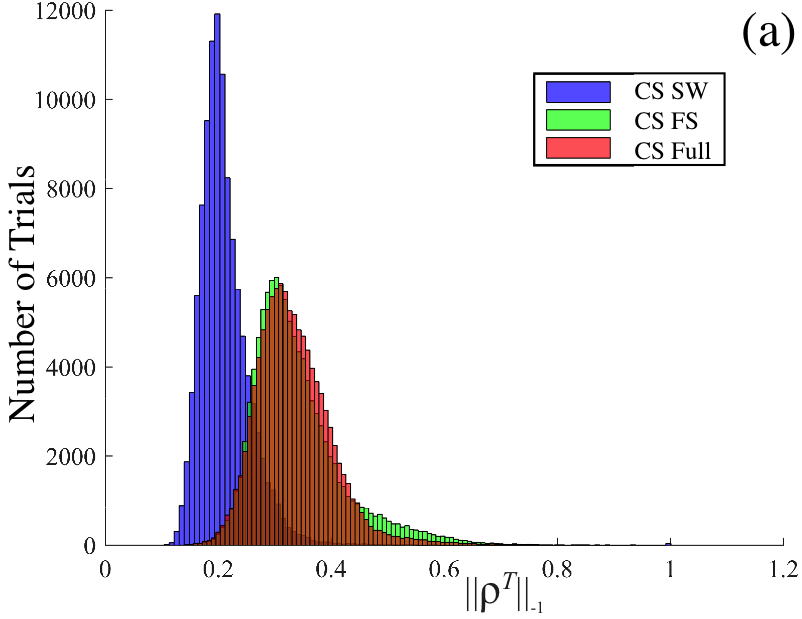}{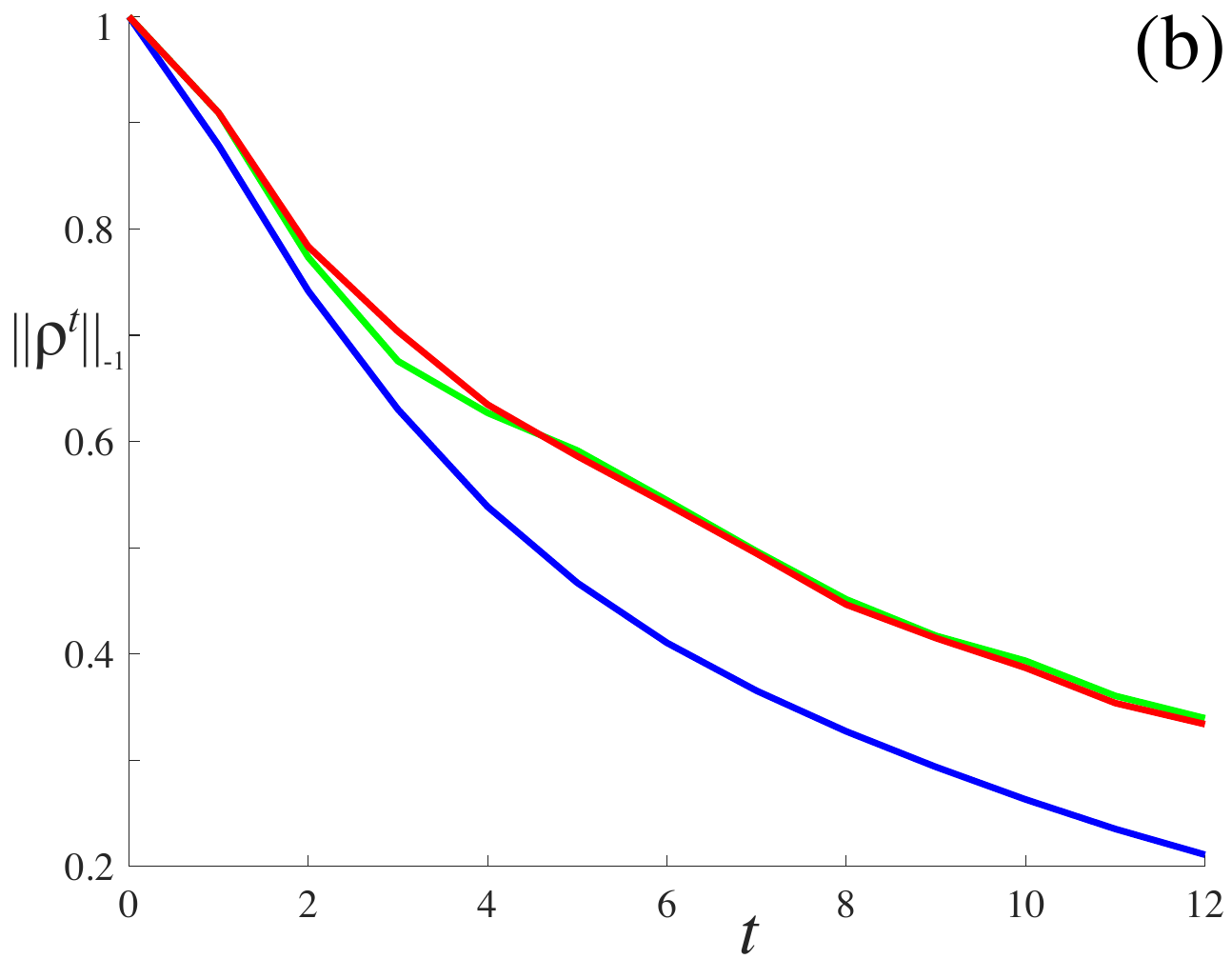}{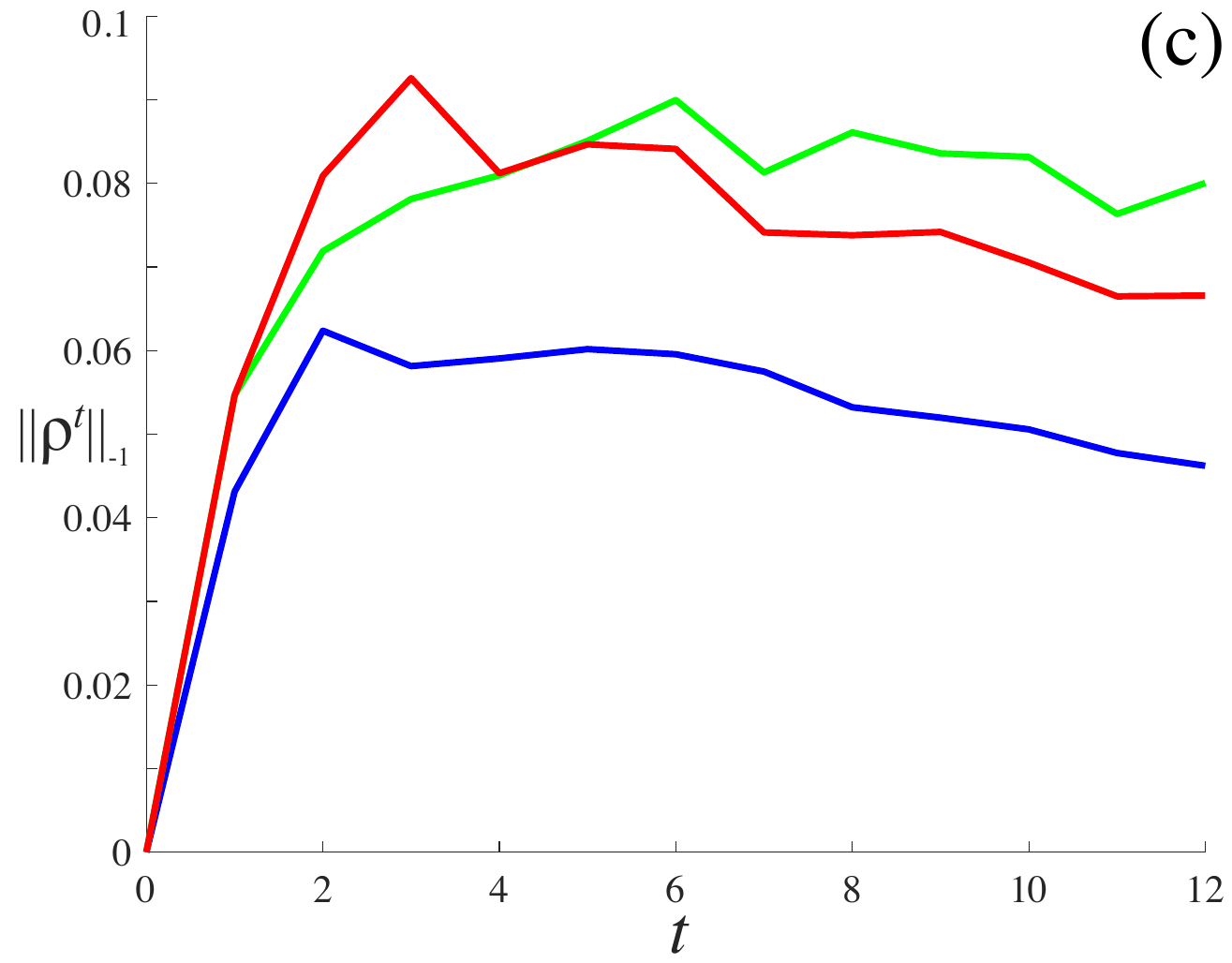}{(a) Distribution of $\Sob{\rho^T}$ using the three shearing techniques, for random choices of \Eq{TestConcentration} where $(u,v)\in[0,1)^2$, $\eps\in[0.1,5.0]$, and $n\in\{-10,10\}^2$. We use energy $E_H = 0.5$ and $T=12$. (b) Means and (c) standard deviations of $\Sob{\rho^t}$ for the CS SW, CS Full, and CS FS solvers.}{HarperShearFullHist}{2in}

Figure~\ref{fig:HarperShearFullHist} shows the performance of the algorithm over $10^5$ random choices of initial conditions \Eq{TestConcentration} normalized such that $\Sob{\rho^0}=1$. The minimum, mean, and standard deviation of $\Sob{\rho^T}$ for each distribution depicted in \Fig{HarperShearFullHist} can be found in \Tbl{HarperTableIC}. The scheme with sign switching beats out the one with fixed signs slightly in terms of mean final mix-norm. However, the SW minimizer outperforms both analytic solvers. The standard deviation of mix-norms using the sign switching methods is smaller than that for fixed signs. Indeed, the distribution of mix-norms for the fixed sign method has a right tail that extends nearly to $\Sob{\rho^T}  = 1$, suggesting that for some initial conditions, the solver performs extremely poorly and results in virtually no mixing. Using sign switching helps to reduce the size of this tail.

\begin{table}[h!t]
\centerline{
\begin{tabular}{l|c|c|c}
 & \multicolumn{3}{|c}{$\Sob{\rho^T}$}\\
 & {Optimal} & {Mean} & {Std.~Dev.}\\
\hline
Elliptic Constant & 0.1191 & 0.4905 & 0.1924\\
CS SW & 0.1085 & 0.2110 & 0.0462\\
CS FS & 0.1449 & 0.3395 & 0.0801\\
CS Full & 0.1449 & 0.3338 & 0.0666\\
\end{tabular}
}
\caption{{\footnotesize  Optimal, mean and standard deviations of the mix-norms at time $T=12$ for the Harper map from the distributions of \Fig{HarperShearFullHist}(a). The initial conditions are normalized such that $\Sob{\rho^0} = 1$.}
       \label{tbl:HarperTableIC}}
\end{table}

\section{Conclusions}\label{sec:Con}

We investigated the design of mixers by studying the optimization of stirring behavior of a two-dimensional device modeled by a finite number of nonlinear shears, under a constraint of fixed energy. The minus-one Sobolev seminorm,
one example of a ``mix-norm", provides an cost-effective way to measure mixing efficiency without explicitly computing diffusion. We computed the results of stirring by evaluating the Perron-Frobenius operator on a uniform grid, assuming that the initial density profile is known analytically. This allowed us to avoid the aliasing errors and numerically induced diffusivity of other methods. One of the most important insights is that effective mixing is highly dependent on the behavior of the lower order Fourier modes: in order to construct a good stirring scheme, one must target the low-order modes with the highest amplitudes.

For the first model, a composition of Chirikov standard maps, concentrating the nonlinear shear energy in a small number of steps was shown to give an efficient mixing scheme. For this map, the horizontal shear is assumed fixed, and using the energy for vertical shearing early in the protocol takes maximal advantage of the shift in Fourier coefficients to higher mode numbers.

For the Harper map model both vertical and horizontal shears can be controlled, allowing for more diverse stirring mechanisms and more interesting behaviors. A simplified scheme assumed that each step corresponds to either a horizontal or a vertical shear with constant energy. Within this reduced parameter space, we devised an algorithm that combines insight on the action of the shears in Fourier space with knowledge of the character of the fixed points, to produce a near-optimal stirring protocol. The method was shown to remain effective when applied to a variety of initial density profiles.

\Tbl{HarperTable} compares the optimal stirring protocols produced by the different methods when when applied to the initial condition $\Eq{InitialGaussian}$, and \Tbl{HarperTableIC} compares the overall performance of solvers for varying initial conditions \Eq{TestConcentration}. 
Our results suggest that both consistency and change---contradictory requirements---are necessary to produce an optimal stirring protocol. While the former leads to stretching near saddle points, the latter can result in enhanced folding. Both were required to obtain the optimal protocol for the Harper map. With only stretching, the mix-norm tends to level off; but with too much folding, stirring can be reversed, causing the mix-norm to grow. An optimal protocol will tend to use hyperbolic behavior if a smooth portion of the density is concentrated near a fixed point---if a nearby fixed point is elliptic, mixing efficiency can be decreased even though there will be (weak) shearing.

Though it is clear that both stretching and folding are required to design an effective mixer, it is less clear how to best combine these processes to optimize mixing. Observing stretching and folding in a wider variety of mixing problems may lead to more general results about how one should balance the two mechanisms to induce mixing.

The models that we studied here have their fixed points at fixed locations. Another stirring mechanism to investigate would be one that allows these locations to vary. For instance, one could introduce variable phases into the shears to control the locations of the fixed points at every step.
 We hope to use results gained from studying these simple, theoretical models to aid investigations of more physically applicable, three-dimensional models such as the partitioned pipe and rotated arc mixers \cite{Galaktionov03,Speetjens14}.

It would be worthwhile to compare the $H^{-1}$ norm to a model that implements diffusion. The effectiveness of our measure depends on the assumption that diffusion is small, i.e., that the P\'eclet number is large. In the future, we plan to include diffusion to get a sense of the errors induced by using the mix-norm for a purely advective model when the P\'eclet number is finite. 

\bibliographystyle{alpha}
\bibliography{mixing}

\end{document}